\documentclass[tightenlines,twocolumn,superscriptaddress,preprintnumbers,amssymb,nofootinbib]{revtex4-2}

\usepackage{amsmath, amsthm, amssymb, mathrsfs, bm, mathtools} 
\usepackage{graphics}
\usepackage{epsfig}
\usepackage{tabularx}
\usepackage{multirow}
\usepackage{graphicx}
\usepackage{wrapfig}
\usepackage{makecell}
\usepackage{color}
\usepackage{tikz}
\usepackage[hidelinks,colorlinks=true,linkcolor=blue,citecolor=blue]{hyperref}

\RequirePackage{subfigure}

\newcommand{\beq}{\begin{equation}}
\newcommand{\eeq}{\end{equation}}
\newcommand{\bqa}{\begin{eqnarray}}
\newcommand{\eqa}{\end{eqnarray}}

\newcommand{\ms}{\overline{\text{\tiny MS}}}

\newcommand{\Tr}{{\mathrm{Tr}}}
\newcommand{\cep}{\text{\tiny CEP}}
\newcommand{\tcp}{\text{\tiny TCP}}

\newcommand{\x}{x}
\newcommand{\y}{y}

\begin{document}

\title{Comparative analysis of critical regions: The renormalized quark-meson model under Polyakov loop, quark back-reaction, and vector interaction effects}

\author{Akanksha Tripathi}
\email{akankshatripathi330@gmail.com}
\affiliation{Department of Physics, University of Allahabad, Prayagraj, India-211002}
\author{Suraj Kumar Rai}
\email{surajrai050@gmail.com}
\affiliation{Department of Physics, Acharya Narendra Deo Kisan P.G. College, Babhnan Gonda-271313; Maa Pateswari University Balrampur,India-271201}
\author{Vivek Kumar Tiwari}
\email{vivekkrt@gmail.com}
\affiliation{Department of Physics, University of Allahabad, Prayagraj, India-211002}
\date{\today}

\begin{abstract}

The critical regions enveloping the critical end point (CEP) in the $\mu$-$T$ plane are mapped by computing the contours of normalized quark number susceptibility within the on-shell renormalized 2+1 flavor quark-meson (RQM) and Polyakov-loop enhanced renormalized Polyakov-quark-meson (RPQM) models for $m_\sigma = 400$ and 500 MeV. The apparent precision for the results of CEP coordinates merely reflects numerical binning of two decimal places rather than the effect of including full thermal and vacuum quantum fluctuations, as direct meson loops are omitted under a large-$N_c$ justification where quark-loop fluctuations dominate. The renormalized 't Hooft coupling $c$ becomes substantially stronger in the RQM model when the meson self-energies due to quark loops are computed using the pole masses of mesons and parameters are fixed on-shell in Ref.~\cite{vkkr23} after a consistent treatment of quark one-loop vacuum fluctuations, while the light and strange chiral symmetry breaking strengths also become weaker. We evaluate the impact of these novel features on critical fluctuations. Furthermore, the improved PolyLog-glue form of the Polyakov loop potential from Ref.~\cite{Haas} is employed to isolate the effects of the quark back-reaction on critical fluctuations, and the results are contrasted against back-reaction-free outcomes obtained using the logarithmic potential. Utilizing inputs from large-$N_c$ standard chiral perturbation theory, phase diagrams are also computed in the light chiral limit ($m_\pi = 0$), quantifying the proximity of the tricritical point (TCP) to the CEP. The critical regions from the RQM/RPQM models are compared with those reported in Ref.~\cite{schafwag12}, where curvature masses are used for parameter fixing. Phase diagrams incorporating vector interactions in the RQM/RPQM model reveal that the CEP and first-order transition survive up to a robust coupling of $g_\omega = 2.79$, rendering them highly relevant for compact star equations of state and astrophysical phenomenology.

\end{abstract}
\keywords{Dense QCD,
chiral transition,}
\maketitle
\section{Introduction}
Strong interaction theory known as the Quantum Chromodynamics (QCD),~has entirely different properties in different energy regimes.~The quarks and gluons,~as the basic units of QCD,~interact through weak coupling constant of color charge exchange for extremely high energies where  perturbation theory works.~The quark and gluons remain confined inside the hadrons in the low energy non-perturbative vacuum of QCD when the QCD coupling constant becomes large.~Hadrons being the low energy degrees of freedom of QCD,~get dissolved into their quark and gluon constituents for extremely high densities and temperatures, and such QCD phase transition~\cite{Cabibbo75,SveLer,Mull,Ortms,Riske} yields the color conducting plasma of quarks and gluons (QGP).~The QCD phase transition studies are highly relevant for understanding the early universe evolution,~the structure of compact stars,~and the nature of the strong interaction.~Since the theory is non-perturbative,~the theoretical constructions of QCD phase transition rely mostly  upon the two important properties of QCD called the phenomenon of spontaneous breaking of chiral symmetry and color charge confinement.~The first principal Lattice QCD simulations performed at zero baryon densities~\cite{AliKhan:2001ek,Digal:01,Fodor:03,Allton:05,Karsch:05,Aoki:06,Cheng:06,Cheng:08,JLange,ABAZ,Ejiri.Lat}~reveal important information and insights for the mapping of QCD phase structure in terms of  chiral symmetry broken and restored phases as well as color charge confined and deconfined phases.~But the LQCD simulations get severely compromised at non-zero densities due to the fermion sign problem \cite{Karsch:02}.~The QCD-like models~\cite{Alf,Fukhat,FejosI,FejosII,FbRenk,Fejos3,Fejos4},~constructed in terms of effective degrees of freedom respecting different QCD symmetries,~constitute the required framework for the study of QCD phase diagram and associated critical phenomenon.

The effective theory framework of the 2+1 flavor quark-meson (QM) model~\cite{Schaefer:09} couples the light $u, d$ quarks and the $s$ quark to the nine scalar and nine pseudoscalar mesons of the three-flavor linear sigma model~\cite{Ortman,Rischke:00,Lenagh}. The 1~GeV mass of the heavy $\eta'$ meson is due to the explicit breaking of the axial $U_A(1)$ symmetry, caused by instanton effects at the quantum level~\cite{tHooft:76prl}. The extensively used Nambu--Jona-Lasinio (NJL) model~\cite{costaA,costaB,fuku08} provides an equivalent description where mesons are not treated as fundamental fields in the Lagrangian, but are instead generated through multi-quark point interactions. Furthermore, these chiral models are coupled to a constant background $SU_c(N)$ gauge field~\cite{SveLer,Polyakov:78plb,benji,BankUka,Pisarski:00prd,fuku,Vkt:06} to mimic the physics of QCD color confinement in a statistical sense. Adding the free energy density from gluons, in the form of a Polyakov-loop potential~\cite{ratti,Roesnr}, to the QM model effective potential yields the unified framework of the Polyakov-loop-augmented quark-meson (PQM) model~\cite{SchaPQM2F,SchaPQM3F,Schaefer:09wspax,Mao,TiPQM3F}. Within this framework, the physics of both chiral symmetry breaking/restoration and the confinement-deconfinement transition can be studied. Recent studies in Refs.~\cite{Haas,Redlo,BielichP,Herbst,THerbst2} have improved the Polyakov-loop potential from a pure gauge potential to an unquenched glue potential, where the effects of the quark back-reaction are also included. This improvement gives rise to a linkage between the confinement-deconfinement and chiral phase transitions at low temperatures and large chemical potentials~\cite{BielichP}.

Lattice QCD~\cite{Wupertal2010,WB2014,HotQCD2014} establishes that the $\mu=0$ transition at the physical point is a crossover, while effective theories~\cite{rob} predict a first-order transition at lower $T$ and higher $\mu$. Thus, the $\mu-T$ plane phase diagram must feature a critical endpoint (CEP)~\cite{Asak,Bard,Berg} where the crossover terminates into a first-order line. The static critical fluctuations at this CEP belong to the 3D Ising $Z(2)$ universality class~\cite{Hatta,Fuji,KFuku}. In the light chiral limit ($m_{ud}=0$), the $\mu=0$ transition becomes second-order, falling into the 3D $O(4)$ universality class~\cite{rob}. Consequently, the existence of a CEP at the physical point implies a tricritical point (TCP) in the light chiral limit, where the $O(4)$ second-order line merges into a first-order line. A similar TCP occurs in the $\mu-T$ plane for (2+1)-flavor QCD when the strange quark mass is kept physical while the light quarks are massless. The mass dependence of the chiral phase transition order is conventionally mapped via the Columbia plot in the $m_{ud}$--$m_s$ (or $m_\pi$--$m_K$) plane~\cite{columb,vktChpt1,vktChpt2}. At $\mu=0$, the physical point lies in a smooth crossover region bounded by a small-mass first-order region that terminates on a second-order line of $Z(2)$ critical points. In the standard scenario--assuming the axial $U_A(1)$ symmetry remains broken--this first-order region originates from the $SU(3)$ chiral limit ($m_{ud}=m_s=0$) and terminates at a TCP on the massless light-quark axis. For non-zero $m_{ud}$, the first-order transition ends at a line of $Z(2)$ critical endpoints. As $\mu$ increases, the first-order region expands, shifting the $Z(2)$ boundaries to higher masses and ultimately giving rise to the CEP at the physical point. Alternatively, exotic axial $U_A(1)$ restoration dynamics in the two-flavor case could trigger a first-order transition across the entire light-quark mass spectrum, radically altering the critical surfaces~\cite{rob,Cute}. Conversely, a recent lattice study by Cuteri et al.~\cite{Cute} finds a second-order chiral transition in the $N_f=3$ chiral limit. Dyson-Schwinger approaches~\cite{bernhardt23}, conformal bootstrap studies~\cite{kousvos22}, and other lattice studies similarly report a second-order transition in the small-mass region. If this first-order region shrinks or vanishes entirely in the continuum limit, the physical point CEP at higher $\mu$ would not be influenced by a TCP in the light chiral limit.

Early signatures of the CEP were proposed and discussed in Refs.~\cite{Son,Misha3,Kraja2,Jeon,Ejik}. To investigate the QCD phase diagram and search for the CEP, the multi-step Beam Energy Scan (BES) program was launched at RHIC (BNL, USA) in 2010~\cite{AdamAg,Odyniec:2019XZ}. The first two phases, BES-I and BES-II, concluded in 2014 and 2021, respectively~\cite{Cliu}. Expected CEP signatures, such as the baryon density susceptibility and factorial cumulants of proton multiplicity fluctuations, were experimentally measured in BES-II~\cite{APandav,MishaConf}. The theoretical aspects of CEP physics are comprehensively reviewed in Refs.~\cite{Bzda,Ldu}. Identifying the CEP relies on large event-by-event fluctuations of conserved quantities---such as net charge, baryon number, and strangeness~\cite{MishaPRL,Krajamisha,bedang,Misha11}---whose experimentally measurable higher-order moments link directly to thermodynamic susceptibilities. Tracking this signal requires a deep understanding of the nonequilibrium dynamics of non-Gaussian fluctuations near the  CEP~\cite{xin,Akam,vovchen,xan,Pradeep1,Parot,Pradeep2}. Consequently, a new approach based on the maximum entropy principle has been developed to address the freeze-out of these non-Gaussian fluctuations both in and out of equilibrium~\cite{Pradeep1,Pradeep2}.

Various aspects of the QCD phase transition have been extensively studied using chiral models~\cite{Rischke:00,Lenagh,Roder,fuku11,grahl,jakobi,Herpay:05,Herpay:06,Herpay:07,Kovacs:2006ym,kahara,Bowman:2008kc,Fejos,Jakovac:2010uy,koch,marko,GFejo} and two- and (2+1)-flavor QM/PQM models~\cite{Schaefer:09,SchaPQM2F,SchaPQM3F,Schaefer:09wspax,Mao,TiPQM3F,scav,mocsy,bj,Schaefer:2006ds}. Under the standard mean-field approximation (s-MFA), neglecting quark one-loop vacuum fluctuations incorrectly predicts a first-order chiral transition at zero baryon density in the chiral limit, contradicting theoretical arguments~\cite{rob,hjss}. This inconsistency was resolved in Ref.~\cite{vac} by including these vacuum fluctuations in the two-flavor QM model. While fermionic vacuum corrections had been explored earlier in finite-temperature and density Yukawa theory~\cite{Fraga1,Fraga2,Fraga3}, subsequent studies confirmed that the quark one-loop vacuum term strongly influences QCD thermodynamics and phase structure within QM/PQM models~\cite{vac,Gatto,Anna,lars,guptiw,chatmoh1,vkkr12,schafwag12,TranAnd,Dima,vkkt13,Weise1,Herbst,Weyrich,kovacs,
zacchi1,zacchi2,Rai,Weise3}. However, these earlier studies fixed their parameters using curvature masses, derived from the second derivative of the effective potential at its minimum. Because the effective potential generates $n$-point functions at zero external momentum, curvature masses only account for vacuum corrections to meson self-energies at zero momentum, rendering this parameter-fixing approach inconsistent for QM/PQM models with the vacuum term (QMVT/PQMVT). Instead, multiple studies emphasize that meson pole masses are the true physical, gauge-invariant quantities~\cite{Kobes,Rebhan,laine,BubaCar,fix1,Naylor,Adhiand1}. Consequently, a consistent treatment of quark one-loop vacuum fluctuations was recently developed for the two-flavor~\cite{Adhiand1,Adhiand2,Adhiand3,asmuAnd,RaiTiw22,raiti23} and (2+1)-flavor~\cite{vkkr23,skrvkt24,vktChpt1,vktChpt2,Joand,Gholami} QM/PQM models. In these on-shell renormalized (RQM) and Polyakov-loop-enhanced (RPQM) models, parameters are properly fixed by matching counter-terms in the on-shell scheme with those in the modified minimal subtraction ($\overline{\text{MS}}$) scheme, relating the mass parameter and running couplings directly to the physical pole masses of the $\pi, K, \eta, \eta'$, and $\sigma$ mesons.




In the RQM model~\cite{vkkr23}, the 't Hooft coupling $c$ is significantly enhanced because the condensate-dependent part of the $U_A(1)$ anomaly term is modified when quark-loop meson self-energies are calculated using pole masses. Furthermore, the explicit symmetry-breaking strengths $h_x$ and $h_y$ are weakened by small and relatively large amounts, respectively. Consequently, the CEP shifts upward in the RQM phase diagram~\cite{vkkr23,skrvkt24}. Here, unlike in QMVT models~\cite{guptiw, vkkr12, chatmoh1, schafwag12}, the quark one-loop vacuum fluctuations exert only a moderate smoothing effect on the chiral phase transition. This upward shift is further amplified by the Polyakov-loop potential in the RPQM model~\cite{skrvkt24}. Located at higher temperatures than the CEP in Schaefer et al.'s curvature-mass-parameterized PQMVT model~\cite{schafwag12}, the RPQM model CEP~\cite{skrvkt24} aligns  closer to recent theoretical consensus. Various techniques--including lattice-based Padé resummations of the Taylor expansion in $\mu_B=3\mu_f$~\cite{Clark}, conformal maps~\cite{Basar}, hybrid lattice/gauge-gravity approaches~\cite{Hipp}, and the functional renormalization group~\cite{Lugao}--all consistently locate the critical point within the same region of the QCD phase diagram: $T_c \sim 100\text{--}110$~MeV and $\mu_B \sim 420\text{--}650$~MeV (corresponding to $\mu_\text{CEP} \sim 140\text{--}217$~MeV)~\cite{MishaConf}. Given these findings, it is important to map the critical regions around the CEP in both the RQM and RPQM models.

Extending the recent work in Ref.~\cite{skrvkt24}, the present study pursues three primary goals. Our first objective will be to compute the critical regions around the RPQM model critical endpoints (CEP) in the already existing phase diagrams in Ref.~\cite{skrvkt24} for $m_\sigma=400$ and $500$ MeV. The critical regions around the CEP will be mapped by computing contours of enhanced normalized quark number susceptibilities. We will compute the RQM/RPQM phase diagrams afresh for the light chiral limit ($m_\pi=0$) and locate the tricritical point (TCP) for $m_\sigma=400$ and $500$ MeV. We will employ both the Log form (without quark back-reaction) and the improved PolyLog-glue form (with quark back-reaction) of the Polyakov-loop potential in the RPQM model. For the light chiral limit ($m_\pi=0$, $m_K=496$ MeV), large-$N_c$ standard chiral perturbation theory inputs~\cite{herrPLB,Escribano,vktChpt1} will be used to determine the modified pion and kaon decay constants, $f_\pi$ and $f_K$. We will systematically analyze how the size and shape of the critical regions change as $m_\sigma$ varies from $400$ to $500$ MeV across different model settings. By evaluating the proximity of the TCP to the CEP, we will determine whether the TCP influences the critical fluctuations around the CEP. The isolated effect of properly treated quark one-loop vacuum corrections will become apparent by comparing the RQM model results with those of the QM model, while the combined effect of these vacuum corrections and the Polyakov-loop potential will be obtained by comparing the results of the RQM and RPQM models. Our second goal will be to evaluate how different treatments of quark one-loop vacuum fluctuations affect the size and shape of the critical regions. To this end, we will compare our RQM/RPQM results against the curvature-mass-parameterized QMVT/PQMVT findings of Schaefer et al. for $m_\sigma=400$ MeV~\cite{schafwag12}, which utilized the Log Polyakov-loop potential with $T_0=270$ MeV for the pure Yang-Mills $SU_c(3)$ gauge theory.

The significant impact of the repulsive vector-type interactions on the CEP was shown in early NJL model studies~\cite{Asak}.~Yukawa coupling of quarks (nucleons) to vector mesons gives repulsive forces~\cite{Hatsvect}, whereas coupling to scalar mesons is attractive.~Some effective model studies reported  appearance of two and  multiple CEPs \cite{Bowman:2008kc,koch,Kitaz,Hastprl} due to the effect of
vector interactions.~Since the strength of vector interaction is not known a priori, several of the NJL/PNJL model studies \cite{fuku08,Kitaz,HAbuki,Kunihiro,Csasaki,KJ4019,Sakai,Bratovic,Blaschke,Hell,HellII} varied the vector to scalar coupling strength ratio as a free parameter, \(r = G_V/G_S \equiv 0 \text{ to }1\), and explored its effect on the position of the CEP in the phase diagram. Our third goal is to study the effects of vector interactions positions of CEP and the extent of critical fluctuations around the CEPs in the 2+1 flavor RQM/RPQM  model.~We will be computing different  phase diagrams for the RQM and RPQM model with and without quark back reaction in the Polyakov loop potential when the strengths of  vector coupling $g_{\omega}$ are successively increased.~Since the RQM model has the propensity to generate larger extent of first order transition line in the phase diagram,~it will be very interesting to find the critical vector coupling  strength for which phase transition becomes crossover in the whole $\mu-T$ plane. Determining the critical vector coupling strength would have important implications for the compact star equation of state and related astrophysical phenomenology. We will primarily use \(m_\sigma = 500\) MeV in this work because the RPQM model \cite{skrvkt24} with this value closely reproduces the LQCD data for temperature variations of thermodynamic quantities and the scaled chiral condensate (\(\Delta _{ls}\)).~Results with \(m_\sigma = 400\) MeV are also computed for comparing with those in Ref.~\cite{schafwag12}.

The paper is organized as follows.~Section~\ref{sec:II} gives a brief recapitulation of the QM/PQM/RQM/RPQM model.~The incorporation of vector interaction is described in Section~\ref{sec:IIA},~whereas Section~\ref{sec:IIB} contains different forms of Polyakov-loop potentials.~Section~\ref{sec:IIC} gives the RPQM model grand potential.~The light chiral limit formula and fixing of model parameters are discussed in section~\ref{sec:IID}.~Section~\ref{sec:III} contains results and discussion.~The phase diagrams,~CEP and TCP are discussed in section~\ref{sec:IIIA},~size and shape of the critical regions in different scenarios are compared in the section~\ref{sec:IIIB},~RQM/RPQM model critical regions are compared with those of the  QMVT/PQMVT model in the section~\ref{sec:IIIC}.~The effect of different strengths of vector interaction on phase diagram, CEP position and regions of critical fluctuations around CEP are described in \ref{sec:IIID}. ~Section~\ref{sec:IV} presents the summary. 


\section{Model Recapitulation}
\label{sec:II}
The Polyakov loop augmented quark meson (PQM) model \cite{TiPQM3F,Schaefer:09wspax,Mao,SchaPQM3F} is obtained when three flavor of quarks are coupled to the $SU_V(3) \times SU_A(3)$ symmetric meson fields and the 
temporal component of gauge field represented by the Polyakov loop 
potential.~The Polyakov loop field $\Phi$ is defined by the thermal expectation value of color trace of Wilson loop in temporal 
direction  as :
\begin{equation}
\Phi(\vec{x}) = \frac{1}{N_c} \langle \Tr_c L(\vec{x})\rangle, \qquad \qquad  \bar\Phi(\vec{x}) =
\frac{1}{N_c} \langle \Tr_c L^{\dagger}(\vec{x}) \rangle
\end{equation}
here $L(\vec{x})$ is a $SU_c(3)$ color gauge group matrix in its fundamental representation.
\begin{equation}
\label{eq:Ploop}
L(\vec{x})=\mathcal{P}\mathrm{exp}\left[i\int_0^{\beta}d \tau
A_0(\vec{x},\tau)\right]
\end{equation}
Where $\mathcal{P}$ is path ordering,  $A_0$ is the temporal component of vector 
field and $\beta = T^{-1}$ \cite{Polyakov:78plb}.~As in the 
Ref. \cite{ratti,Roesnr},~the homogeneous Polyakov loop fields $\Phi(\vec{x})=\Phi$=constant and
$\bar\Phi(\vec{x})=\bar\Phi$=constant.
 
The Lagrangian of Polyakov quark-meson (PQM) model contains the Lagrangian of quark meson model and the Polyakov loop potential ${\cal U} \left( \Phi, \bar\Phi, T \right)$ as :
\bqa
{\cal L_{PQM}}&=&{\cal L_{QM}}-{\cal U} \big( \Phi , \bar\Phi , T \big).
\label{lag:PQM}
\eqa
The Lagrangian  of QM model \cite{Rischke:00,Schaefer:09,TiPQM3F} is the following:
\bqa
\label{lag}
{\cal L_{QM}}&=&\bar{\psi}[i\gamma^\mu \partial_\mu- g\; T_a\big( \sigma_a 
+ i\gamma_5 \pi_a\big) ] \psi+\cal{L(M)}\;.
\eqa

The derivative $\partial_\mu$ in the Eq.~(\ref{lag}) will be replaced by the covariant derivative $ D_{\mu}$ when the expression of the ${\cal L_{QM}}$ is substituted in the Eq.~(\ref{lag:PQM}) to define the PQM model Lagrangian.~Quarks couple with the uniform temporal background gauge field in the PQM model as the following $D_{\mu} = \partial_{\mu} -i A_{\mu}$ 
and  $A_{\mu} = \delta_{\mu 0} A_0$ (Polyakov gauge), where $A_{\mu} = g_s A^{a}_{\mu} \lambda^{a}/2$ with vector potential $A^{a}_{\mu}$ for color gauge fields.~$g_s$ is the $SU_c(3)$ gauge coupling and $\lambda_a$ ($a=1 \cdots 8$) are Gell-Mann matrices in the color space.~The Lagrangian for the $SU_V(3) \times SU_A(3)$ symmetric meson fields is the following.
\bqa
\nonumber
\label{lagM}
\cal{L(M)}&=&\text{Tr} (\partial_\mu {\cal{M}}^{\dagger}\partial^\mu {\cal{M}}-m^{2}({\cal{M}}^{\dagger}{\cal{M}}))\\ \nonumber
&&-\lambda_1\left[\text{Tr}({\cal{M}}^{\dagger}{\cal{M}})\right]^2-\lambda_2\text{Tr}({\cal{M}}^{\dagger}{\cal{M}})^2\\ 
&&+c[\text{det}{\cal{M}}+\text{det}{\cal{M}}^\dagger]+\text{Tr}\left[H({\cal{M}}+{\cal{M}}^\dagger)\right].\
\eqa
The flavor blind Yukawa coupling $g$ couples the flavor triplet quark fields $\psi$ (color $N_c$-plet  Dirac spinor) to the nine scalar and pseudo-scalar meson $\xi$ fields $\sigma_a (\pi_a$) of $3\times3$ complex matrix ${\cal{M}}=T_{a} \xi_{a}=T_{a}(\sigma_{a}+i\pi_{a})$.~$T_{a}=\frac{\lambda_{a}}{2}$ where $\lambda_a$ ($a=0,1 \cdots 8$) are Gell-Mann matrices with $\lambda_0=\sqrt{\frac{2}{3}}{\mathbb I}_{3\times3}$.~The $\xi$ field  develops the non-zero vacuum expectation value (VEV) $\overline{\xi}$ in the $0$ and $8$ directions.~The $SU_L(3) \times SU_R(3)$ chiral symmetry is broken spontaneously  by the 
condensates $\bar{\sigma_0}$ and $\bar{\sigma_8}$ and it is broken explicitly also by 
the external fields $H= T_{a} h_{a}$ with $h_0$, $h_8  \neq 0$.~The change from the singlet octet $(0,8)$  to the non-strange strange basis $(\x,\y)$ gives $\x ( h_{x})= \sqrt{\frac{2}{3}}\bar{\sigma}_0 (h_{0}) +\frac{1}{\sqrt{3}} \bar{\sigma}_8(h_{8})$ and $ \y (h_{y}) = \frac{1}{\sqrt{3}}\bar{\sigma}_0 (h_{0})-\sqrt{\frac{2}{3}}\bar{\sigma}_8 (h_{8})$.~Considering the thermal and quantum fluctuations of the quarks/anti-quarks and treating the mesons at mean field level,~the QM model grand potential \cite{Schaefer:09,TiPQM3F},~is obtained after adding the vacuum effective potential $ U(\x,\y) $ with the quark/anti-quark  contribution $\Omega_{q\bar{q}}$ at finite temperature $T$ and quark chemical potential $\mu_{f} (f=u,d,s)$.
\bqa
\label{Grandpxy}
&&\Omega_{\rm MF }(T,\mu)=U(\x,\y)+\Omega_{q\bar{q}} (T,\mu;\x,\y)\;. \\ \nonumber \\ 
&&\Omega_{q\bar{q}}(T,\mu;\x,\y)=\Omega_{q\bar{q}}^{vac}+\Omega_{q\bar{q}}^{T,\mu}(x,y). \\ \nonumber
\label{eq:mesop}
&&U(\x,\y)=\frac{m^{2}}{2}\left(\x^{2} +
\y^{2}\right) -h_{x} \x -h_{y} \y
- \frac{c}{2 \sqrt{2}} \x^2 \y \\  
&&+ \frac{\lambda_{1}}{2} \x^{2} \y^{2}+
\frac{1}{8}\left(2 \lambda_{1} +
\lambda_{2}\right)\x^{4} 
+\frac{1}{8}\left(2 \lambda_{1} +
2\lambda_{2}\right) \y^{4}\;. \\ \nonumber \\
\label{vac1}
&&\Omega_{q\bar{q}}^{vac} =- 2 N_c\sum_f  \int \frac{d^3 p}{(2\pi)^3} \ E_q \  \theta( \Lambda_c^2 - \vec{p}^{2})\;.\\ \nonumber \\
\label{vac2}
\nonumber
&&\Omega_{q\bar{q}}^{T,\mu}(x,y)=- 2 N_c \sum_{f=u,d,s} \int \frac{d^3 p}{(2\pi)^3} T \left[ \ln g_f^{+}+\ln g_f^{-}\right].\; \\  \\
\label{GrandQM}
&&\Omega_{\rm QM }(T,\mu,x,y)=U(\x,\y)+\Omega_{q\bar{q}}^{T,\mu}(x,y)\;. 
\eqa
The $ g^{\pm}_f = \left[1+e^{-E_{f}^{\pm}/T}\right] $ where $E_{f}^{\pm} =E_f \mp \mu_{f}$ and $E_f=\sqrt{p^2 + m{_f}{^2}}$ is the quark/anti-quark energy.~The light (strange) quark mass  $m_{u/d}=\frac{g\x}{2}$ ($m_{s}=\frac{g\y}{\sqrt{2}}$) and $\mu_{u}=\mu_{d}=\mu_{s}=\mu$.~In the standard mean field approximation (s-MFA) where  Dirac sea contribution is neglected,~the quark one-loop vacuum term with ultraviolet cut-off $\Lambda_c$ in Eq.~(\ref{vac1}) is dropped in the quark meson (QM) model grand effective potential in Eq.~(\ref{GrandQM}).~Several studies have used the minimal subtraction scheme to regularize the  quark one-loop vacuum divergences after including the vacuum fluctuations in the extended mean field approximation (e-MFA) \cite{vac, lars,guptiw, vkkr12, chatmoh1, schafwag12,  vkkt13, Rai} but the vacuum effective potential 
$\Omega_{vac} (\x,\y)=U(\x,\y)+\Omega_{q\bar{q}}^{vac}$
in their treatment,~turns inconsistent as they fix the model parameters using curvature masses of mesons which are obtained by taking the
double derivatives of the effective potential with respect
to the different fields at its minimum.~The above mentioned inconsistency  becomes apparent when one notes that the calculation of the curvature masses,~involves the evaluation of the meson self-energies at zero momentum because the effective potential is the generator of the n-point functions of the theory at vanishing external momenta~\cite {laine,Adhiand1,Adhiand2,Adhiand3,asmuAnd,RaiTiw22,raiti23,vkkr23,skrvkt24}.~The Refs.~\cite{Kobes, Rebhan} has emphasized that the pole definition of the meson mass is the physical and gauge invariant one.

The consistent e-MFA RQM model effective potential  
will be used in the present work which is derived in our very recent works \cite{vkkr23,skrvkt24} after relating the counter-terms in the $\overline{\text{MS}}$ scheme to those in the on-shell (OS) scheme \cite{Adhiand1,Adhiand2,Adhiand3,asmuAnd,RaiTiw22,raiti23}.~The relations between the renormalized parameters of both the schemes are determined when the physical quantities ( the on-shell pole masses of the $m_{\pi}, m_{K}, m_{\eta},m_{\eta^{\prime}} \ \text{and} \ m_{\sigma}$,~the pion and kaon decay  constants $f_{\pi}$ and $f_{K}$ )  are put into the relation of the $\overline{\text{MS}}$ running  couplings and mass parameter.~These relations are used as input when the effective potential is calculated using the modified minimal subtraction procedure.~After the cancellation of the  $ 1 / \epsilon$ divergences,~the vacuum effective potential 
$ \Omega_{vac}=U(x_{\ms},y_{\ms})+\Omega^{q,vac}_{\ms}+\delta U(x_{\ms},y_{\ms})$ in the $\overline{\text{MS}}$ scheme has been rewritten in Ref. \cite{vkkr23} in terms of the scale $\Lambda$ independent constituent quark mass parameters $\Delta_{x}=\frac{g_{\ms} \ x_{\ms}}{2}$ and $\Delta_{y}=\frac{g_{\ms} \ y_{\ms}}{\sqrt{2}}$ as the following.
\begin{align}
\label{OmegDelxy}
\nonumber 
&\Omega_{\rm vac}(\Delta_{x},\Delta_{y})=\frac{m^2_0}{g^2_0}(2\Delta_{x}^2+\Delta_{y}^2)-2\frac{h_{x0}}{g_0}\Delta_{x}-\sqrt{2}\frac{h_{y0}}{g_0}\Delta_{y}\\ \nonumber 
&-2\frac{c_{0}}{g^3_{0}}\Delta_{x}^2 \ \Delta_{y}+4\frac{\lambda_{10}}{g^4_{0}}\Delta_{x}^2 \ \Delta_{y}^2+2\frac{(2 \lambda_{10}+\lambda_{20})}{g^4_{0}}  \Delta_{x}^{4} \ \\ \nonumber 
&+\frac{(\lambda_{10}+\lambda_{20})}{g^4_{0}}  \Delta_{y}^{4}+
\frac{2N_c\Delta_{x}^4}{(4\pi)^2}\left[\frac{3}{2}+\ln\left(\frac{\Lambda^2}{m_{u}^2}\right)+\ln\left(\frac{m_{u}^2}{\Delta_{x}^2}\right)\right] \\ 
&+\frac{N_c\Delta_{y}^4}{(4\pi)^2}\left[\frac{3}{2}+\ln\left(\frac{\Lambda^2}{m_{u}^2}\right)+\ln\left(\frac{m_{u}^2}{\Delta_{y}^2}\right)\right]\;.  
\end{align}
The scale $\Lambda_0$ gets fixed by requiring that the minimum of the RQM model effective potential does not shift from that of the QM model~\cite{vkkr23,skrvkt24} as :
\bqa
\label{Sclcond}
&\ln\left(\frac{\Lambda^2_0}{m_u^2}\right)+\mathcal{C}(m^2_\pi)+m^2_\pi \mathcal{C}^{\prime}(m^2_\pi)=0\;. 
\eqa
The terms $\mathcal{C}(m^2_\pi) $ and $ \mathcal{C}^{\prime}(m^2_\pi) $ and the derivations for the renormalized parameters  $m^{2}_{0}=(m^{2}+m^2_{\text{\tiny{FIN}}})$,~$h_{x0}=(h_{x}+h_{x\text{\tiny{FIN}}})$,~$h_{y0}=(h_{y}+h_{y\text{\tiny{FIN}}}) $,~$\lambda_{10}=(\lambda_{1}+\lambda_{1\text{\tiny{FIN}}})$,~$\lambda_{20}=(\lambda_{2} +\lambda_{2\text{\tiny{FIN}}})$ and $c_{0}=(c+c_{\text{\tiny{FINTOT}}})$ are given in detail in Refs. \cite{vkkr23,skrvkt24}.~The $m^2_{\text{\tiny{FIN}}}$,
$ h_{x\text{\tiny{FIN}}} $, $ h_{y\text{\tiny{FIN}}} $, $\lambda_{1\text{\tiny{FIN}}} $, $ \lambda_{2\text{\tiny{FIN}}}$ and 
$c_{\text{\tiny{FINTOT}}}$ are the finite on-shell corrections in the parameters at the scale $\Lambda_0$.~The experimental values of the pseudo-scalar meson masses $m_{\pi}$, $m_{K}$, $m_{\eta}$, $m_{\eta^{\prime}} $ ($m_{\eta}^2+m_{\eta^{\prime}}^2$), the scalar $\sigma$ mass $m_{\sigma}$ and the $f_{\pi}$, $f_K$ as input determine  the tree level QM model quartic couplings  $\lambda_1$, $\lambda_2$,~mass parameter  $m^2$,~$h_x$, $h_y$ and the coefficient $c$ of the t'Hooft determinant term for the $U_A(1)$ axial anomaly \cite{Rischke:00,Schaefer:09}.~The parameter determination is explained in Refs.~\citep{vkkr23,skrvkt24} also.

~The dressing of the meson propagator in the on-shell scheme gives rise to the renormalization of $f_{\pi}$, $f_{K}$ and $g$,~but they do not change as the $g_{\ms}=g_{ren}=g_{0}=g$, $x_{\ms}=x$, $y_{\ms}=y$ at $\Lambda_0$.~Further  $x_{\ms}=f_{\pi,ren}=f_\pi$ and $y_{\ms}=\frac{2f_{K,ren}-f_{\pi,ren}}{\sqrt{2}}$= $\frac{2f_K-f_\pi}{\sqrt{2}}$ at the minimum.~Using $\Delta_{x}=\frac{g \ x}{2}$ and $\Delta_{y}=\frac{g \ y}{\sqrt{2}}$,~the vacuum effective potential in the Eq.~(\ref{OmegDelxy}) can be  written in terms of the $x$ and $y$ as :
\begin{align}
\label{vacRQM}
\nonumber
&\Omega_{vac}^{\rm RQM}(\x,\y)=\frac{(m^{2}+m^2_{\text{\tiny{FIN}}})}{2} \ \left(\x^{2} +
\y^{2}\right)-(h_{x}+h_{x\text{\tiny{FIN}}}) \ \x 
\\ \nonumber 
&-(h_{y}+h_{y\text{\tiny{FIN}}}) \y -\frac{(c+c_{\text{\tiny{FINTOT}}})}{2 \sqrt{2}} \x^2 \y + \frac{(\lambda_{1}+\lambda_{1\text{\tiny{FIN}}})}{2} \x^{2} \y^{2}
\\ \nonumber
&+\frac{\left\{2(\lambda_{1} +\lambda_{1\text{\tiny{FIN}}})+
( \lambda_{2} +\lambda_{2\text{\tiny{FIN}}})\right\}\x^{4}}{8}+\left( \lambda_{1} +\lambda_{1\text{\tiny{FIN}}}+\lambda_{2}
 \right.  \\  \nonumber
& \left. +\lambda_{2\text{\tiny{FIN}}}\right)\frac{ \y^{4}}{4} +\frac{N_c g^4 (\x^4+2\y^4)}{8(4\pi)^2} \left[\frac{3}{2}-\mathcal{C}(m^2_\pi)-m^2_\pi \mathcal{C}^{\prime}(m^2_\pi)\right]\ \\
&-\frac{N_c g^4 }{8(4\pi)^2} \left[\x^4 \ln\left(\frac{\x^2}{f_{\pi}^2}\right)+2\y^4 \ln\left(\frac{2 \  \y^2}{f_{\pi}^2}\right) \right]. \\ \nonumber  \\ 
\label{grandRQM}
&\Omega_{\rm RQM }(T,\mu,x,y)=\Omega_{vac}^{\rm RQM}(\x,\y)
+\Omega_{q\bar{q}}^{T,\mu}(x,y)\;.
\end{align}
The $T$ and  $\mu$ dependence of the $ \x$ and $ \y$ is obtained after searching the minima of grand effective potential in the  Eq.~(\ref{grandRQM}) as  $\frac{\partial \Omega_{\rm RQM}}{\partial \x}= \frac{\partial \Omega_{\rm RQM}}{\partial \y}=0$.
~The expressions of  curvature masses  of the scalar and pseudo-scalar mesons for the RQM model are derived in Refs.~\cite{vkkr23,vktChpt1}.~The equations of motion $\frac{\partial \Omega_{vac}^{\rm RQM}}{\partial \x}=0= \frac{\partial \Omega_{vac}^{\rm RQM}}{\partial \y}$ obtained from the Eq.~(\ref{vacRQM}),~give the following expressions for the renormalized chiral symmetry breaking strengths $h_{x0}$ and $h_{y0}$.
\bqa
\label{hx0}
&h_{x0}=m_{\pi,c}^2 \ f_{\pi}. \\
\label{hy0}
&h_{y0}=\biggl(\sqrt{2} \ f_K \ m^2_{K,c}-\frac{f_{\pi}}{\sqrt{2}} \ m^2_{\pi,c}\biggr).
\eqa 
The expressions of the pion and kaon curvature masses $m_{\pi,c}$ and $m_{K,c}$ are the following.
\begin{align}
\label{mpicr}
&m_{\pi,c}^2=m_{\pi}^2\biggl\{ 1-\frac{N_{c}g^2}{4\pi^2} \ m_{\pi}^2 \ \mathcal{C}^{\prime}(m^2_\pi,m_u)  \biggr\}. \\ \nonumber
\label{kcr}
&m_{K,c}^2= m_{K}^2 \biggl[ 1-\frac{N_{c}g^2}{4\pi^2} \biggl\{  \mathcal{C}(m^2_\pi,m_u)+m^2_\pi \ \mathcal{C}^{\prime}(m^2_\pi,m_u)  \\ \nonumber
&\hspace*{.8cm}-\biggl(1-\frac{(m_s-m_u)^2}{m_K^2}\biggr)\ \mathcal{C}(m^2_K,m_u,m_s)+ \\  &\hspace*{.8cm}\biggl(1-\frac{f_\pi}{f_K}\biggr) \ \biggl(\frac{m^2_u-m_s^2+2m_s^2\ln(\frac{m_s}{m_u})}{m^2_K}\biggr)\biggr\}\biggr]. 
\end{align}
The $\pi \text{ and } K$ curvature masses  $m_{\pi;c}= 135.95$ MeV and $m_{K;c}= 467.99$ MeV are, respectively, smaller by 2.05 MeV and 28.01 MeV from their corresponding pole masses  $m_{\pi}= 138$ MeV and $m_{K}= 496$  MeV \citep{vkkr23,skrvkt24}.

\subsection{Incorporating Vector Interaction}
\label{sec:IIA}
The Yukawa coupling of the scalar mesons to the quarks gives attractive interaction.~When the quarks are coupled to the vector mesons by a Yukawa type interaction term,~one gets the repulsive force.~The vector interactions are included in the 2+1 flavor linear sigma model by augmenting the meson Lagrangian of the Eq.(\ref{lagM}) with  the following Lagrangian for the vector mesons~\cite{Feroni, Beisit}.

\begin{align}
\mathcal{L}_{V}=-\frac{1}{4}F^{\mu\nu}F_{\mu\nu}+\frac{m_{v}^{2}}{2}V^{a\mu}V_{\mu}^{a}-g_{V}^{a}\bar{\psi}\gamma^{\mu}T^{a}\psi
V_{\mu}^{a}.
\label{vector_lag}
\end{align}
The vector meson fields are \ensuremath{V_{\mu}^{a}} where $a=0,...,8$ and \ensuremath{T^{a}} are the $SU(3)$ generators.~The vector nature of the mesons are signified by the Lorentz index \ensuremath{\mu}.~The first term of the Eq. \eqref{vector_lag} is the kinetic term whereas the second one is the mass term for the vector particles.~The Yukawa type coupling \ensuremath{g_{V}^{a}} in the third term couples the three flavor of quark fields \ensuremath{\psi} to the vector mesons.

~The meson fields are treated as classical fields in the mean filed approximation and assume VEV.~Recall that the finite VEV is taken for \ensuremath{\sigma_{0}} and \ensuremath{\sigma_{8}} while the \ensuremath{\sigma_{3}}=0 since the spontaneous breaking of the iso-spin has not been considered in the present work.~The rotational symmetry requires that the spatial components of the vector meson fields become zero while the time components of vector fields $\bar{V}_{0}^{0}$ and $\bar{V}_{0}^{8}$ in the 0 and 8 direction,~are non-zero.~The mixing of the singlet-octet (0,8) fields,~decouples the non-strange and strange quark sector as : 

\begin{align}
\omega & \equiv \sqrt{\frac{2}{3}}\bar{V}_{0}^{0}+\frac{1}{\sqrt{3}}\bar{V}_{0}^{8}\ ;   \quad \phi & \equiv \frac{1}{\sqrt{3}}\bar{V}_{0}^{0}-\sqrt{\frac{2}{3}}\bar{V}_{0}^{8}.
\end{align}
The  effective vector coupling constants after rotation of fields into each other,~change as the following
\begin{equation}
\frac{g_V}{2}=g_{\omega } = \frac{g_{\phi}}{\sqrt{2}}
\end{equation}
~Since the fields  occur in combination with the Gell-Mann matrices,~the physical fields
are to be presented along with the following definitions~\cite{Beisit}.
\begin{equation}
\chi^{\omega}=\left(\begin{array}{ccc}
1 & 0 & 0\\
0 & 1 & 0\\
0 & 0 & 0
\end{array}\right)\ ; \quad \chi^{\phi}=\left(\begin{array}{ccc}
0 & 0 & 0\\
0 & 0 & 0\\
0 & 0 & 1
\end{array}\right).
\end{equation}
The time component of vector fields lead to the redefinitions of chemical potentials.~The modified effective  chemical potentials for the non-strange and strange quarks $\tilde{\mu}_{u(d)} $ and $\tilde{ \mu}_{s}$ are defined by the diagonal elements of the matrix given below.
\begin{equation}
\tilde{\mu}=\left(\begin{array}{ccc}
\tilde{\mu}_{u} & 0 & 0\\
0 & \tilde{\mu}_{d} & 0 \\
0 & 0 & \tilde{\mu}_{s}\\
\end{array}\right)
=\left(\begin{array}{ccc}
\mu - g_{\omega}\omega & 0 & 0 \\
0 & \mu-g_{\omega}\omega &0 \\
0 & 0  & \mu-g_{\phi}\phi \\
 \end{array}\right).
\end{equation}
The meson potential in the Eq.~(\ref{lag}) gets rewritten with the above notations and definitions.
The mean field Lagrangian for the 2+1 flavor QM model with vector interactions takes the following form.
\begin{align}
\nonumber
\mathcal{L_{QM:\text{V}}}= & \bar{\psi} \bigg[ i {\not}\partial+ \mu \gamma ^0 -
g\left(\sigma_{0}\frac{\lambda^{0}}{2}+\sigma_{8}\frac{\lambda^{8}}{2}\right) \\  \nonumber
&-g_{\omega}  \gamma^{0} \omega   \chi^{\omega}  -g_{\phi}  \gamma^{0} \phi   \chi^{\phi} \ \biggr] \psi -U(\x,\y)\\ 
&+\frac{m_{\omega}^{2}}{2}\omega^{2}+\frac{m_{\phi}^{2}}{2}\phi^{2}. 
\end{align}
The $U(\x,\y)$ is given by the Eq.(\ref{eq:mesop}) in  terms of light and strange condensates $\x \text{ and } \y$.~The grand canonical potential for the QM model for the above Lagrangian with the vector interaction under the s-MFA,~becomes  
\begin{align}
\label{QMVect}
\nonumber
\Omega_{\rm QM : V }(T,\mu,x,y)=& U(\x,\y)+\Omega_{q\bar{q}:\text{V}}^{T,\mu}(x,y)\\
&-\frac{m_{\omega}^{2}}{2}\omega^{2}-\frac{m_{\phi}^{2}}{2}\phi^{2}\;. 
\end{align}
\begin{align}
\label{VectFT}
&\Omega_{q\bar{q}:\text{V}}^{T,\mu}(x,y)=- 2 N_{c} \sum_{f=u,d,s} \int \frac{d^3 p}{(2\pi)^3} T \left[ \ln \tilde{g}_f^{+}+\ln \tilde{g}_f^{-}\right].\;
\end{align}
The $ \tilde{g}^{\pm}_f = \left[1+e^{-\tilde{E}_{f}^{\pm}/T}\right] $ where $\tilde{E}_{f}^{\pm} =E_f \mp \tilde{\mu}_{f}$ with $E_f=\sqrt{p^2 + m{_f}{^2}}$  and the $m_{u}=m_{d}=\frac{g\x}{2}$ ; $m_{s}=\frac{g\y}{\sqrt{2}}$.~The effective chemical potential is $\tilde{\mu}_{u}=\tilde{\mu}_{d}=(\mu-g_{\omega}\omega)$  for the light quarks and $\tilde{\mu}_{s}=(\mu-g_{\phi}\phi)$ for the strange quarks.~Note that only the temperature and chemical potential dependent quark-antiquark contributions  get modified in the grand potential  after including the vector interactions.~The vacuum part $\Omega_{vac}^{\rm RQM}$ of the RQM model grand potential in Eq.~(\ref{grandRQM}) and its renormalized parameters do not change in the presence of vector interaction.~The RQM model grand potential with vector interaction is written as :    
\begin{align}
\label{RQMVect}
\nonumber
\Omega_{\rm RQM : V }(T,\mu,x,y)=&\Omega_{vac}^{\rm RQM}(\x,\y)+\Omega_{q\bar{q}:\text{V}}^{T,\mu}(x,y) \\ &-\frac{m_{\omega}^{2}}{2}\omega^{2}-\frac{m_{\phi}^{2}}{2}\phi^{2}\;. 
\end{align}
The values of the $\x, \ \y$ condensates and vector meson fields $\omega, \phi$ for the QM and RQM  model are computed after solving the respective gap equations under the s-MFA and e-MFA.
\begin{align}
\frac{\partial \Omega_{\rm QM : V }}{\partial x}=\frac{\partial\Omega_{\rm QM : V }}{\partial y}
=\frac{\partial\Omega_{\rm QM : V }}{\partial\omega}=\frac{\partial\Omega_{\rm QM : V }}{\partial\text{\ensuremath{\phi}}}
=0.  
\label{gap:QMV}  
\end{align}
\begin{align}
\frac{\partial \Omega_{\rm RQM : V }}{\partial x}=\frac{\partial\Omega_{\rm RQM : V }}{\partial y}
=\frac{\partial\Omega_{\rm RQM : V }}{\partial\omega}=\frac{\partial\Omega_{\rm RQM : V }}{\partial\text{\ensuremath{\phi}}}
=0.
\label{gap:RQMV}  
\end{align}

The masses of the vector meson are taken as $m_{\omega}=783$ MeV and $m_{\phi}=1019$ MeV.

\subsection{Different forms of Polyakov-loop potential}
\label{sec:IIB}
The confinement-deconfinement phase transition has been studied in the literature using different forms of the Polyakov-loop effective potential $\mathcal{U}(\Phi,\bar{\Phi},T)$.~Its simplest form is constructed  by finding a potential which respects all the given symmetries and accounts for the spontaneously broken $Z(3)$ symmetry for the system in the deconfined phase \cite{SveLer,benji,BankUka}.~The following polynomial form  constitutes the minimal content of the Polyakov-loop potential~\cite{ratti}. 

\bqa
\label{plykov_poly}
\hspace{-0.5 cm}\frac{\mathcal{U_{\rm Poly}}}{T^4}&=&-\frac{b_2(T)}{2}\Phi\bar{\Phi}-\frac{b_3}{6}(\Phi^3+\bar{\Phi}^3)+\frac{b_4}{4}(\Phi\bar{\Phi})^2\;,
\eqa
the coefficients of the Eq.~(\ref{plykov_poly}) are given by
\bqa
b_2(T)=a_0+a_1\left(\frac{T_0}{T}\right)+a_2\left(\frac{T_0}{T}\right)^2+a_3\left(\frac{T_0}{T}\right)^3\;,
\eqa
where $a_0=6.75$, $a_1=-1.95$, $a_2=2.625$, $a_3=-7.44$, $b_3=0.75$ and $b_4=7.5$ .\\

The parameter $T_0$ in the Polyakov loop potential  defines the critical temperature scale for the deconfinement phase transition in a pure $SU(3)$ Yang-Mills gauge theory. Physically, it represents the energy scale at which the global $Z_3$ center symmetry of the pure gauge sector is spontaneously broken. While it is fixed at $T_0 = 270 \text{ MeV}$ for a pure gluonic system, the introduction of dynamical quarks acts as an external background field that explicitly breaks the $Z_3$ symmetry and screens the strong interaction of color charges.

The above form is improved by adding the contribution from  the integration of the $SU(3)$ group volume in the generating functional for the Euclidean action.~The Haar measure is used to perform this integration which takes the form of a Jacobian determinant.~Its logarithm is added as an effective potential to the action in the generating functional.~ The potential gets bounded from below for the large $\Phi$ and $\bar{\Phi}$  by the positive coefficient of the logarithm term.~The logarithmic form of the Polyakov-loop potential~\cite{fuku,Roesnr} has the following expression.
\bqa
\label{plykov_log}
\nonumber
\hspace{-0.5 cm}\frac{\mathcal{U_{\rm Log}}}{T^4}&=&b(T)\ln[1-6\Phi\bar{\Phi}+4(\Phi^3+\bar{\Phi}^3)-3(\Phi\bar{\Phi})^2]\;\\&&-\frac{1}{2}a(T)\Phi\bar{\Phi}\;.
\eqa
The parameters of the polynomial and log form of the Polyakov-loop potential were determined \cite{ratti,Roesnr} by fitting the lattice data for pressure, entropy density  as well as energy density and the evolution of Polyakov-loop $\Phi$ on the lattice in pure gauge theory.~The coefficients of the Eq.~(\ref{plykov_log}) are the following \cite{Roesnr},
\bqa
&&a(T)=a_0+a_1\left(\frac{T_0}{T}\right)+a_2\left(\frac{T_0}{T}\right)^2\;,\\
&&b(T)=b_3\left(\frac{T_0}{T}\right)^3\;,
\eqa
where $a_0=3.51$, $a_1=-2.47$, $a_2=15.2$, $b_3=-1.75$. 
It is pointed  out that the log potential has qualitative consistency with the leading order result of the strong-coupling expansion \cite{JLange}.~Furthermore,~since the potential diverges for $\Phi$,~$\bar{\Phi} \longrightarrow$  1,~the Polyakov-loop always remains smaller than 1 and approaches this value asymptotically as $T \longrightarrow \infty$.

The new Polyakov-loop effective potential was  constructed  in the Ref.~\cite{Redlo} after incorporating the Polyakov-loop fluctuations as well.~Its  parameters are  adjusted such that the longitudinal as well as the transverse susceptibilities are also reproduced in addition to the other existing lattice data.~The following new expression of the PolyLog Polyakov-loop potential gets constructed in this work after the logarithmic term is added to the polynomial form of the Polyakov-loop potential.
\bqa
\label{plykov_polylog}
\nonumber
\hspace{-0.5 cm}\frac{\mathcal{U_{\rm PolyLog}}}{T^4}&=&b(T)\ln[1-6\Phi\bar{\Phi}+4(\Phi^3+\bar{\Phi}^3)-3(\Phi\bar{\Phi})^2]\;\\ \nonumber
&&+a_2(T)\Phi\bar{\Phi}+a_3(T)(\Phi^3+\bar{\Phi}^3)+a_4(T)(\Phi\bar{\Phi})^2.\\
\eqa
In the PolyLog parametrization,~the coefficients in the Eq.~(\ref{plykov_polylog}) are defined as :
\bqa
&&a_i(T)=\frac{a^{(i)}_0+a^{(i)}_1\left(\frac{T_0}{T}\right)+a^{(i)}_2\left(\frac{T_0}{T}\right)^2}{1+a^{(i)}_3\left(\frac{T_0}{T}\right)+a_4^{(i)}\left(\frac{T_0}{T}\right)^2}\;\\
&&b(T)=b_0\left(\frac{T_0}{T}\right)^{b_1}\left[1-e^{b_2\left(\frac{T_0}{T}\right)^{b_3}}\right].
\eqa
\\
The  summary of the parameters is given in the Table~\ref{tab:plglg}.
\begin{table}[!htbp]
    \caption{Parameters of the PolyLog Polyakov-loop potential have been taken from the Ref.~\cite{Redlo}.}
        \label{tab:plglg}
    \resizebox{0.48\textwidth}{!}{
    \begin{tabular}{p{1.5cm} p{1.5cm} p{1.5cm} p{1.5cm} p{1.5cm} p{1.5cm} p{0.5 cm}}
      \toprule 
      PolyLog& $a^{(2)}_0$ & $a^{(2)}_1$ & $a^{(2)}_2$ & $a^{(2)}_3$ & $a^{(2)}_4$&\\
      & 22.07 & -75.7 & 45.03385 & 2.77173 & 3.56403&\\
      & $a^{(3)}_0$ & $a^{(3)}_1$ & $a^{(3)}_2$ & $a^{(3)}_3$ & $a^{(3)}_4$&\\
      &-25.39805&57.019&-44.7298&3.08718&6.72812&\\
      & $a^{(4)}_0$ & $a^{(4)}_1$ & $a^{(4)}_2$ & $a^{(4)}_3$ & $a^{(4)}_4$&\\
      &27.0885&-56.0859&71.2225&2.9715&6.61433&\\
      &$b_0$&$b_1$&$b_2$&$b_3$& & \\
      &-0.32665&5.8559&-82.9823&3.0& &  \\
      \hline      
    \end{tabular}}
\end{table}

The deconfinement phase transition is first order with $T_c^{\rm YM}=T_0=270$ MeV for the pure gauge Yang-Mills theory.~The first order transition becomes a crossover in presence of the dynamical quarks.~The $T_0$ depends on the number of quark flavors and chemical potential in the full dynamical QCD~\cite{SchaPQM2F,Haas,kovacs,BielichP,THerbst2} because it is linked to the mass-scale $\Lambda_{\rm QCD}$ that gets modified by the effect of the fermionic matter fields.~The $T_0 \longrightarrow T_0(N_f,\mu)$ has the following form of the number of flavor and  chemical potential dependence.
\bqa
\label{t0_mu}
T_0(N_f,\mu)=\hat{T} \ e^{-1/(\alpha_0 b(N_f,\mu))}\;,
\eqa
with 
\bqa
b(N_f,\mu)=\frac{1}{6\pi}(11N_c-2N_f)-b_\mu\frac{\mu^2}{(\hat{\gamma}\hat{T})^2}.
\eqa
here the $\hat{T}$ parameter is fixed at the scale $\tau$,  $\hat{T}=T_\tau=1.77$ GeV and $\alpha_0=\alpha(\Lambda)$ at a UV scale $\Lambda$.~The $T_0(N_f=0)$ = 270 MeV gives  $\alpha_0$ = 0.304 and $b_\mu\simeq\frac{16}{\pi}N_f$. The parameter $\hat{\gamma}$ governs the curvature of $T_0(\mu)$ with the systematic error estimation range $0.7\lesssim\hat{\gamma}\lesssim1$ \cite{SchaPQM2F,THerbst2}.~Massive flavors lead to suppression factors of the order $T_0^2/(T^2_0 +m^2)$ in the $\beta$-function.~For 2+1 flavors and a current strange quark mass $m_s\sim$ 150 MeV,~one obtains $T_0(2+ 1)$=187 MeV\cite{SchaPQM2F,THerbst2}.


The Polyakov-loop potential gets replaced by the QCD glue potential after the back reaction of the quarks is included in the full QCD with dynamical quarks.~The Ref.~\cite{Haas} applied the FRG equations to the QCD and compared the pure gauge potential $\mathcal{U}_{\rm YM}$ to the ``glue'' potential $\mathcal{U}_{\rm glue}$ where quark polarization was included in the gluon propagator and they found significant differences between the two potentials.~However,~it was observed that the two potentials are of the same shape and they can be mapped into each other by relating the temperatures of the two systems,~$T_{\rm YM}$ and $T_{\rm glue}$.~Denoting the previous equations of the Polyakov-loop potential by $\mathcal{U}_{\rm YM}$,~the improved Polyakov-loop potential $\mathcal{U}_{\rm glue}$ is constructed as \cite{Haas}  :    

\bqa
\frac{\mathcal{U}_{\rm glue}}{T^4_{\rm glue}}(\Phi, \bar{\Phi}, T_{\rm glue})&=&\frac{\mathcal{U}_{\rm YM}}{T^4_{\rm YM}}(\Phi, \bar{\Phi}, T_{\rm YM})
\eqa
here the temperature $T_{\rm glue}$ is related to $T_{\rm YM}$ as 
\bqa
\frac{T_{\rm YM}-T^{\rm YM}_{\rm c}}{T^{\rm YM}_{\rm c}}=0.57 \
\frac{T_{\rm glue}-T^{\rm glue}_{\rm c}}{T^{\rm glue}_{\rm c}}
\eqa
The $T^{\rm glue}_{\rm c}$ is the transition temperature for the unquenched case.~The coefficient 0.57 comes from the comparison of the two effective potentials.~$T^{\rm glue}_{\rm c}$ lies within a range $T^{\rm glue}_{\rm c} \in [180,270]$.~In practice,~one  uses the replacement $T  \longrightarrow T^{\rm YM}_{\rm c}(1+0.57(\frac{T}{T^{\rm glue}_{\rm c}}-1))$ in the right-hand side of the Polyakov-loop potentials where $T_0$ means $T^{\rm YM}_{\rm c}$ and ($T\sim T_{\rm YM}$) on the left side of the arrow while ($T\sim T_{\rm glue}$) on the right side \cite{skrvkt24}.~We have taken $T^{\rm glue}_{\rm c}$ and $T_0$ both fixed at 187 MeV in our calculations for the 2+1 quark flavor as in the Ref.~\cite{SchaPQM2F,Haas}.

When the effects of the Polyakov loop are correctly incorporated into QCD models, the sign problem arises in the form of the thermodynamic potential becoming complex in the presence of the Polyakov loop background at nonzero chemical potential~\cite{Roesnr,RosenNPA,Mintz,Hansen:2019lnf}. Therefore, a straightforward minimization of the complex grand potential becomes meaningless.~However, even for a complex Euclidean action, one can still search for the configuration with the largest weight in the path integral and refer to this as the mean-field configuration~\cite{Roesnr}.~In our RPQM model setup, $\Phi$ and $\bar{\Phi}$  are treated as independent real variables which are obtained by minimizing the real part of the grand potential, determined from the saddle-point of the mean-field free energy\cite{Hansen:2019lnf}. Thus, the sign problem is bypassed in the present mean-field treatment, as is customarily done in a multitude of previous works ~\cite{fuku08,SchaPQM2F,SchaPQM3F,Haas,Redlo,BielichP,THerbst2,TiPQM3F,guptiw,vkkr12,chatmoh1,schafwag12,
vkkt13,kovacs,Rai,asmuAnd,raiti23,skrvkt24}.~Since a stable saddle point is needed for the mean-field approximation to work properly,~Ref.~\cite{Hidaka}, while using a phase reweighting technique to circumvent the complex action of $SU_c(3)$ gauge theory, suggested that this stability should be enforced by the Gauss law constraint on the number density of the gauge charge.
Charge conjugation (C) symmetry is respected at $\mu = 0$, resulting in $\Phi = \bar{\Phi}$ and ensuring that the free energies of a heavy test quark and anti-quark are equal. However, a net quark density at $\mu > 0$ explicitly breaks C-symmetry, causing the color charge screening of a heavy test quark to differ from that of a test anti-quark. Therefore, assigning different real expectation values ($\Phi \neq \bar{\Phi}$) accurately captures this physical reality, since $\Phi \propto e^{-\beta F_q}$ and $\bar{\Phi} \propto e^{-\beta F_{\bar{q}}}$~\cite{SveLer,KarshPRL,Dumitru}. Furthermore, because the free energies $F_q$ and $F_{\bar{q}}$ differ from one another at finite density, the two Polyakov loop fields must necessarily take different, yet still real, values.~Note that the fluctuations beyond mean field are at the origin of $\Phi \neq \bar{\Phi}$~\cite{RosenNPA} but the effects of the fluctuations at finite chemical potential, turn out not to be of major qualitative importance in determining the phase diagram~\cite{Roesnr}

Important and interesting developments have recently emerged from a series of extensive investigations~\cite{NishimuraI, NishimuraII,Tanizaki}, where the authors pushed the boundaries of the sign problem from a mere bypass to a rigorous solution. The mean-field sign problem was effectively bypassed in Ref.~\cite{NishimuraI} by moving the background gauge fields off the real axis. They showed that the charge conjugation and complex conjugation ($\mathcal{CK}$) symmetry of finite-density QCD guarantees that the thermodynamic potential remains strictly real at these complex saddle points, establishing a computationally viable foundation at finite density. These complex saddle points naturally lead to $\langle \mathrm{Tr} P \rangle \neq \langle \mathrm{Tr} P^{\dagger} \rangle$ (equivalent to $\Phi \neq \bar{\Phi}$), a physical result that is exceedingly difficult to obtain by restricting the fields to the real axis. Following an examination of a class of models, their subsequent work in Ref.~\cite{NishimuraII} suggests that only a modest excursion into the complex plane is required. Because the saddle point remains close to the real axis--as indicated by small values of the parameter $\psi$ and the correspondingly small differences between $\mathrm{Tr} P$ and $\mathrm{Tr} P^{\dagger}$--they point out that this poses a promising prospect for lattice simulation efforts. By computing the fluctuations in the form of a non-Hermitian mass matrix (the second derivatives of the effective potential with respect to the Polyakov loop variables) around the complex saddle points, they found that due to $\mathcal{CK}$ symmetry, the eigenvalues of this matrix can suddenly transition from purely real to complex conjugate pairs. Consequently, a new physical regime of disorder lines emerges at high density where spatial correlation functions between quarks do not merely decay exponentially, but begin to oscillate spatially--a feature entirely invisible to standard real-axis mean-field studies. Instead of merely evaluating the thermodynamics at a single isolated saddle point, they finally performed an exact contour integration over the complex plane using Lefschetz thimbles in Ref.~\cite{Tanizaki}. They proved that when integrating over pairs of $\mathcal{CK}$-symmetric thimbles, all residual imaginary phases interfere and cancel perfectly.~They conclude that the Lefschetz-thimble decomposition is suitable for the saddle-point analysis and the real partition function can be computed in a systematic way. This solves the sign problem appearing in the MFA when the classical action is complex. Further development in Ref.~\cite{Mori} has shown how the path optimization method in an effective QCD model can  control the model sign problem.




\subsection{RPQM Model}
\label{sec:IIC}
The Polyakov-loop potential becomes active only at non-zero temperatures,~hence it does not affect the vacuum parameters of the chiral part of the effective potential.~The finite temperature and chemical potential dependent quark-antiquark  contribution $\Omega_{q\bar{q}}^{T,\mu}$ in the Eq.~(\ref{vac2}) ,~gets modified by the Polyakov loop potential.
\bqa
\label{FTpoly}
\Omega_{q\bar{q}}^{T,\mu}(\x,\y,\Phi,\bar{\Phi})=- 2 \sum_q \int \frac{d^3 p}{(2\pi)^3} T \left[ \ln g_f^{+}+\ln g_f^{-}\right].\;
\eqa
The $\Omega_{q\bar{q}}^{T,\mu}(\x,\y,\Phi,\bar{\Phi})=\Omega_{q\bar{q}}(T,\mu;x,\y,\Phi,\bar{\Phi})$ and new $ g_{f}^{\pm}$ are different from those in the  Eq.~(\ref{vac2}).
\bqa
\hspace{-1 cm}g_{f}^{+}&=&\left[1+3\Phi e^{-E_{f}^{+}/T}+3\bar{\Phi}e^{-2E_{f}^{+}/T}+e^{-3E_{f}^{+}/T}\right]\;,\qquad\\
g_{f}^{-}&=&\left[1+3\bar{\Phi} e^{-E_{f}^{-}/T}+3\Phi e^{-2E_{f}^{-}/T}+e^{-3E_{f}^{-}/T}\right]\;.
\eqa
$E_{f}^{\pm} =E_f \mp \mu_{f} $; $E_f=\sqrt{p^2 + m{_f}{^2}}$ with $f=u,d,s$ and  $\mu_{u}=\mu_{d}=\mu_{s}=\mu$.~The $(T,\mu)$ dependent part of the grand potential with Ployakov-loop potential gets modified further by the  vector interaction effects to give :
\bqa
\label{FTpolyV}
\Omega_{q\bar{q}:\text{V}}^{T,\mu}(\x,\y,\Phi,\bar{\Phi})=- 2 \sum_q \int \frac{d^3 p}{(2\pi)^3} T \left[ \ln \tilde{g}_f^{+}+\ln \tilde{g}_f^{-}\right].\;
\eqa
The $\Omega_{q\bar{q}:\text{V}}^{T,\mu}(\x,\y,\Phi,\bar{\Phi})=\Omega_{q\bar{q}:\text{V}}(T,\mu;x,\y,\Phi,\bar{\Phi})$ and 
\bqa
\hspace{-1 cm}\tilde{g}_{f}^{+}&=&\left[1+3\Phi e^{-\tilde{E}_{f}^{+}/T}+3\bar{\Phi}e^{-2\tilde{E}_{f}^{+}/T}+e^{-3\tilde{E}_{f}^{+}/T}\right]\;,\qquad\\
\tilde{g}_{f}^{-}&=&\left[1+3\bar{\Phi} e^{-\tilde{E}_{f}^{-}/T}+3\Phi e^{-2\tilde{E}_{f}^{-}/T}+e^{-3\tilde{E}_{f}^{-}/T}\right]\;.
\eqa
where $\tilde{E}_{f}^{\pm} =E_f \mp \tilde{\mu}_{f}$.~The $\tilde{\mu}_{u}=\tilde{\mu}_{d}=(\mu-g_{\omega}\omega)$  for the light quarks and $\tilde{\mu}_{s}=(\mu-g_{\phi}\phi)$ for the strange quarks.

The grand thermodynamic potential of the renormalized Polyakov loop enhanced quark meson (RPQM) model~\cite{skrvkt24} is obtained when the PQM model Lagrangian of the Eq.~(\ref{lag:PQM}) is used and the RQM model vacuum effective potential $\Omega_{vac}^{\rm RQM}(\x,\y)$ in the Eq.~(\ref{vacRQM}) is added to the thermal contributions  of quarks-anti-quarks in the presence of the Polyakov loop potential. 
\begin{align}
\label{rpqmomega}
\nonumber
\hspace{-1 cm}\Omega_{RPQM}(T,\mu;\x,\y,\Phi,\bar{\Phi})=&\Omega_{vac}^{\rm RQM}(\x,\y)+{\cal U} \big( \Phi , \bar\Phi , T \big) \\
&+\Omega_{q\bar{q}} (T,\mu;\x,\y,\Phi,\bar{\Phi}) 
\end{align} 

When the effects of vector interactions are taken into account,~the RPQM model grand thermodynamic potential takes the following form.

\begin{align}
\label{rpqmomV}
\nonumber
\hspace{-0.15 cm}\Omega_{RPQM:\text{V}}(T,\mu;\x,\y,\Phi,\bar{\Phi})=&\Omega_{vac}^{\rm RQM}(\x,\y)-\frac{m_{\omega}^{2}\omega^{2}}{2}-\frac{m_{\phi}^{2}\phi^{2}}{2}\; \\ \nonumber
&+\Omega_{q\bar{q}:\text{V}} (T,\mu;\x,\y,\Phi,\bar{\Phi}) \\ 
&+{\cal U} \big( \Phi , \bar\Phi , T \big).
\end{align}

When quark one-loop vacuum contribution is dropped in the s-MFA,one gets the PQM model grand thermodynamic potential  after adding the meson contribution in Eq.~(\ref{eq:mesop}) to the thermal contributions of quarks-anti-quarks in the presence of the Polyakov loop potential as the following.
\bqa
\label{pqmomega}
\nonumber
\Omega_{PQM}(T,\mu;\x,\y,\Phi,\bar{\Phi})&=&U(\x,\y)+ {\cal U} \big( \Phi , \bar\Phi , T \big) \\
&&+\Omega_{q\bar{q}} (T,\mu;\x,\y,\Phi,\bar{\Phi}) 
\eqa

The PQM model grand thermodynamic potential 
with the effects of vector interactions is written as :

\begin{align}
\label{pqmomV}
\nonumber
\Omega_{PQM:\text{V}}(T,\mu;\x,\y,\Phi,\bar{\Phi})=&U(\x,\y)-\frac{m_{\omega}^{2}\omega^{2}}{2}-\frac{m_{\phi}^{2}\phi^{2}}{2}\; \\ \nonumber
&+\Omega_{q\bar{q}:\text{V}} (T,\mu;\x,\y,\Phi,\bar{\Phi}) \\
&+{\cal U} \big( \Phi , \bar\Phi , T \big).
\end{align} 

\begin{table*}[!htbp]
    \caption{Different Model Parameters at the physical point and in the light chiral limit study.}
    \label{tab:table2}
    \begin{tabular}{p{0.1\textwidth} p{0.1\textwidth} p{0.1\textwidth} p{0.1\textwidth} p{0.123\textwidth} p{0.1\textwidth} p{0.1\textwidth} p{0.123\textwidth} p{0.1\textwidth}}
    \toprule 
      Model& $m_{\sigma}$&$m_{\pi}(\text{MeV})$&$\lambda_2$&$c(\text{MeV}^2)$ &$\lambda_1$&$m^2(\text{MeV}^2)$& $h_x(\text{MeV}^3)$ & $h_y(\text{MeV}^3)$\\
      \hline 
      &$400$&$138$&46.43 &4801.82 &-5.89 &$(494.549)^2$ &$(120.73)^3$ &$(336.43)^3$ \\
       QM& $500$&  $138$&$46.43$ &$4801.82$ &$-2.69$&$(434.305)^2$ &$(120.73)^3$ &$(336.43)^3$\\
       &$500$&$0$&$44.62$ &$4966.63$ &$-0.96 $&$(411.663)^2$ &$\phantom{text}0$&$(340.05)^3$ \\
      \hline
      \phantom{text} & \phantom{text}&\phantom{text}&$\lambda_{20}$&$c_{0}(\text{MeV}^2)$ &$\lambda_{10}$&$m^2_{0}(\text{MeV}^2)$& $h_{x0}(\text{MeV}^3)$ & $h_{y0}(\text{MeV}^3)$\\
      \hline
      RQM&$400$&$138$&34.88 &7269.20 &1.45 &$(442.447)^2 $&$(119.53)^3$ &$(323.32)^3$ \\
      &$400$&$\quad0$&$34.28$ &7485.73 &2.03 &$(437.090)^2 $&$\qquad 0$ &$(329.83)^3$ \\
      &$500$&$138$&$34.88 $&$7269.20$ &$3.676 $&$ (396.075)^2$&$(119.53)^3$ &$(323.32)^3$ \\
      
      &$500$&$\quad0$&$34.28$&$7485.73$&$4.622 $&$(382.143)^2$&$\qquad 0$ &$(329.83)^3$\\
      
          \hline 
      
    \end{tabular}
\end{table*}

The $(T, \ \mu)$ dependence of the light, strange condensates $\x \text{ , } \y$ and Polyakov loop condensates~$\Phi$, $\bar{\Phi}$ are obtained in the RPQM and PQM model by optimizing the global minimum of the grand potential respectively in the Eq.~(\ref{rpqmomega}) and Eq.~(\ref{pqmomega}) . 
\bqa
\nonumber
\frac{\partial{\Omega_{RPQM/PQM}}}{\partial
      {\x}}&=&\frac{\partial \Omega_{RPQM/PQM}}{\partial{\y}}  
      =\frac{\partial \Omega_{RPQM/PQM}}{\partial\Phi}\\ &&=\frac{\partial \Omega_{RPQM/PQM}}{\partial\bar{\Phi}} =0
\label{EoMMF3}
\eqa

Search and optimization of the global minimum of the grand potential respectively in the Eq.~(\ref{rpqmomV}) and Eq.~(\ref{pqmomV}) gives the $(T \ , \mu)$ dependence of the $\x,\y,\Phi,\bar{\Phi}, \omega \text{ and } \phi$ for the RPQM and PQM model when the effects of vector interactions are present.
  
\begin{align}
\nonumber
&\frac{\partial{\Omega_{(RPQM:\text{V}/PQM:\text{V})}}}{\partial
      {\x}}=\frac{\partial \Omega_{(RPQM:\text{V}/PQM:\text{V})}}{\partial{\y}}=  
      \\ \nonumber &\frac{\partial \Omega_{(RPQM:\text{V}/PQM:\text{V})}}{\partial \omega}=\frac{\partial \Omega_{(RPQM:\text{V}/PQM:\text{V})}}{\partial \phi}=\\&\frac{\partial \Omega_{(RPQM:\text{V}/PQM:\text{V})}}{\partial\Phi}=\frac{\partial \Omega_{(RPQM:\text{V}/PQM:\text{V})}}{\partial\bar{\Phi}}=0
\label{EoMMF6V}
\end{align}

The 2+1 flavor RPQM and PQM model phase diagram with Log form and PolyLog-glue form of the Polyakov loop potential,~different thermodynamic quantities and critical end point have been calculated and presented in the Ref.\cite{skrvkt24}.~We will be using the above described 2+1 flavor RQM and RPQM model and extend our work to calculate the quark number susceptibilities for evaluating the critical regions around the CEP in the presence as well as absence of the effect of vector interactions in different model scenarios. 

\subsection{Light Chiral Limit}
\label{sec:IID}
We have used the  standard large $N_c$,~$U(3)$ chiral 
perturbation theory (ChPT) inputs as in the Ref.~\cite{herrPLB,Escribano,vktChpt1} to find the pion and kaon decay constants $f_{\pi}$ and $f_{K}$ and the parameters of the QM/RQM model for the light chiral limit.~Taking the values $(f_{\pi},f_{K})=(92.4,113.0216)$ MeV at the physical point for $(m_{\pi},m_{K})=(138,496)$ MeV and using  the following set of equations,
\bqa
\label{fpiU3}
f_{\pi}&=&f \Biggl( 1+4\frac{L_{5}}{f^2} m_{\pi}^{2} \Biggr)\;. \\ 
\label{fkU3}
f_{K}&=&f \Biggl( 1+4\frac{L_{5}}{f^2} m_{K}^{2} \Biggr)\;. 
\eqa
one gets the $f=90.67$ MeV and the chiral constant $L_{5}=2.06 \times 10^{-3}$~\cite{herrPLB,Escribano,vktChpt1}.~Chiral constants do not change when $(m_{\pi},m_K)$ are reduced to study the chiral limit while one move away from the physical point.~The $f_{\pi}=f=90.67$ MeV while $f_{K}=113.0216$ MeV in the light chiral limit when $m_{\pi}=0$ and $m_{K}=496$ MeV.

~For the physical point,~RQM (QM) model parameters  $\lambda_{20} (\lambda_{2})$, $c_{0}(c)$, $h_{x0}(h_{x})$, $h_{y0}(h_{y})$, $m_{0}^2(m^2)$  and $\lambda_{10} (\lambda_{1})$,~are determined as in Refs.~\cite{vkkr23,skrvkt24,vktChpt1} taking the experimental values of the $\pi,K,\eta \text{ and } \eta'$ masses as $(m_{\pi},m_{K},m_{\eta},m_{\eta'})=(138,496,547.5,957.78)$ MeV and choosing the scalar $\sigma$ meson mass as $m_{\sigma}=500$ MeV whereas the $f_{\pi}=92.4$ MeV and $f_{K}=113.0216$ MeV.~Given the experimental masses $(m_{\eta},m_{\eta'})=(547.5,957.78)$ MeV,~one needs only the $M_{\eta}=\sqrt{(m_{\eta}^2+m_{\eta'}^2)}=1103.22$ MeV to find the parameters.~One gets $m_{\eta}=538.98$ MeV and $m_{\eta'}=962.60$  MeV in the QM model after using its parameters determined at the physical point with $M_{\eta}=1103.22$ MeV.~Note that for the above $M_{\eta}$,~when the pole masses $m_{\eta}=527.58$ MeV and $m_{\eta'}=968.89$ MeV are used as input in the RQM model for computing the $\eta,\eta'$ self energy corrections and the renormalized parameters,~the same pole masses for the $\eta$ and $\eta'$ are reproduced in the output after adding the corresponding self energy corrections to their respective masses calculated from the new set of renormalized parameters~\cite{vkkr23,skrvkt24}.~The ChPT parameters $L_{8}, \ v_{31} \text{ and } v_{02} $ in the present study are obtained by requiring that the physical point value $M_{\eta}=1103.22$ MeV gets reproduced when the ChPT Eq.~(49) and Eq.~(50) of the Refs.~\cite{vktChpt1} are used for calculating the $m_{\eta} \text{ and } m_{\eta'}$.~The  numerical values of the parameters get fixed as the following.
\bqa
\nonumber
L_{8}=1.425 \times 10^{-3}, v_{31} = -0.164 \text{ and } v_{02}=-29.4392f^2 \\
\eqa
Note that with the above values of parameters at the physical point,~the ChPT expressions in Eqs. (49) and (50) of Ref.~\cite{vktChpt1} give the $\eta$ and $\eta'$ masses as $m_{\eta}=527.88$ MeV and $m_{\eta'}=968.73$ which are almost equal  to the corresponding masses found in the RQM model.
~With the above set of $U(3)$ ChPT parameters,~the ChPT expressions in the Eqs. (49) and (50) of Ref.~\cite{vktChpt1} give the $m_{\eta}=527.34$ MeV and $m_{\eta'}=977.68$ MeV for the light chiral limit when $m_{\pi}=0$ and $m_{K}=496$ MeV.~The RQM/QM model parameters,~for the light chiral limit,~are obtained by using the $\pi,K,\eta \text{ and } \eta'$ masses as $(m_{\pi},m_{K},m_{\eta},m_{\eta'})=(0,496,527.34,977.68)$ MeV with the $f_{\pi}=90.67$ MeV,~$f_{K}=113.0216$ MeV and $m_{\sigma}=500$ MeV.~The QM and RQM model parameters at the physical point and also for the light chiral limit are given in Table~\ref{tab:table2}.

\section{Results and Discussion}
\label{sec:III}

\begin{figure*}[htb]
\subfigure[]{
\label{fig1a} 
\begin{minipage}[b]{0.47\textwidth}
\centering \includegraphics[width=\linewidth]{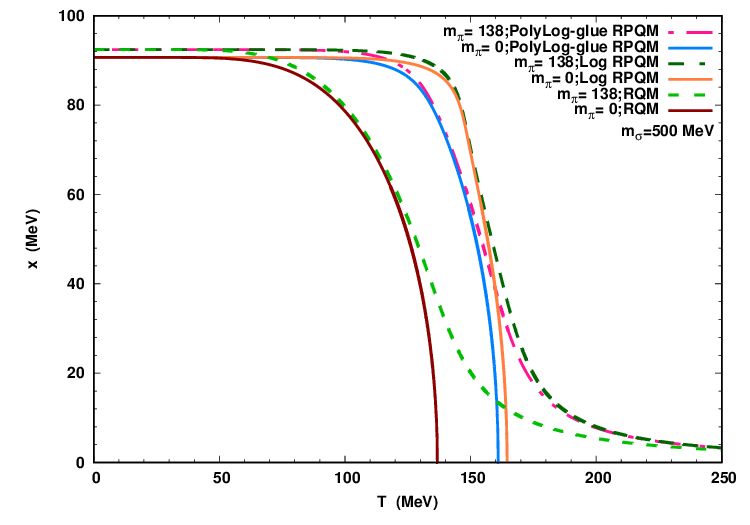}
\end{minipage}}
\hfill
\subfigure[]{
\label{fig1b} 
\begin{minipage}[b]{0.47\textwidth}
\centering \includegraphics[width=\linewidth]{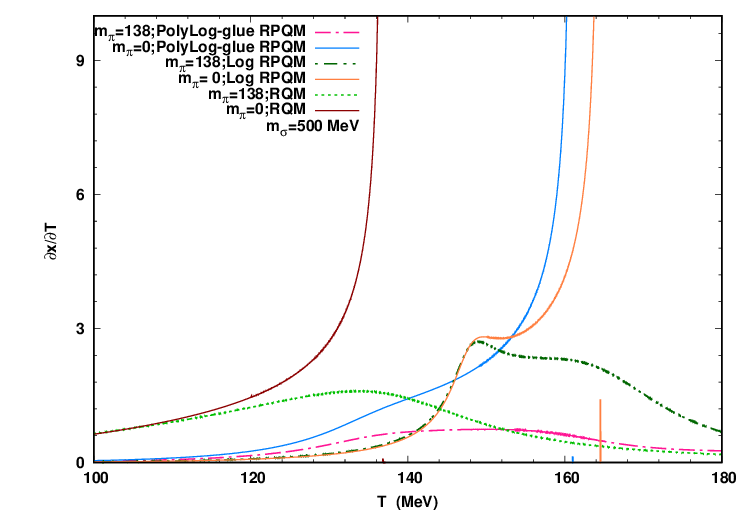}
\end{minipage}}
\caption{Temperature variations of  the non-strange condensate $x$ in [a] and it's derivative $\partial x/ \partial T$ in [b] for the RQM,~Log RPQM and PolyLog-glue RPQM model for the physical point $m_{\pi}=138$ MeV and the light chiral limit $m_{\pi}=0$ with $m_K=496$ MeV.}
\label{fig:mini:fig1}
\end{figure*}

The temperature variations of the subtracted chiral condensate $\Delta_{ls}$, calculated in the PolyLog-glue RPQM model for the $\mu=0$ chiral crossover transition, show the best agreement in Fig.~3(a) of Ref.~\cite{skrvkt24} with the Wuppertal-Budapest collaboration lattice QCD (WBQCD-I) data for $\Delta_{ls}$ \cite{Wupertal2010}. It is also important to note that the pseudo-critical temperature $T_{\chi} \equiv 155 \pm 2$ MeV for the $\mu=0$ chiral crossover transition, as obtained from different LQCD groups, is reproduced quite well in the PolyLog-glue RPQM model, where it has been found that $T_{c}^{\chi}=157.3$ MeV (see Table IV of Ref.~\cite{skrvkt24}). We emphasize that the above-described RPQM model results, which show close agreement with the LQCD results, were computed by fixing the mass of the scalar $\sigma$ meson at $m_{\sigma}=500$ MeV.  Furthermore, the temperature variations of the thermodynamic quantities--pressure, entropy density, energy density, interaction measure, specific heat, speed of sound, and the ratio $P/\epsilon$--for the crossover transition at $\mu=0$ MeV and $m_{\sigma}=500$ MeV have also been computed with and without the quark back-reaction in the Polyakov-loop potential for the present 2+1 flavor RPQM model setup in Ref.~\cite{skrvkt24}. A detailed comparison of the results in the above work with the latest 2+1 flavor lattice QCD data of the Wuppertal-Budapest and HotQCD collaborations \cite{WB2014,HotQCD2014} showed that the lattice data for the pressure, entropy density, and energy density are in close agreement with the respective temperature variations obtained in the PolyLog-glue RPQM model calculation. The PolyLog-glue RPQM model results for the temperature variations of the interaction measure (its peak structure and curve for $T>1.5\ T_c$), the specific heat, the speed of sound squared $c^2_s$, and the ratio $P/\epsilon$ (for $T>1.4\ T_c$) show a close resemblance to the WBLQCD-II \cite{WB2014} and HotQCD-I \cite{HotQCD2014} LQCD data.

It is worth mentioning that the connection between the scalar $\sigma$ meson state occurring in Quark-Meson (QM) model Lagrangians and the physical $\sigma/f_0(500)$ state is not entirely clear. The physical $\sigma/f_0(500)$ state, as quoted by the Particle Data Group (PDG), is a broad resonance state. Rigorous dispersive analyses of $\pi\pi$ scattering amplitudes locate the model-independent $f_0(500)$ pole deep on the second Riemann sheet, with a mass of $m_\sigma = 449^{+22}_{-16}$ MeV and a broad width of $\Gamma_\sigma = 550 \pm 24$ MeV \cite{Pelaez}. This broad nature allows for phenomenological flexibility when defining the field in an effective Lagrangian. Since the results of the renormalized Polyakov-loop-extended quark-meson (RPQM) model for the physical point parameters provide a good fit to current Lattice QCD data when the $\sigma$ meson mass is taken as $m_\sigma = 500$ MeV \cite{skrvkt24}, our primary choice for the $\sigma$ mass in the various model scenarios of this work is $m_\sigma = 500$ MeV. This value is used to compute: (1) the contours of the critical fluctuations of the critical endpoint (CEP), (2) the light chiral limit phase diagrams, and (3) the effects of different vector coupling strengths on the phase diagram, its CEP, and the extent of critical fluctuations around it. We also consider a second choice for the $\sigma$ meson mass, $m_\sigma = 400$ MeV, to compare our results regarding the size and shape of the critical regions around the CEP with the findings of Schaefer et al.\ in the QMVT/PQMVT model setup \cite{schafwag12}. In their work, they used $m_\sigma = 400$ MeV alongside the logarithmic form of the Polyakov-loop potential and a parameter $T_0 = 270$ MeV for the pure Yang-Mills $SU_c(3)$ gauge theory. Because Schaefer et al.\ used curvature masses of mesons to fix the model parameters, comparing our results with their work reveals how different methods of treating the quark one-loop vacuum fluctuations in the QM/PQM model affect the size and shape of the critical regions around the CEP. Furthermore, computations with this second mass choice allow us to investigate how varying the $\sigma$ meson mass influences the regions of critical fluctuations.

\begin{table*}[!htbp]
    \caption{Pseudo-critical temperature, Critical temperature for $m_{\pi}=0$, Critical  end points and Tricritical points.}
    \label{tab:table3}
    \begin{tabular}{p{1.2cm}| p{3cm}|p{3cm} |p{3cm}|p{2.6cm} |p{2.6cm}}
      \toprule 
       $\quad m_{\sigma}$ &\hphantom{textt} Models&$T^\chi_c(\mu=0,m_\pi=138)$ \hphantom{textext} MeV& $T_c(\mu=0,m_\pi=0)$ \hphantom{textext} MeV&$(\mu_{\cep},T_{\cep})$ MeV& $(\mu_{\tcp},T_{\tcp})$ MeV \\ 
      
 \hline
 \hphantom{text}& RQM
&\qquad$121.1$&\qquad$122.4$ & $\quad(243.12,37.03)$ & $(194.6,65.9)$ \\ 
\hphantom{text}& QMVT Ref.\cite{schafwag12}
&\qquad$144.7$&\qquad$...$ & $\quad(286.0,32.0)$ & $...$ \\ 

 \hphantom{text} & Log RPQM &\qquad$146.4$&\qquad$154.4$&\quad$(230.5,93.8)$& $(164.21,126.41)$\\
 & PolyLog-glue RPQM  &\qquad$145.6$&\qquad$149.1$&\quad$(227.6,73.6)$& $(172.4,107.28)$ \\ 
                  
   400 &  RPQM-\text I &\qquad$194.1$&\qquad$193.2$&\quad$(237.8,104.23)$& $(168.7,158.57)$\\
   
    \hphantom{text}&  RPQM-\text II &\qquad$172.9$&\qquad$173.4$&\quad$(228.54,86.47)$& $(163.28,133.47)$ \\ 
    \hphantom{text}& PQMVT-\text I Ref.\cite{schafwag12}
&\qquad$205.0$&\qquad$...$ & $\quad(283.0,90.0)$ & $...$ \\ 
\hphantom{text}& PQMVT-\text II Ref.\cite{schafwag12}
&\qquad$205.0$&\qquad$...$ & $\quad(280.0,83.0)$ & $...$ \\ 
\hline

     & QM&\qquad$129.0$ & \qquad $123.2$& $\quad(165.24,97.52)$ & \qquad ...  \\ 
    \hphantom{text}   & Log PQM &\qquad$159.3$&\qquad 155.7&\quad$(146.6,124.7)$ &\qquad ... \\
     \hphantom{text} & PolyLog-glue PQM&\qquad$155.2$ &\qquad 150.9&\quad $(129.6,137.7)$& \qquad ... \\
      500 & RQM &\qquad$133.6$&\qquad$136.8$ & $\quad(265.42,38.71)$ & $(220.15,71.52)$ \\
                     
       \hphantom{text} & Log RPQM &\qquad$148.9$&\qquad$164.5$&\quad$(252.0,94.6)$& $(194.3,126.53)$\\

     \hphantom{text}  &  PolyLog-glue RPQM &\qquad$157.3$&\qquad$160.9$&\quad$(252.7,70.9)$& $(202.20,108.14)$ \\
      \hline      
            
      \hline 
    \end{tabular}
\end{table*}


The temperature variations of the non-strange order parameter in the light chiral limit $m_{\pi}=0$ at $\mu=0$ when $m_{\sigma}=500$ MeV for the RQM, Log RPQM and PolyLog-glue RPQM model,~are compared in  Fig.~\ref{fig1a},~with the corresponding  temperature variations computed in the respective models for the parameters determined at the physical point.~The line types for different cases in models are labeled.~The vertically dropping non-strange order parameter temperature variations,~in the light chiral limit signify the second order chiral phase transitions which become crossover transitions for the $m_{\pi}=138$ MeV at the physical point in the RQM, Log RPQM and PolyLog-glue RPQM model as the corresponding temperature variations, of the non-strange condensate,~ develop continuously decreasing tail like structure.~The temperature variations of the $\partial x / \partial T$ in the light chiral limit  of the RQM, Log RPQM and PolyLog-glue RPQM model when $m_{\sigma}=500$ MeV,~become divergent and discontinuous in the Fig.~\ref{fig1b},~respectively at the  critical temperatures $T_{c}=136.8, 164.5 \text{ and } 160.9$ MeV of the second order chiral phase transitions.~The $\partial x / \partial T$  temperature variations for the physical point in the RQM, Log RPQM and PolyLog-glue RPQM model,~show a peak respectively at the pseudo-critical temperatures $T_{c}^{\chi}=133.6, 148.9 \text{ and } 157.3$ MeV~\cite{skrvkt24} of the chiral crossover transitions.~Note that the $T_{c}^{\chi}$ for the crossover transitions in the above models with on-shell renormalized parameters,~are smaller than the corresponding critical temperatures $T_{c}$ of the second order chiral transitions.~The temperature axis chiral crossover transitions at $\mu=0$,~occurring respectively at the pseudo-critical temperatures $T_{c}^{\chi}=129.0, 159.3 \text{ and } 155.2$ MeV~in the QM,~Log PQM and PolyLog-glue PQM model with the physical point parameters,~are significantly sharper because the quark one-loop vacuum correction terms are dropped under the standard mean field approximation (s-MFA)~\cite{Schaefer:09,TiPQM3F,vkkr23,skrvkt24}.~When parameters in  the light chiral limit are used for the $m_{\sigma}=500$ MeV,~the respective temperature variations of the order parameters at $\mu=0$ in the QM,~Log PQM and PolyLog-glue PQM model under the s-MFA,~show a discontinuous jump signifying the first order chiral  phase transition (plot not shown in the Fig.~\ref{fig1a}) respectively at the  critical temperatures $T_{c}=123.2, 155.7 \text{ and } 150.9$ MeV (see Table~\ref{tab:table3}).~One should note that the theoretical arguments~\cite{rob},~have ruled out any occurrence of a first order chiral phase transition at $\mu=0$ on the temperature axis,~in the light chiral limit of the chiral models.~The above mentioned pathology of finding a first order chiral phase transition  on the temperature axis at $\mu=0$,~in the light chiral limit of the QM/PQM model,~gets cured when the properly on-shell renormalized quark one-loop vacuum corrections are included in the RQM/RPQM model~\cite{vkkr23,skrvkt24} and one gets a second order chiral phase transition (belonging to the $O(4)$ universality class) in consonance with the  prediction of the Ref.~\cite{rob}.

\begin{figure*}[htb]
\subfigure[]{
\label{fig2a} 
\begin{minipage}[b]{0.48\textwidth}
\centering \includegraphics[width=\linewidth]{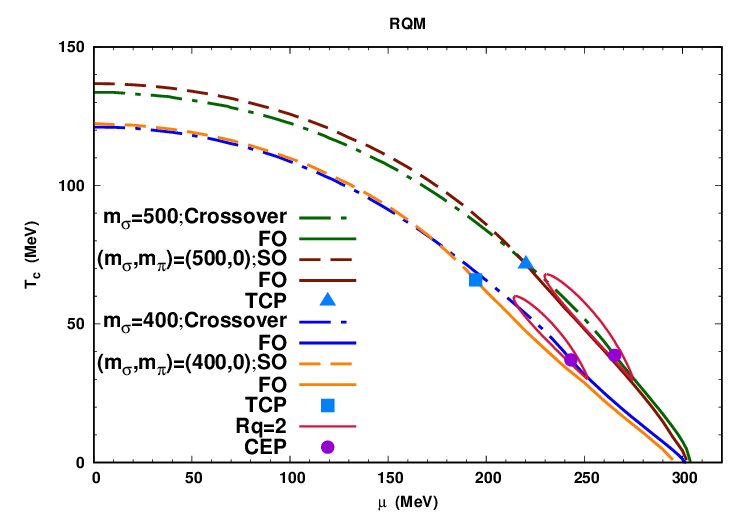}
\end{minipage}}
\hfill
\subfigure[]{
\label{fig2b} 
\begin{minipage}[b]{0.48\textwidth}
\centering \includegraphics[width=\linewidth]{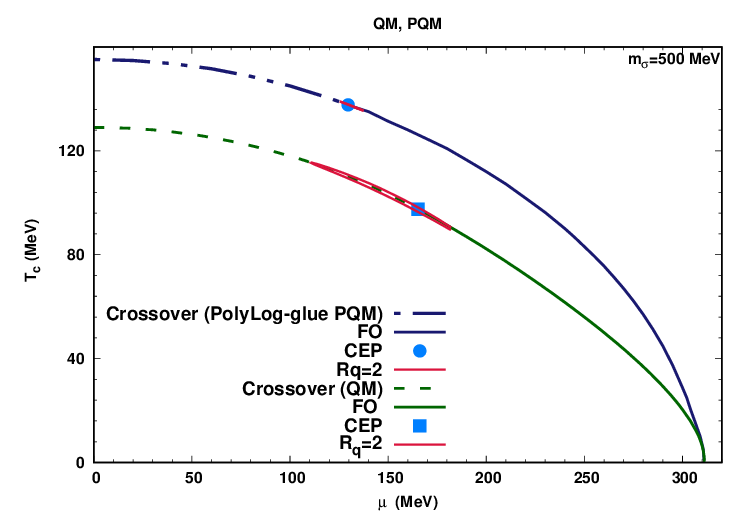}
\end{minipage}}
\caption{The RQM model phase diagrams for the $m_{\sigma}=400 \text{ and } 500$ MeV computed for the physical point and light chiral limit parameters,~are presented in the left panel (a) while the right panel (b) presents the phase diagrams for the QM model and PolyLog-glue PQM model when the $m_{\sigma}=500$ MeV at the physical point.~Line types and position of the critical end points (CEP) and tricritical points (TCP) are labeled.~The lines enveloping the CEP of different phase diagrams are the contours of constant quark number susceptibility ratio $R_q=2$.}
\label{fig:mini:fig2}
\end{figure*}

\subsection{Phase diagrams : TCP and CEP}
\label{sec:IIIA}

The $2+1$ flavor RQM model phase diagrams when $m_{\sigma}=400 \text{ and }500$ MeV computed for the light chiral limit $m_\pi =0$ parameters and the physical point $m_\pi =138$ parameters,~are shown in Fig.~\ref{fig2a} whereas the  $2+1$ flavor QM and PolyLog-glue PQM model phase diagrams when $m_{\sigma}=500$ MeV computed with the parameters determined at physical point,~are presented in  Fig.~\ref{fig2b}.~The line types in the phase diagrams are labeled.~The dark red (yellow) color dashed line for $m_{\sigma}=500 (400)$ MeV which is the locus of the $O(4)$ critical points for the second order chiral phase transition in the light chiral limit of the RQM model,~ends at the tricritical  point (TCP) $(\mu_{\tcp},T_{\tcp})=(220.15,71.52)((194.6,65.9) ) $  MeV denoted by the solid sky blue triangle (square) and the first order chiral phase transition afterward is denoted by the deep red (yellow) color solid line in Fig.~\ref{fig2a}.~The experimental information for determining the model parameters are available only at  physical point.~One needs to know the values of $f_{\pi},~f_{K},~m_{\eta},~m_{\eta'}$ when $m_{\pi}=0,m_{K}=496$ MeV for fixing the model parameters.~The above mentioned inputs to find the model parameters in the light chiral limit,~are obtained by using the consistent method of large $N_{c}$, $U(3)$ chiral perturbation theory which is described briefly in the section~\ref{sec:IID}.~For the physical point parameters in the RQM model,~the dark green (blue) dash dot line for the $m_{\sigma}=500 (400)$ MeV depicting the crossover transition in  Fig.~\ref{fig2a},~ends at the second order critical point of the $Z(2)$ universality class known as the critical end point (CEP).~This CEP for  $m_{\sigma}=500 (400)$ MeV at $(\mu_{\cep},T_{\cep})=(265.42,38.71)((243.12,37.03))$  MeV is denoted by the solid circle in dark violet color and first order transition afterward is shown by the solid line in dark green (blue) color.~Since the pseudo-critical (critical) temperature $T_{c}^{\chi}(T_{c})$ for the crossover (second order) chiral transition occurring at $\mu=0$ for $m_{\sigma}=400$ MeV is smaller by 12.5 (14.4) MeV than the corresponding  $T_{c}^{\chi}(T_{c})$ for the chiral crossover (second order) transition  when  $m_{\sigma}=500$ MeV,~the phase diagrams for the $m_{\sigma}=400$ MeV in the $\mu-T $ plane of  Fig.\ref{fig2a},~are lying  below the corresponding phase diagrams for $m_{\sigma}=500$ MeV.

\begin{figure*}[htb]
\subfigure[]{
\label{fig3a} 
\begin{minipage}[b]{0.48\textwidth}
\centering \includegraphics[width=\linewidth]{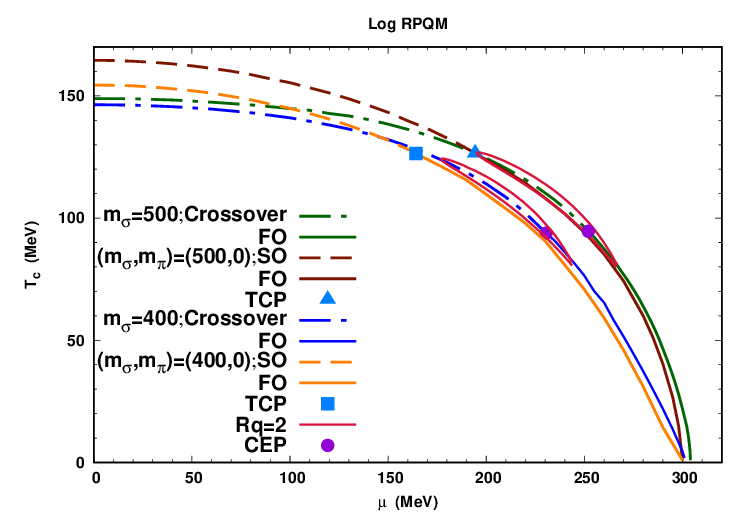}
\end{minipage}}
\hfill
\subfigure[]{
\label{fig3b} 
\begin{minipage}[b]{0.48\textwidth}
\centering \includegraphics[width=\linewidth]{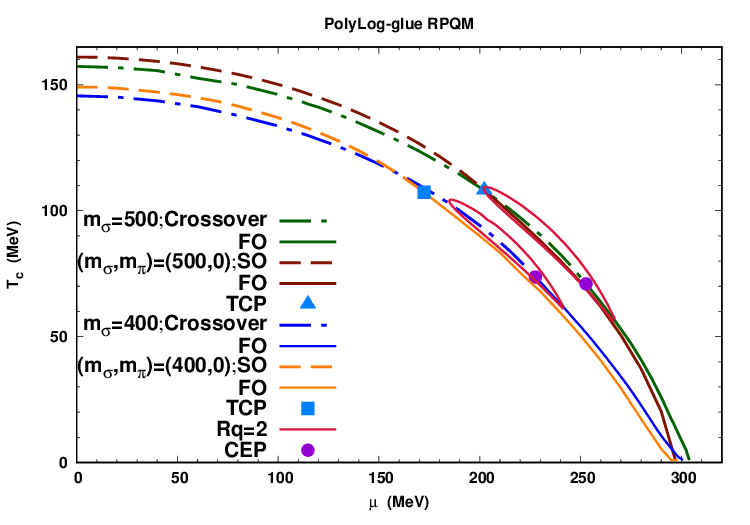}
\end{minipage}}
\caption{The phase diagrams for the $m_{\sigma}=400 \text{ and } 500$ MeV computed for the physical point and light chiral limit parameters,~are presented in the left panel (a) for the Log RPQM model and right panel (b) for the PolyLog-glue RPQM model.~Line types and position of the critical end points (CEP) and tricritical points (TCP) are labeled in the Figs.~The lines enveloping the CEP of different phase diagrams are the contours of constant quark number susceptibility ratio $R_q=2$.}
\label{fig:mini:fig3} 
\end{figure*}

~The CEP in the phase diagram is identified by finding the point in the  $\mu-T$ plane at which the quark number susceptibility diverges.~The significant enhancement of the quark number susceptibility in the regions of the phase diagram where the CEP lies,~can be used as a quantitative measure for mapping the critical regions around the CEP \cite{Asak,Bard,Berg,Hatta,Son,Misha3,Jeon,Schaefer:2006ds}.~The ratio of the quark number susceptibility computed from the models to the quark number susceptibility obtained from the free quark gas,~will be used to quantify the strength of critical fluctuations around the CEP.~This ratio represents normalized quark number susceptibility and it  is defined as the following.
\bqa
R_{q} = \frac {\chi_q} { \chi_{q}^{free}}
\eqa
The quark number susceptibility is  defined as : 
\bqa
\chi_{q}= -\frac{d^2 \ \Omega_{\rm RQM/RPQM}}{d {\mu}^2} 
\eqa
\bqa
\lim_{m_q \to 0} \chi(T,\mu) = \frac{\nu_q}{6} [T^2 + \frac{3\mu^2}{\pi^2}] = \chi^{free}_q .
\eqa

Here, $\nu_q=2{N_c}{N_f}$,~$N_c=3$ for three color charges and $N_f=3$ ($f=u,d,s$) for the 2+1 flavor in the RQM/RPQM model. Use of finite difference method for numerical computation of derivatives upto first order gives consistent results. But 
second order numerical differentiation using finite difference method becomes full of fluctuations. These fluctuations occur more often when number of  chemical potential dependent implicit variables increase. To circumvent these fluctuations, we use semi-analytic method \citep{Ghosh:2014,Gholami:2025} to compute second order differential of grand potential as, 
\begin{equation}
\label{Der}
\frac{\partial \Omega}{\partial X_i}=0
\end{equation}
\begin{align}
\nonumber
\frac{d \Omega}{d \mu}&=\frac{\partial \Omega }{\partial \mu}+\sum_i \frac{\partial \Omega}{\partial X_i}\frac{d X_i}{d \mu}\Big|_{X_i=\bar{X}_i} \\
&=\frac{\partial \Omega }{\partial \mu}
\end{align}
due to the equation (\ref{Der}).
\begin{align}
\frac{d^2\Omega}{d^2\mu}&=\frac{\partial^2\Omega}{\partial^2 \mu}+\sum_i \frac{\partial^2\Omega}{\partial \mu \partial X_i}\frac{d X_i}{d \mu}\Big|_{X_i=\bar{X}_i}
\end{align}
Here $X=\{\x,\y,\Phi,\bar{\Phi},\omega,\phi \}$ and $\bar{X}$ are  values of condensate.

The solid line in crimson red color enveloping the CEP for both $m_{\sigma}=500 \text{ and } 400$ MeV cases of the RQM model in  Fig.~\ref{fig2a},~is the contour of the  constant ratio $R_{q}=2$ representing the locus of the points in the $\mu-T$ plane,~on which the quark number susceptibility is double of the free quark gas susceptibility.~The CEP lies higher up on the temperature axis at ($\mu_{\cep},T_{\cep}$)=($165.24,97.52$) MeV in the phase diagram of the QM model under the s-MFA when $m_{\sigma}=500$ MeV in Fig.~\ref{fig2b}.~The constant $R_{q}=2$ susceptibility contour for the QM model in Fig.~\ref{fig2b},~is very small and extremely thin in the temperature direction. It is  small  in the chemical potential direction also.~When compared to the CEP position in the QM model and the critical region around it,~the RQM model CEP in  Fig.~\ref{fig2a},~not only shifts to higher $\mu$ and lower $T$ (lower right region of the $\mu-T$ plane) due to the effect of quark one-loop vacuum fluctuations but it also has a large,~broad and well developed critical region surrounding it.~Two different $\mu-T$ planes of the phase diagrams corresponding to  $m_{\pi}=0$ and  $m_{\pi}=138$ MeV,~are plotted in the same two dimensional Fig.~\ref{fig2a}.~The three dimensions of the $\mu-T-m_{\pi}$ coordinates constitute the real  physical picture.~The $R_{q}=2$ susceptibility contour in the $\mu-T$ plane for  $m_{\pi}=138$ MeV will have its projection (shadow) contour below in the $\mu-T$ plane for $m_{\pi}=0$.~The proximity of the TCP from the above mentioned projected contour,~would affect the critical fluctuations around the CEP in the $\mu-T$ plane above for  $m_{\pi}=138$ MeV.~The TCP of the RQM model when $m_{\sigma}=500$ MeV,~lies close to the boundary of  $R_{q}=2$ susceptibility contour around the CEP while the TCP for  $m_{\sigma}=400$ MeV,~is located noticeably away from the corresponding   $R_{q}=2$ contour  in  Fig.~\ref{fig2a}.~It means that the hidden tricritical point is likely to have some effect on the physics of critical fluctuations around the CEP~\cite{Hatta,xiong} for the case of $m_{\sigma}=500$ MeV and this effect will be negligible when  $m_{\sigma}=400$ MeV.~Polyakov loop potential is known to lift both the $\mu=0$ chiral crossover transition location and the CEP position,~up on the temperature axis in the $\mu-T$ plane when it gets combined with the  chiral effective potential of the QM model ~\cite{TiPQM3F,skrvkt24}.~For the PolyLog-glue PQM model in  Fig.~\ref{fig2b},~CEP position is significantly up on the temperature axis at ($\mu_{\cep},T_{\cep}$)=($137.7,129.6$) MeV but  $R_{q}=2$  contour around it,~is very much pinched and extremely thin similar to what is already reported in the PQM model calculations \citep{vkkr12,schafwag12}.

\begin{figure*}[htb]
\subfigure[]{
\label{fig4a} 
\begin{minipage}[b]{0.47\textwidth}
\centering \includegraphics[width=\linewidth]{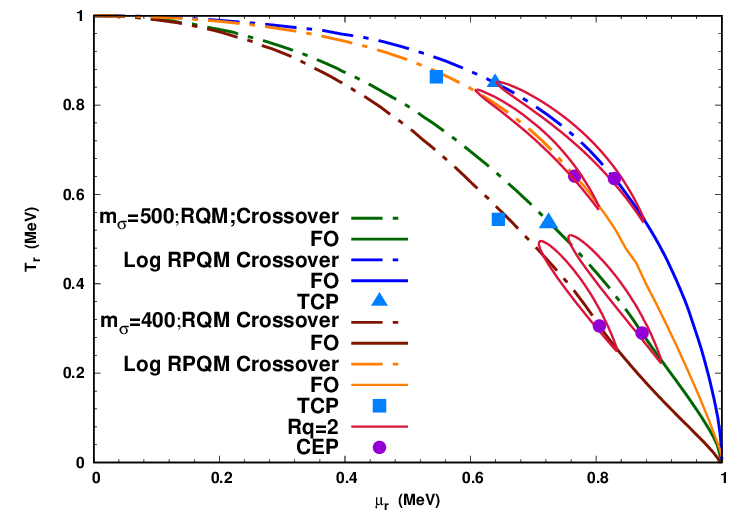}
\end{minipage}}
\hfill
\subfigure[]{
\label{fig4b} 
\begin{minipage}[b]{0.50\textwidth}
\centering \includegraphics[width=\linewidth]{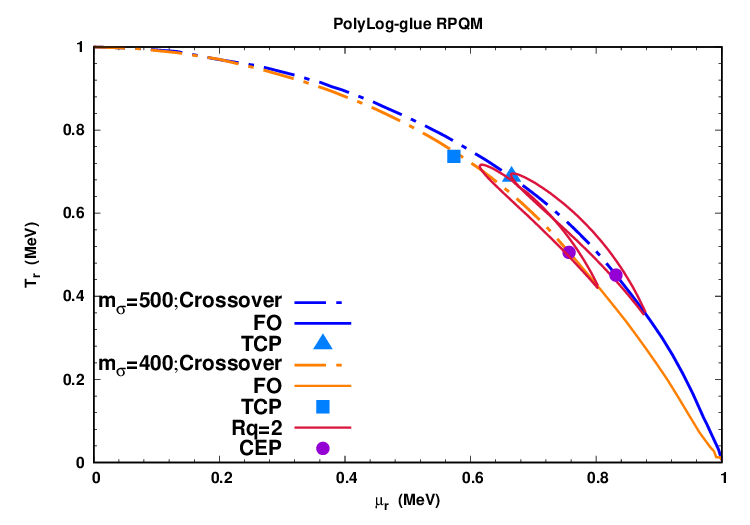}
\end{minipage}}
\caption{The phase diagrams for the physical point and light chiral limit parameters when the $m_{\sigma}=400 \text{ and } 500$ MeV are presented in the reduced temperature and chemical potential $\mu_{r}-T_{r}$ plane in the left panel (a) for the RQM and Log RPQM model and right panel (b) for the PolyLog-glue RPQM model.~Other features of the phase diagrams are same as in the Fig.\ref{fig:mini:fig3}.~The $T_{r}=\frac{T}{T_{c}^{\chi}}$ is obtained after dividing the  temperature $T$ by the pseudo-critical temperature $T_{c}^{\chi}$ at which the chiral crossover transition occurs on the temperature axis at $\mu=0$.~The $\mu_{r}=\frac{\mu}{\mu_{0}}$ is defined after dividing  the chemical potential $\mu$ by the highest chemical potential $\mu_{0}$ at which first order chiral phase transition on the $\mu$ axis occurs at $T=0$.}
\label{fig:mini:fig4} 
\end{figure*}

In comparison to the RQM model CEP  of Fig.~\ref{fig2a},~the position  of CEP for the Log and the PolyLog-glue RPQM model when  $m_{\sigma}=500(400)$ MeV,~shifts up respectively to the point ($\mu_{\cep},T_{\cep}$)=($252.0,94.6$) (($230.5,93.8$)) MeV in  Fig.~\ref{fig3a} and the point ($\mu_{\cep},T_{\cep}$)=($252.7,70.9$)(($227.6,73.6$)) MeV in Fig.~\ref{fig3b}.~The Polyakov loop potential with no quark back reaction in the Log RPQM model,~generates compression in the temperature direction of the critical region which gets large elongation in the chemical potential direction,~therefore the $R_{q}=2$ susceptibility contour looks pinched near the CEP in Fig.~\ref{fig3a} which is similar to what is found in the Ref.~\cite{schafwag12}.~The critical region is significantly large in the Log RPQM model,~but narrower than that of the RQM model in the temperature direction.~The Polyakov loop potential caused pinching of the $R_{q}=2$ contour near the CEP,~becomes small in  Fig.~\ref{fig3b} due to the smoothing effect of the quark back reaction in the PolyLog-glue RPQM model and one finds a large and broader (somewhat similar to the RQM model) critical region.~The above feature gets explained when one notes that the confinement-deconfinement transition  gets linked with the chiral phase transitions even at small temperatures and large chemical potentials ~\cite{BielichP} due to the presence of the quark back reaction in the Polyakov loop potential.~It is worth mentioning that for the light chiral limit phase diagram of the PolyLog-glue RPQM model with \(m_{\sigma}=500\) MeV, the tricritical point at \((\mu_{\mathrm{TCP}}, T_{\mathrm{TCP}}) = (202.2, 108.14)\) MeV lies just inside the boundary of the \(R_{q}=2\) contour in Fig.~\ref{fig3b}.~But for $m_{\sigma}=400$ MeV in the PolyLog-glue RPQM model,~the tricritical point at ($\mu_{\tcp},T_{\tcp}$)=($172.4,107.28$) MeV gets located well outside the $R_{q}=2$  contour.~The TCP at ($\mu_{\tcp},T_{\tcp}$)=($194.3,126.53$) MeV when $m_{\sigma}=500$ MeV for the light chiral limit phase diagram of the Log RPQM model,~also lies just on the boundary of the $R_{q}=2$ susceptibility contour in Fig.~\ref{fig3a}  but when $m_{\sigma}=400$ MeV in the Log RPQM model,~the TCP at ($\mu_{\tcp},T_{\tcp}$)=($164.21,126.41$) MeV,~gets located noticeably away from the $R_{q}=2$ contour.~Thus for $m_{\sigma}=500$ MeV,~the critical fluctuations around the CEP will be influenced by the presence of TCP in both the scenarios whether the quark back reaction is,~absent in the Log RPQM model or present in the  PolyLog-glue RPQM model but the influence of TCP would be negligible on the critical fluctuations around the CEP if  $m_{\sigma}=400$ MeV in both the versions of the RPQM model as the TCP lies  outside of the $R_{q}=2$ susceptibility contour.

The comparative effects of several factors on the phase diagrams, the CEP and TCP positions, and the extent of critical fluctuations around the CEP are analyzed in this work. These factors include: (1) the \(\sigma \)-meson mass, (2) quark one-loop vacuum fluctuations, (3) the Polyakov-loop potential, (4) the quark back-reaction in the improved Polyakov-loop potential, and (5) varying strengths of the vector interaction, which are discussed in the final subsection of the results. Higher values of \(m_{\sigma }\), the inclusion of vacuum fluctuations, and the choice of Polyakov-loop potentials all lead to a higher chiral crossover temperature, \(T_{c}^{\chi }\), at \(\mu=0\) (as detailed in Table~\ref{tab:table3}), shifting the entire phase diagram upward in the \(\mu\text{--}T\) plane \cite{vkkr23,skrvkt24}. Specifically, higher \(m_{\sigma }\) values and vacuum corrections shift the CEP toward larger \(\mu \) and lower \(T\) (to the right of the phase diagram). Conversely, the different Polyakov-loop potentials shift the CEP toward higher temperatures and lower chemical potentials (to the left) \cite{schafwag12,skrvkt24}. To isolate and clear up the distinct role of each factor, we present the phase diagrams in the reduced temperature (\(T_{r}\)) and reduced chemical potential (\(\mu _{r}\)) plane in Fig.~\ref{fig:mini:fig4}.~The reduced temperature and chemical potential are defined as \(T_{r} = T / T_{c}^{\chi}\) and \(\mu_{r} = \mu / \mu_{0}\), respectively. Here, \(T_{c}^{\chi }\) is the chiral crossover temperature at \(\mu=0\), and \(\mu _{0}\) is the critical chemical potential for the first-order phase transition at \(T=0\). Figures~\ref{fig4a} and \ref{fig4b} show that the curvatures of the crossover and first-order transition lines change when \(m_{\sigma }\) increases from 400 to 500 MeV in both the RQM and RPQM models. In the Log RPQM model without quark back-reaction (Fig.~\ref{fig4a}), the Polyakov-loop potential alters these curvatures significantly; the entire phase diagram shifts upward and exhibits pronounced bending in the upper-right region of the \(\mu_{r}\text{--}T_{r}\) plane. Conversely, when the quark back-reaction is included via the PolyLog-glue Polyakov-loop potential (Fig.~\ref{fig4b}), the transition lines display a smooth, uniform, and moderate modification compared to the reference RQM model. Consequently, the overall phase diagram for this case shifts only moderately upward in the \(\mu_{r}\text{--}T_{r}\) plane.

\begin{figure*}[htb]
\subfigure[]{
\label{fig5a} 
\begin{minipage}[b]{0.48\textwidth}
\centering \includegraphics[width=\linewidth]{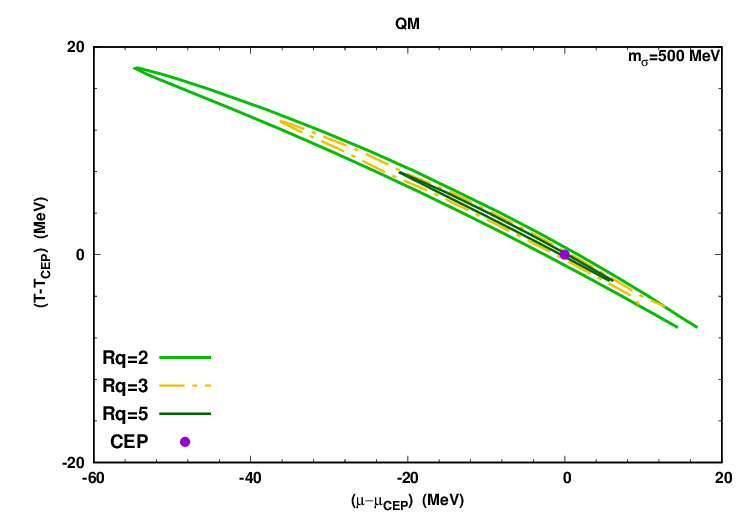}
\end{minipage}}
\hfill
\subfigure[]{
\label{fig5b} 
\begin{minipage}[b]{0.48\textwidth}
\centering \includegraphics[width=\linewidth]{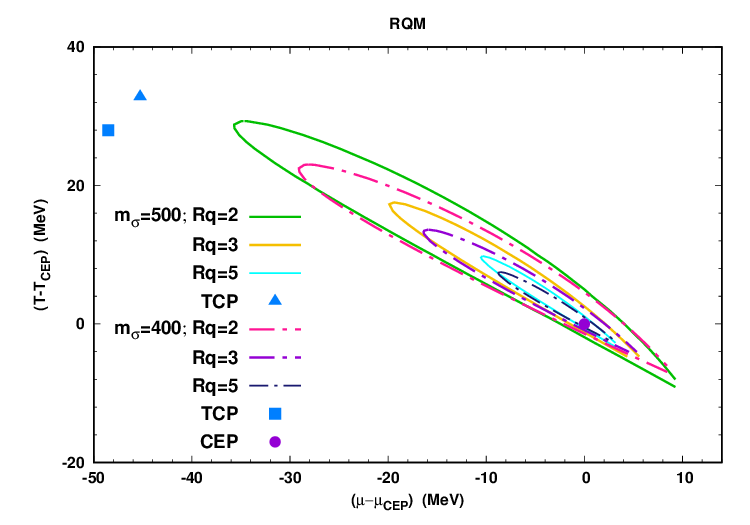}
\end{minipage}}
\caption{The $R_q=2,3$ susceptibility contours for the QM model are plotted in the left panel (a) while the $R_q=2,3,5$ contours for the RQM model are plotted in the right panel (b).}
\label{fig:mini:fig5}
\end{figure*} 

In the reference RQM model (Fig.~\ref{fig4a}), increasing the \(\sigma \)-meson mass from 400 to 500 MeV shifts the CEP from \((\mu_{r,\mathrm{CEP}}, T_{r,\mathrm{CEP}}) = (0.80, 0.305)\) to \((0.87, 0.289)\), and the TCP from \((\mu_{r,\mathrm{TCP}}, T_{r,\mathrm{TCP}}) = (0.644, 0.544)\) to \((0.724, 0.535)\). This represents a clear shift toward higher \(\mu _{r}\) and slightly lower \(T_{r}\).~As shown in Fig.~\ref{fig4a}, the Log form of the Polyakov-loop potential in the RPQM model significantly elevates the CEP temperature. For \(m_{\sigma} = 400\) (500) MeV, the CEP shifts to \((\mu_{r,\mathrm{CEP}}, T_{r,\mathrm{CEP}}) = (0.765, 0.640)\) \([(0.828, 0.635)]\), more than doubling its \(T_{r}\) value relative to the RQM baseline. A similarly large upward shift is observed for the TCP, which moves to \((\mu_{r,\mathrm{TCP}}, T_{r,\mathrm{TCP}}) = (0.545, 0.863)\) \([(0.639, 0.849)]\) for \(m_{\sigma} = 400\) (500) MeV.~In contrast, when the quark back-reaction is included via the PolyLog-glue Polyakov-loop potential (Fig.~\ref{fig4b}), the upward shift of the CEP is more moderate. The CEP moves to \((\mu_{r,\mathrm{CEP}}, T_{r,\mathrm{CEP}}) = (0.757, 0.505)\) \([(0.831, 0.450)]\) for \(m_{\sigma} = 400\) [500] MeV, which is roughly 1.5 times its RQM value. Under the same framework, the TCP is located at \((\mu_{r,\mathrm{TCP}}, T_{r,\mathrm{TCP}}) = (0.573, 0.736)\) \([(0.665, 0.687)]\). This position is noticeably higher than that of the reference RQM model, yet lower than the TCP coordinates found in the Log RPQM model.~Comparing these results to the \(m_{\sigma} = 400\) MeV baseline highlights that increasing the mass to \(m_{\sigma} = 500\) MeV causes the \(R_{q}=2\) susceptibility contours enveloping the CEP to expand along both the temperature and chemical potential axes across all models (RQM, Log RPQM, and PolyLog-glue RPQM). This reinforces our earlier conclusions: the reference RQM contours are well-rounded and broad due to quark one-loop vacuum fluctuations, whereas the contours in the Log RPQM model become compressed and pinched along the temperature axis due to the absence of the quark back-reaction. Crucially, this pinching in the temperature width is significantly moderated by the quark back-reaction in the PolyLog-glue RPQM model, resulting in larger, smoother contours that closely resemble the qualitative behavior of the reference RQM framework.

\begin{figure*}[htb]
\subfigure[]{
\label{fig6a} 
\begin{minipage}[b]{0.48\textwidth}
\centering \includegraphics[width=\linewidth]{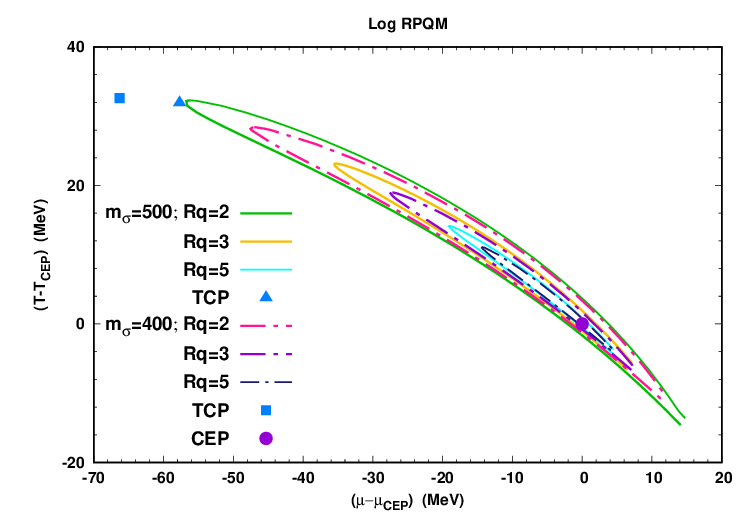}
\end{minipage}}
\hfill
\subfigure[]{
\label{fig6b}
\begin{minipage}[b]{0.48\textwidth}
\centering \includegraphics[width=\linewidth]{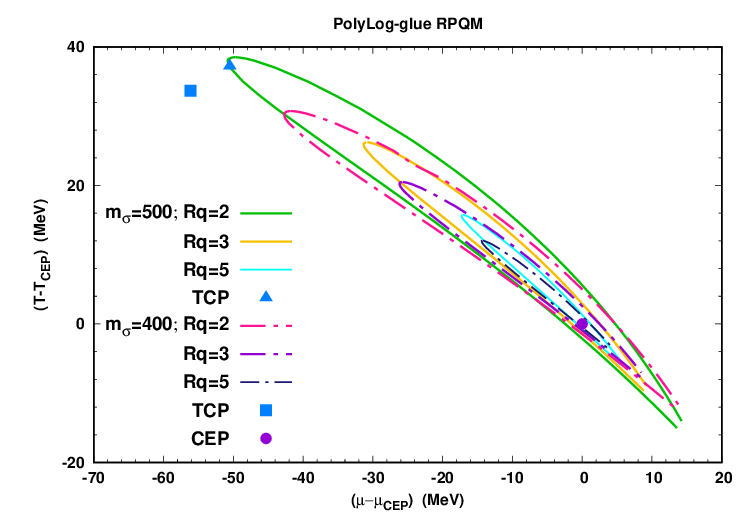}
\end{minipage}}
\caption{The $R_q=2,3,5$ susceptibility contours are plotted in the left panel (a) for the  Log RPQM model while the right panel (b) presents the contour plots for the PolyLog-glue RPQM model.}
\label{fig:mini:fig6} 
\end{figure*}

.

\subsection{Comparing size and shape of critical regions}
\label{sec:IIIB}

The susceptibility contours corresponding to three different constant ratios $R_{q}=2,3 \text{ and } 5$,~have been plotted for the relevant range of temperatures and chemical potentials in the Fig.~\ref{fig:mini:fig5} and Fig.~\ref{fig:mini:fig6} for analyzing the size and shape of the critical regions surrounding  the CEP of the QM,~RQM,~Log RPQM and PolyLog-glue RPQM model.~The extent of QM model $R_{q}=2$ contour in the Fig.~\ref{fig5a} is  25.0 MeV in the temperature direction (18.0 MeV above and 7.0 MeV below the $T_{\cep}$ on the $T$ axis) and 71.7 MeV  in the chemical potential direction (54.8 MeV lower  and 16.9 MeV higher than the $\mu_{\cep}$ on the $\mu$ axis).~For the $R_{q}=3$ contour of the QM model,~the size is very small as its $T$ axis spread is only 17.9 MeV (12.9 MeV above and 5.0 MeV below the $T_{\cep}$) whereas the $\mu$ axis spread is 49.5 MeV (36.2 MeV lower  and 13.3 MeV higher than the $\mu_{\cep}$).~The critical region around the CEP of QM model is quite small and very much compressed in the  temperature direction.

\begin{table*}[!htbp]
    \caption{The $T$ and $\mu$ axis spread of the $R_{q}=2,\bf {3 \text{ and } 5}$ contours in different model scenarios.~The size of the contours above and below $T_{\cep}$ are denoted respectively by the $\Delta T_{a}$ and $\Delta T_{b}$ while the total contour size on $T$ axis is $\Delta T$.~The size of the contours  on the lower side and higher side of the $\mu_{\cep}$ are denoted respectively by the $\Delta \mu_{l}$ and $\Delta \mu_{h}$ while the total contour size on $\mu$ axis is $\Delta \mu$.} 
    \label{tab:table4}
    \begin{tabular}{p{3cm} |p{2.4cm}|p{2.36cm}|p{2.5cm}|p{2.4cm}|p{2.2cm}|p{2.47cm}}
      \toprule 
        \hphantom{texttext} Models & \multicolumn{6}{|c}{Entries below the $\Delta T_{a}$, $\Delta T_{b}$, $\Delta T$, $\Delta\mu_{l}$, $\Delta\mu_{h}$ and $\Delta\mu$ correspond to the susceptibility ratio ${\bf {R_q}}= 2\textbf{(3)(5)}$} \ \\
         & $\Delta T_{a}$ & $\Delta T_{b}$ & $ \Delta T=\Delta T_{a}+\Delta T_{b}$ &  $\Delta\mu_{l}$ &  $\Delta \mu_{h}$& $\Delta \mu=\Delta \mu_{l}+\Delta \mu_{h}$ \\      
\hline
 RQM $(m_\sigma=400)$
&$23.0\textbf{(13.6)(7.5)}$  &$6.9 \textbf{(4.0)(2.6)}$ & $29.9\textbf{(17.6)(10.1)}$ & $29.1\textbf{(16.4)(8.8)}$& $8.4\textbf{(5.3)(2.9)}$ &$37.5\textbf{(21.7)(11.7)}$  \\ 
QMVT Ref.\cite{schafwag12}&  $31.4\textbf{(19.5)(11.5)}$ & $16.0\textbf{(7.7)(4.1)}$ & $47.4\textbf{(27.2)(15.6)}$ & $32.8\textbf{(17.6)(9.1)}$ &$10.1\textbf{(6.3)(3.3)}$ & $42.9\textbf{(23.9)(12.4)}$ \\ 

  Log RPQM &$28.4\textbf{(19.0)(11.1)} $ &$11.0\textbf{(6.8)(4.0)}$ &$39.4\textbf{(25.8)(15.1)}$& $47.5\textbf{(27.5)(14.4)}$&$11.5\textbf{(7.2)(4.3)}$& $59.0\textbf{(34.7)(18.7)}  $\\   
  PolyLog-glue RPQM  & $30.8\textbf{(20.5)(12.1)}$&  $ 12.6\textbf{(8.0)(3.8)}$ & $43.4\textbf{(28.5)(15.9)} $& $ 42.7\textbf{(26.2)(14.4)}$ &$ 13.9\textbf{(8.7)(4.0)}$ &$ 56.6\textbf{(34.9)(18.4)}$  \\ 
               
      RPQM-\text I & $28.7\textbf{(19.3)(11.7)}$ & $8.2\textbf{(5.7)(1.0)} $& $36.9\textbf{(25.0)(12.7)}$ &  $23.5\textbf{(14.6)(8.2)} $ & $5.4\textbf{(3.3)(0.6)}$&$28.9\textbf{(17.9)(8.8)}   $\\
   
       RPQM-\text II &$34.3\textbf{(23.1)(13.8)}$&$14.0\textbf{(7.6)(4.0)}  $&$48.3\textbf{(30.7)(17.8)}$& $ 37.1\textbf{(22.9)(12.9)}$ &$ 12.5\textbf{(6.3)(3.2)}$& $ 49.6\textbf{(29.2)(16.1)}$ \\ 
            
      PQMVT-\text I Ref.\cite{schafwag12} & $49.8\textbf{(33.9)(20.7)}$ &$41.1\textbf{(21.7)(11.2)} $& $90.9\textbf{(55.6)(31.9)} $ & $33.6\textbf{(19.5)(11.1)}$ & $12.6\textbf{(7.9)(4.4)} $ & $46.2\textbf{(27.4)(15.5)}$\\
      PQMVT-\text II Ref.\cite{schafwag12} & $49.9\textbf{(33.9)(20.3)}$ & $36.5\textbf{(19.8)(8.8)}$ & $86.4\textbf{(53.7)(29.1)}$ & $48.6\textbf{(27.5)(14.8)}$ & $15.4\textbf{(10.0)(5.5)}$ & $64.1\textbf{(37.5)(20.3)}$ \\
\hline 
      QM $(m_\sigma=500$) &$ 18.0\textbf{(12.9)}$ & $7.0\textbf{(5.0)}$& $25.0\textbf{(17.9)}$ & $54.8\textbf{(36.2)}$&$16.9\textbf{(13.3)} $&$71.7\textbf{(49.5)}$ \\ 
       RQM &$29.3\textbf{(17.6)(9.8)}$&$9.1\textbf{(5.7)(3.3)}$ &$38.4\textbf{(23.3)(13.1)}$ &$35.7\textbf{(19.9)(10.5)}$& $ 9.3\textbf{(5.6)(3.2)}$&$45.0\textbf{(25.5)(13.7)}$ \\

       Log RPQM &$32.3\textbf{(23.2)(14.2)}$&$14.6\textbf{(6.6)(4.6)}$& $46.9\textbf{(29.8)(18.8)}$& $56.7\textbf{(35.5)(19.1)}$ &$14.7\textbf{(7.1)(4.7)}$ &$ 71.4\textbf{(42.6)(23.8)}$\\             
                          
       PolyLog-glue RPQM &$38.5\textbf{(26.3)(15.8)}$&$15.0\textbf{(9.6)(5.9)}$ &$53.5\textbf{(35.9)(21.7)}$   &$50.8\textbf{(31.0)(17.3)}$ &$ 14.2\textbf{(9.2)(5.5)}$&    $ 65.0\textbf{(40.2)(22.8) } $ \\
      \hline      
            
      \hline 
    \end{tabular}
\end{table*}

~The $R_{q}=2$ contour of the RQM model when the $m_{\sigma}=500(400)$ MeV in the Fig.~\ref{fig5b},~is spread to  38.4 (29.9) MeV  \{29.3(23.0) MeV above and 9.1(6.9) MeV below the $T_{\cep}$\} on the $T$ axis while the corresponding spread on the $\mu$ axis is 45.0(37.5) MeV \{35.7(29.1) MeV lower and 9.3(8.4) MeV higher than the $\mu_{\cep}$)\}.~The smaller size of the $R_{q}=3$ contour for the $m_{\sigma}=500(400)$ MeV is 23.3(17.6) MeV on the $T$ axis \{17.6(13.6) MeV above and 5.7(4.0) MeV below the $T_{\cep}$\} and 25.5(21.7) MeV on the $\mu$ axis \{19.9(16.4) MeV lower and 5.6(5.3) MeV higher than the $\mu_{\cep}$\}.~The smallest spread of the $R_{q}=5$ contour corresponding to  the $m_{\sigma}=500(400)$ MeV,~is 13.1(10.1) MeV on the $T$ axis \{9.8(7.5) MeV above and 3.3(2.6) MeV below the $T_{\cep}$\} and 13.7(11.7) MeV on the $\mu$ axis \{10.5(8.8) MeV lower and 3.2(2.9) MeV higher than the $\mu_{\cep}$\}.~The extensions of the $R_{q}=2,3 \text{ and } 5$ contours for the $m_{\sigma}=400$ MeV along the $T$ axis and $\mu$ axis in the Fig.~\ref{fig5b},~are smaller than the corresponding contour sizes for the  $m_{\sigma}=500$ MeV in the sequential order of (8.5,5.7 and 3.0) MeV on the $T$ axis and (7.5,3.8 and 2.0) MeV on the $\mu$ axis.~One can see that the spreads of the RQM model susceptibility contours for the $m_{\sigma}=500(400)$ MeV in the temperature direction are nearly equivalent to their  spread in the chemical potential direction (only the $R_{q}=2$ contour is about 15 (20) percent larger along the $\mu$ axis (by 6.6 (7.6) MeV) when compared to its size along the $T$ axis).~It is worth pointing out that the consistently treated effect of the quark one-loop vacuum fluctuations in the  improved effective potential of the RQM model,~gives rise to a broad,~somewhat symmetrical critical regions enveloping the CEP,~which are larger in a direction perpendicular to  the crossover transition line.~Furthermore, in the RQM model, the TCP lies closer to the outer \(R_q = 2\) contour boundary when \(m_\sigma = 500\) MeV, whereas it shifts noticeably further away when \(m_\sigma = 400\) MeV. Table IV presents a comparison of the spread of the \(R_q = 2, 3,\) and \(5\) contours along the \(T\) and \(\mu \) axes for \(m_\sigma = 500\,(400)\) MeV across different model scenarios.

In the Fig.~\ref{fig6a},~the Log RPQM model $R_{q}=2$ contour for the $m_{\sigma}=500(400)$ MeV is extended to 46.9 ( 39.4) MeV on the $T$ axis \{32.3(28.4) MeV above and 14.6(11.0) MeV below the $T_{\cep}$\} and 71.4 (59.0) MeV on the $\mu$ axis \{56.7(47.5) MeV lower and 14.7(11.5) MeV higher than the $\mu_{\cep}$\}.~The smaller size of $R_{q}=3$ contour for the $m_{\sigma}=500(400)$ MeV is extended to 29.8(25.8) MeV on the $T$ axis \{23.2(19.0) MeV above and 6.6(6.8) MeV below the $T_{\cep}$\} and 42.6(34.7) MeV on the $\mu$ axis \{35.5(27.5) MeV lower and 7.1(7.2) MeV higher than the $\mu_{\cep}$\}.~The extent of smallest $R_{q}=5$ contour when the $m_{\sigma}=500(400)$ MeV,~is 18.8 (15.1) MeV on the $T$ axis \{14.2(11.1) MeV above and 4.6(4.0) MeV below the $T_{\cep}$\} and 23.8 (18.7) MeV on the $\mu$ axis \{19.1(14.4) MeV lower and 4.7(4.3) MeV higher than the $\mu_{\cep}$\}.~The $T$ and $\mu$ axes extensions of the $R_{q}=2,3 \text{ and } 5$ contours in the Fig.~\ref{fig6a} for the $m_{\sigma}=400$ MeV,~are smaller than the corresponding contour sizes for the  $m_{\sigma}=500$ MeV in the sequential order of (7.5, 4 and 3.7) MeV on the $T$ axis and (12.4, 7.9 and 5.1) MeV on the $\mu$ axis.~The chemical potential direction  stretching of the Log RPQM model $R_{q}=2,3 \text{ and } 5$ contours for the $m_{\sigma}=500(400)$ MeV,~are noticeably larger than their respective extension in the temperature direction,~respectively, by a factor of 1.52 (1.50), 1.43 (1.34) and 1.27 (1.24).~The smoothing effect of quark one-loop vacuum fluctuation that generates broader and larger critical region perpendicular to the crossover transition line in the RQM model,~looks somewhat compromised in the presence of Log Polyakov loop potential and one finds that the width of contours near the CEP,~become narrow  owing to its compression in the temperature direction similar to what is noticed in the Ref.~\cite{schafwag12}.~Note that in the Log RPQM model, when \(m_\sigma = 400\) MeV, the TCP lies noticeably away from the boundary of the \(R_q = 2\) contour, whereas for \(m_\sigma = 500\) MeV, the TCP lies exactly on the boundary of the \(R_q = 2\) contour.

The spread of the PolyLog-glue RPQM model $R_{q}=2$ contour in the Fig.~\ref{fig6b} for the $m_{\sigma}=500(400)$ MeV is 53.5 (43.4) MeV on the $T$ axis \{38.5(30.8) MeV above and 15(12.6) MeV below the $T_{\cep}$\} and 65(56.6) MeV on the $\mu$ axis \{50.8(42.7) MeV lower and 14.2(13.9) MeV higher than the $\mu_{\cep}$\}.~The smaller size of the $R_{q}=3$ contour for the $m_{\sigma}=500(400)$ MeV,~is spread 35.9(28.5) MeV on the $T$ axis \{26.3(20.5) MeV above and 9.6(8) MeV below the $T_{\cep}$\} and 40.2(34.9) MeV on the $\mu$ axis \{31.0(26.2) MeV lower and 9.2(8.7) MeV higher than the $\mu_{\cep}$\}.~The smallest spread of the $R_{q}=5$ contour for the $m_{\sigma}=500(400)$ MeV is 21.7 (15.9) MeV on the $T$ axis \{15.8(12.1) MeV above and 5.9(3.8) MeV below the $T_{\cep}$\} and 22.8 (18.4) MeV on the $\mu$ axis \{17.3(14.4) MeV lower and 5.5(4.0) MeV higher than the $\mu_{\cep}$\}.~The $T$  and $\mu$ axes extensions of the $R_{q}=2,3 \text{ and } 5$ contours for the $m_{\sigma}=400$ MeV in the Fig.~\ref{fig6b},~are smaller than the corresponding contour sizes for the  $m_{\sigma}=500$ MeV in the sequential order of (10.1, 7.4  and 5.8) MeV on the $T$ axis and (8.4, 5.3 and 4.4) MeV on the $\mu$ axis.~The stretching of the $R_{q}=2,3 \text{ and } 5$ contours for the $m_{\sigma}=500(400)$ MeV in the chemical potential direction for the PolyLog-glue RPQM model,~are moderately larger than their respective extension in the temperature direction respectively by a factor of 1.21 (1.3), 1.12 (1.22) and 1.05 (1.16).~We point out that the effect of quark back-reaction in the PolyLog-glue RPQM model gives rise to a smaller spread of the contours in the chemical potential direction, whereas the spread in the temperature direction becomes larger in a similar proportion, when compared with the corresponding spread of contours observed in Fig.~\ref{fig6a} for the Log RPQM model.~Thus, the temperature direction compression in the width of the contours near the CEP becomes smaller due to the smoothing effect of the quark back-reaction in the PolyLog-glue model, where the contours are noticeably broader than the corresponding Log RPQM model contours. Furthermore, in the PolyLog-glue RPQM model, the TCP lies away from the boundary of the \(R_{q}=2\) contour when \(m_{\sigma}=400\) MeV, while it lies just inside the \(R_{q}=2\) contour when \(m_{\sigma}=500\) MeV.

\begin{figure*}[htb]
\subfigure[]{
\label{fig7a} 
\begin{minipage}[b]{0.49\textwidth}
\centering \includegraphics[width=\linewidth]{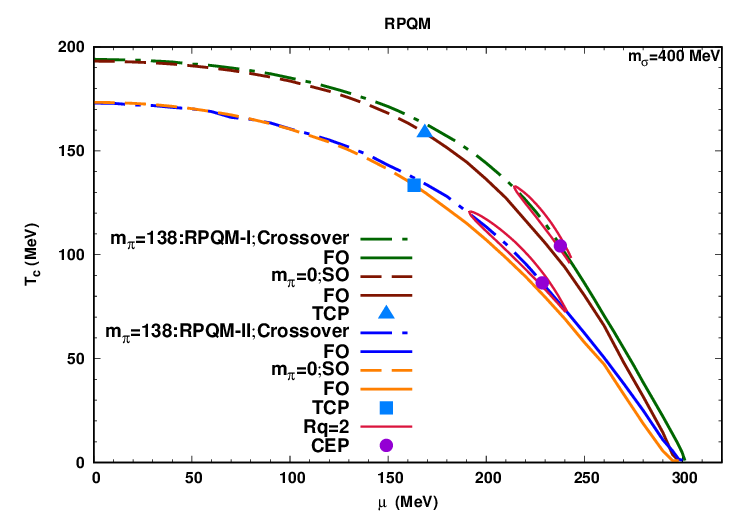}
\end{minipage}}
\hfill
\subfigure[]{
\label{fig7b} 
\begin{minipage}[b]{0.48\textwidth}
\centering \includegraphics[width=\linewidth]{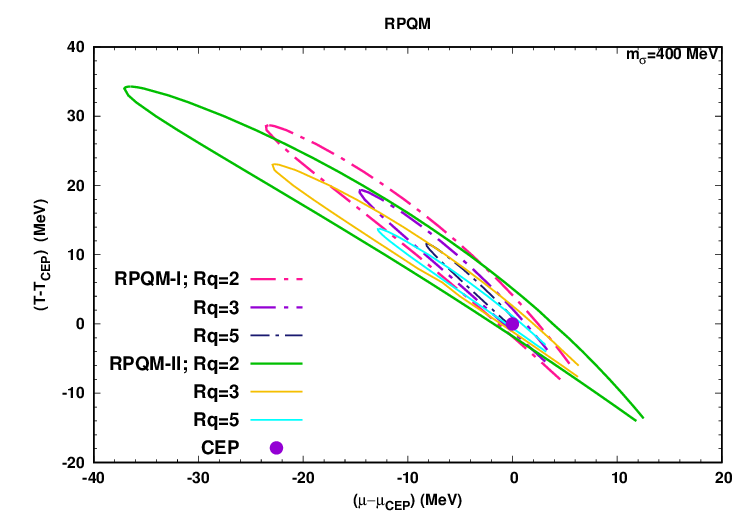}
\end{minipage}}
\caption{The RPQM-I and RPQM-II model phase diagrams for the light chiral limit and physical point parameters are presented in the left panel (a) while the contours for the constant  susceptibility ratio $R_{q}=2,3 \text{ and } 5$ are presented in the right panel (b).~The Log RPQM model in the present work with the $T_{0}=270$ MeV,~$m_{\sigma}=400$ MeV and no quark back reaction,~is termed as the RPQM-I model.~The PolyLog-glue RPQM model with the quark back reaction where the $T_{0}=270$ MeV and $m_{\sigma}=400$ MeV,~has been termed as the RPQM-II model}
\label{fig:mini:fig7}
\end{figure*} 

\subsection{Comparing RQM model critical regions with results of Ref. \cite{schafwag12}}
\label{sec:IIIC}

Recall that after adding  the quark one-loop vacuum fluctuations to the effective potential of the QM model,~several studies ~\cite{guptiw,chatmoh1,vkkr12,schafwag12} have used the curvature masses of the mesons to fix the model parameters.~Since the curvature masses incorporate the quark one-loop vacuum corrections in the meson self energies only at the zero momentum,~the curvature mass based parameter fixing become inconsistent in the above studies of the quark meson/Polyakov quark meson with the vacuum term (QMVT/PQMVT) model.~Recent two and 2+1 flavor RQM/RPQM model studies \cite{RaiTiw22,raiti23,vkkr23,skrvkt24} with the consistent on-shell renormalized  parameters,~have shown that the calculated shift,~in the position of the CEP to the lower right corner in the $\mu-T$ plane of the  phase diagram,~is quite large due to  the overestimation of the effect of quark one-loop corrections in the above mentioned QMVT/PQMVT model studies.
 Therefore, it is important to compare the shape and size of the critical regions around the CEP reported by Schaefer et al. \cite{schafwag12} with those obtained in the present work. Specifically, we compare the earlier 2+1 flavor curvature mass-parameterized QMVT/PQMVT model against our 2+1 flavor RQM/RPQM model. For a direct comparison, both models employ \(m_\sigma = 400\) MeV and a logarithmic Polyakov-loop potential with \(T_0 = 270\) MeV for the pure Yang-Mills \(SU_c(3)\) gauge theory.

~The $R_q=2$ susceptibility contour in the Fig.3(b) of the Ref.~\cite{schafwag12} (after extracting the data) for the QMVT model is extended around its CEP at ($\mu_{\cep},T_{\cep}$)=(286,32) MeV,~to about 31.4 MeV above and 16 MeV below the $T_{\cep}$ on the $T$ axis while the spread of the contour on the $\mu$ axis is 32.8 MeV to the lower left and 10.1 MeV to the higher right side of the $\mu_{\cep}$.~In contrast to the above,~the CEP in our RQM model work is positioned higher up and noticeably left at the ($\mu_{\cep},T_{\cep}$)=(243.12,37.03) MeV around which the $R_{q}=2$  contour (in the Fig.~\ref{fig5b} for the $m_{\sigma}=400$ MeV)~is extended on the $T$ axis to 23.0 MeV above and 6.9 MeV below the $T_{\cep}$ while its $\mu$ axis spread is 29.1 MeV  on the lower left and 8.4 MeV on the higher right side of the $\mu_{\cep}$.~The total size of 29.9 MeV of the RQM model contour on the $T$ axis,~is smaller than the  total  $T$ axis size of 47.4 MeV for the QMVT model contour whereas the RQM model contour total $\mu$ axis size of  37.5 MeV,~is also smaller than the total size of 42.9 MeV on the $\mu$ axis for the QMVT model. Note that, below the CEP along the \(T\)-axis, the \(R_q = 2\) contour of the QMVT model in Fig. 3(b) of Ref.~\cite{schafwag12} is 9.1 MeV larger than that of the RQM model in  Fig.~\ref{fig5b}. Conversely, above the CEP, the QMVT model contour is only 8.4 MeV larger than the corresponding RQM model contour. Below the CEP, the QMVT contour exhibits a narrow, neck-like bending structure. This feature in the work \cite{schafwag12} originates from the large curvature of the first-order transition line below the critical end point \((\mu_{\text{\cep}}, T_{\text{\cep}}) = (286, 32)\) MeV; here, the temperature must drop from 32 MeV to \(T \simeq 0\) within a narrow chemical potential span of \((\mu_0 - \mu_{\text{CEP}}) = 300.13 - 286 = 14.13\) MeV before intersecting the axis at \(\mu_0 = 300.13\) MeV.~In contrast, the RQM model $R_{q}=2$ contour in Fig.~\ref{fig2a} (and also in Fig.~\ref{fig5b}) appears smooth and more symmetric, lacking any narrow bending below its CEP. This difference arises because the curvature of the first-order transition line below \((\mu_{\text{\cep}}, T_{\text{\cep}}) = (243.12, 37.03)\) MeV is noticeably smaller, as the temperature drops from approximately 37 MeV to \(T \simeq 0\) over a significantly larger chemical potential span of \((\mu_0 - \mu_{\text{CEP}}) = 302 - 243 \simeq 59\) MeV before reaching \(\mu_0 = 302\) MeV.

~Furthermore, the RQM model contour closes at $(\mu,T)=(214.02,60.05)$ MeV on the 
upper-left side and  $(\mu,T)=(251.56,30.1)$ MeV on the lower-right side, while the QMVT model contour closes  at $(\mu,T)=(253.2,63.4)$ MeV on the upper-left and at $(\mu,T)=(296.1,16.0)$ MeV on the lower-right side.~Note that the complete RQM model phase diagram in the $\mu-T$ plane of  Fig.~\ref{fig2a} for  $m_{\sigma}=400$ MeV,~would lie below the QMVT model phase diagram in Fig.~1(b) of  Ref.~\cite{schafwag12}, because ($\mu_{0},T_{c}^{\chi}$)=(302,121.1) MeV in the RQM model whereas  ($\mu_{0},T_{c}^{\chi}$)=(300.13,144.1) MeV in the QMVT model.~Recall that  $T_{c}^{\chi}$ is the pseudo-critical temperature of the chiral crossover transition occurring on the temperature axis at $\mu=0$, while $\mu_{0}$ is the  highest chemical potential at which the first order chiral transition occurs at $T \simeq 0$.~In order to see the real difference between the size and shape of the critical regions for the RQM and QMVT models,~one must compare the quantities in terms of the reduced chemical potential $\mu_{r}=\frac{\mu}{\mu_{0}}$ and reduced temperature $T_{r}=\frac{T}{T_{c}^{\chi}}$.~In terms of the reduced variables, the upper-left boundaries of the RQM and QMVT model \(R_q = 2\) contours close at \((\mu_r, T_r) = (0.7086, 0.495)\) and \((\mu_r, T_r) = (0.8436, 0.440)\), respectively.~Conversely, on the lower-right side of the phase diagram, the RQM and QMVT model contours close at \((\mu_r, T_r) = (0.8329, 0.248)\) and \((\mu_r, T_r) = (0.9865, 0.111)\), respectively.~Note also that the RQM and QMVT model reduced CEPs lie at  $(\mu_{r\cep},T_{r\cep})=(0.805,0.306)$ and $(\mu_{r\cep},T_{r\cep})=(0.9529,0.2268)$, respectively.

\begin{figure*}[htb]
\subfigure[]{
\label{fig8a} 
\begin{minipage}[b]{0.50\textwidth}
\centering \includegraphics[width=\linewidth]{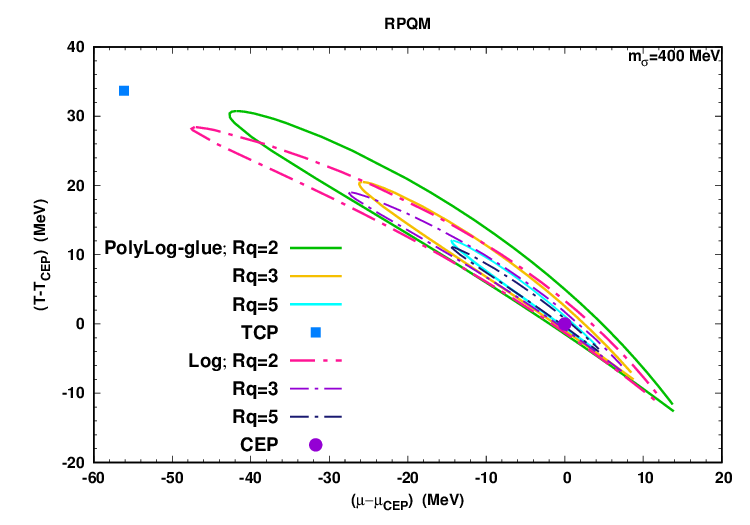}
\end{minipage}}
\hfill
\subfigure[]{
\label{fig8b}
\begin{minipage}[b]{0.48\textwidth}
\centering \includegraphics[width=\linewidth]{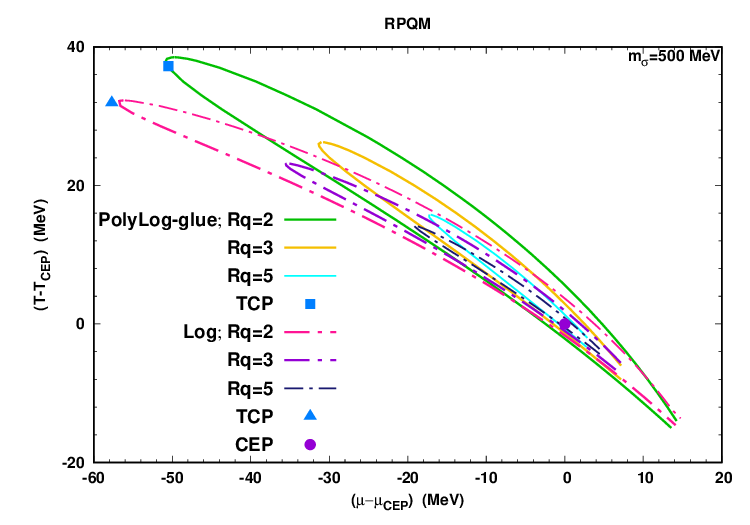}
\end{minipage}}
\caption{With the $T_{0}=T_{c}^{glue}=187$ MeV for the Polyakov loop potentials in the Log and PolyLog-glue RPQM model,~the $R_q=2,3,5$ susceptibility contours are presented in the left panel (a) for the case of $m_{\sigma}=400$ MeV and right panel (b) for the case of $m_{\sigma}=500$ MeV}
\label{fig:mini:fig8} 
\end{figure*}  

The model setting of the Log form of Polyakov-loop potential with a constant $T_0=270$ MeV for the pure Yang-Mills $SU_c(3)$ gauge theory in the work of Schaefer et al. has been termed PQMVT-I for the purpose of comparison. On the other hand, their scenario with the quark matter back-reaction in the Log Polyakov-loop potential is termed PQMVT-II, where $T_{0}(\mu)$ is chemical-potential-dependent. For comparison with the work in Ref.~\cite{schafwag12}, the Log RPQM model in our work with $T_{0}=270$ MeV, $m_{\sigma}=400$ MeV, and no quark back-reaction has been termed the RPQM-I model. The enhancement of the RQM model with the improved PolyLog-glue form of the Polyakov-loop potential~\cite{Haas,Redlo,BielichP} accounts for the quark back-reaction in our work; this model setting with the parameter $T_{c}^{glue}=T_{0}=270$ MeV has been termed the RPQM-II model.~The CEP in our RPQM-I (RPQM-II) model without (with) the quark back-reaction in Fig.~\ref{fig7a} is located significantly further to the left and noticeably higher up in the phase diagram at $(\mu_\cep,T_\cep)=(237.8\,(228.54),104.23\,(86.47))$~MeV compared to the PQMVT-I (PQMVT-II) model CEP without (with) the quark back-reaction, which lies at $(\mu_\cep,T_\cep)=(283\,(280),90\,(83))$~MeV in Fig.~1(b) of Ref.~\cite{schafwag12}.

   After extracting the data from Fig.~3(a) of Ref.~\cite{schafwag12}, one finds that the $R_q=2$ susceptibility contour for the PQMVT-I (PQMVT-II) model extends around its CEP along the $T$-axis to about 49.81 (49.96) MeV above and 41.13 (36.48) MeV below $T_{\cep}$. Along the $\mu$-axis, the spread of the contour is 33.63 (48.61) MeV to the left and 12.56 (15.45) MeV to the right of $\mu_{\cep}$. In contrast, Fig.~\ref{fig7b} of the present work shows that the RPQM-I (RPQM-II) model susceptibility $R_{q}=2$ contour extends along the $T$-axis to 28.7 (34.3) MeV above and 8.2 (14) MeV below $T_{\cep}$, while its spread along the $\mu$-axis is 23.5 (37.1) MeV to the left and 5.4 (12.5) MeV to the right of $\mu_{\cep}$.~The total $T$-axis extension of the RPQM-I (RPQM-II) model contour is 36.9 (48.3) MeV, which is significantly smaller than the total size of 90.94 (86.44) MeV for the corresponding PQMVT-I (PQMVT-II) contour. Similarly, the total $\mu$-axis extension of the RPQM-I (RPQM-II) contour is 28.9 (49.6) MeV, which is roughly 17.3 (14.5) MeV smaller than the total size of 46.2 (64.1) MeV observed for the PQMVT-I (PQMVT-II) contour. The reason for this structural variation is that the RPQM-I (RPQM-II) critical regions are centered around a CEP located at a significantly lower chemical potential ($\mu_\cep=237.8\,(228.54)$ MeV) in the phase diagram, whereas the corresponding temperatures are moderately higher ($T_\cep=104.23\,(86.47)$ MeV).~Furthermore, to gain a proper perspective on this difference, one should note that $T_{c}^{\chi}=194.1$ MeV at $\mu=0$ for the RPQM-I phase diagram in Fig.~\ref{fig7a}, and its CEP sits noticeably higher in temperature and further to the left (lower $\mu$) compared to the CEP in Fig.~1(b) of the PQMVT-I model, where $T_{c}^{\chi}=205$ MeV at $\mu=0$. While this detailed comparative analysis is explicitly provided here for the \(R_{q}=2\) contour, one can see a similar pattern of comparative differences for the \(R_{q}=3 \text{ and } 5\) contours by looking at the numbers for the RPQM-I (RPQM-II) model in Table~\ref{tab:table4} and comparing those with the corresponding numbers after extracting the data for the PQMVT-I (PQMVT-II) model from Fig.~3(a) of Ref.~\cite{schafwag12}.
   
The total size of 90.94 MeV along the $T$-axis (with a large extension of 41.1 MeV below $T_{\cep}$) for the PQMVT-I model contour with no quark back-reaction is nearly double its $\mu$-axis total size of 46.19 MeV. Furthermore, the width of the contour becomes successively narrower and more compressed in the temperature direction below the CEP in Fig.~3(a) of Ref.~\cite{schafwag12}. Compared to the QMVT model, the significantly enhanced critical region with a relatively compressed width for the PQMVT-I model was attributed to the observations that (1) the Polyakov loop modifies the quark determinant mostly at moderate chemical potentials, and (2) the PQMVT-I model CEP is located at a significantly higher temperature ($T_{\cep}=90$ MeV), whereas the CEP in both the QMVT and PQMVT models is positioned at larger chemical potentials around $280$--$285$ MeV~\cite{schafwag12}. The PQMVT model contour, with its successively decreasing width below $T_{\cep}$, exhibits a pronounced, neck-like narrow structure. This features originates from the large curvature of the first-order transition line below the CEP $(\mu_{\cep}, T_{\cep}) = (283, 90)$ MeV in Fig.~3(a) of Ref.~\cite{schafwag12}, as the high temperature of 90 MeV must drop to $T \simeq 0$ within a narrow chemical potential span of only $(\mu_{0}-\mu_{\cep}) = 299.54 - 283 = 16.54$ MeV before the first-order transition line reaches its maximum value of $\mu_{0}=299.54$ MeV at $T \simeq 0$ along the chemical potential axis.~Note that since the total $T$-axis contour size of 46.3 MeV is only 9.9 MeV larger than the $\mu$-axis total contour size of 36.4 MeV, the RPQM-I model $R_{q}=2$ contour (for no quark back-reaction) looks more symmetric and smoother. Its narrowing below the CEP in Fig.~\ref{fig7b} is quite small when compared to the RQM model $R_{q}=2$ contour for $m_{\sigma}=400$ MeV in Fig.~\ref{fig5b}. Furthermore, in contrast to the PQMVT-I model contour, the RPQM-I model contour is significantly smaller in the $T$ direction, appearing smooth and more symmetric with no narrow bending below its $T_{\cep}$. This difference arises because the curvature of the first-order transition line below the CEP $(\mu_{\cep}, T_{\cep}) = (237.8, 104.23)$ MeV is noticeably smaller in Fig.~\ref{fig7a}; here, the relatively high temperature of approximately 104 MeV must decrease to $T \simeq 0$ over a significantly larger chemical potential span of $(\mu_{0}-\mu_{\cep}) = 301.2 - 237.8 \simeq 63.4$ MeV before the first-order transition line reaches its maximum value of $\mu_{0}=301.2$ MeV at $T \simeq 0$ along the chemical potential axis.

The presence of the differently treated quark back-reaction in the PolyLog-glue Polyakov-loop potential causes a significant smoothing effect in the RPQM-II model. This is evident because the total $T$-axis size of 48.3 MeV for the $R_{q}=2$ contour is nearly equal to its total $\mu$-axis size of 49.6 MeV in Fig.~\ref{fig7b}, resulting in an overall contour that looks quite symmetric and smooth, where narrowing below the CEP disappears. In contrast, Ref.~\cite{schafwag12} implemented a $\mu$-dependence in $T_{0}$ to account for the quark back-reaction in the PQMVT-II model; there, the total $T$-axis size of 86.44 MeV (with a still large extension of 36.48 MeV below $T_{\cep}$) for the $R_{q}=2$ contour is approximately 1.35 times its total $\mu$-axis size of 64.06 MeV.~The width of the PQMVT-II model contour also becomes successively more compressed below its CEP, similar to the PQMVT-I scenario, but the narrowing in the contour shape is less pronounced than what is observed for the PQMVT-I model. Note also that in the PQMVT-II model, $T_{c}^{\chi}=205$ MeV at $\mu=0$, which is identical to that in the PQMVT-I model with no quark back-reaction.

The RPQM-I (RPQM-II) model $R_{q}=2$ contour closes at $(\mu,T)=(214.3\,(191.47),132.93\,(120.75))$ MeV on the upper-left side and $(\mu,T)=(243.23\,(241.06),96.0\,(72.47))$ MeV on the lower-right side, while the PQMVT-I (PQMVT-II) model $R_{q}=2$ contour closes at $(\mu,T)=(249.37\,(231.39),139.81\,(129.96))$ MeV on the upper-left and at $(\mu,T)=(295.56\,(295.45),48.87\,(43.52))$ MeV on the lower-right side. Note that the complete RPQM-I (RPQM-II) model phase diagram for $m_{\sigma}=400$ MeV in the $\mu-T$ plane of Fig.~\ref{fig7a} would lie below the PQMVT-I (PQMVT-II) model phase diagram in Fig.~1(b) of Ref.~\cite{schafwag12}, because ($\mu_{0},T_{c}^{\chi}$)=(301.2\,(300.0),194.1\,(179.2)) MeV in the RPQM-I (RPQM-II) model whereas ($\mu_{0},T_{c}^{\chi}$)=(299.54\,(299.54),205\,(205)) MeV in the PQMVT-I (PQMVT-II) model.~By comparing the quantities in terms of the reduced chemical potential $\mu_{r}$ and reduced temperature $T_{r}$, one can see the real difference between the size and shape of the critical regions for the RPQM-I (RPQM-II) and PQMVT-I (PQMVT-II) models. In terms of these reduced variables, the RPQM-I (RPQM-II) and PQMVT-I (PQMVT-II) model $R_{q}=2$ contours close at $(\mu_{r},T_{r})=(0.7114\,(0.638),0.6848\,(0.6738))$ and $(\mu_{r},T_{r})=(0.832\,(0.772),0.682\,(0.633))$, respectively, on the upper-left side. On the lower-right side of the phase diagram, the RPQM-I (RPQM-II) and PQMVT-I (PQMVT-II) model contours close at $(\mu_{r},T_{r})=(0.807\,(0.803),0.4945\,(0.404))$ and $(\mu_{r},T_{r})=(0.986\,(0.986),0.238\,(0.212))$, respectively. Here, one should also take note of the CEP locations for the RPQM-I (RPQM-II) and PQMVT-I (PQMVT-II) models in terms of reduced variables, which lie at $(\mu_{r\cep},T_{r\cep})=(0.7895\,(0.761),0.5369\,(0.5))$ and $(\mu_{r\cep},T_{r\cep})=(0.944\,(0.934),0.439\,(0.404))$, respectively.

Note that the positions of the PolyLog-glue RPQM-II model contours (with quark back-reaction) along the temperature axis lie below the Log-RPQM-I model contours, despite the former having a larger $T$- and $\mu$-axis spread in Fig.~\ref{fig7b}. A similar temperature-direction positioning is observed in Fig.~3(a) of Ref.~\cite{schafwag12}, where the PQMVT-II model contours with quark back-reaction lie below the PQMVT-I contours, even though their spread is slightly larger along the $T$-axis with more elongation along the $\mu$-axis. However, this positioning along the temperature axis reverses in Fig.~\ref{fig8a} when $T_{0}=187$ MeV in the Polyakov-loop potentials for $2+1$ flavors of dynamical quarks; here, the Log-RPQM model contours with no quark back-reaction lie below the PolyLog-glue RPQM model contours in the temperature direction.~Furthermore, the Log-RPQM model contours exhibit a larger $\mu$-axis and a slightly smaller $T$-axis spread compared to the PolyLog-glue model contours. These distinct features of the Log-RPQM and PolyLog-glue RPQM models become even more pronounced for the case of $m_{\sigma}=500$ MeV in Fig.~\ref{fig8b}.

\begin{table*}[!htbp]
    \caption{ The positions of critical end points in different model scenarios when the strength of vector interaction coupling $(g_\omega)$ is varied for the $m_{\sigma}=500$ MeV.~The $(g_\omega=0)$ shows the earlier  cases when vector interaction is absent. }
    \label{tab:table5}
    \begin{tabular}{p{2.6cm}| p{2.4cm}|p{2.4cm} |p{2.4cm}|p{2.4cm}|p{2.4cm}|p{2.4cm}}
         
      \hline
        \hphantom{text}   Models   & \multicolumn{5}{|c}{$(\mu_{\cep},T_{\cep})$ MeV} \ \\
      
	   &$g_{\omega} = 0.0$& $g_{\omega} = 1.0$& $g_{\omega}=1.5$& $g_{\omega}=2.0$ &$g_{\omega}=2.5$ &$g_{\omega}=2.75$ \\
	   \hline
       RQM & $(265.42,38.71)$& $(273.5,34.05)$ & $(282.49,28.23)$ & $(292.83,20.33)$ & $(302.17,10.68)$ & $(305.83,4.06)$ \\
     
      Log RPQM & $(252.0,94.6)$ & $(264.11,86.0)$ & $(277.02,74.1)$ & $(290.91,55.5)$ & $(301.82,31.1)$ & $(305.71,13.5)$ \\
      
       Polylog-glue RQM & $(252.7,70.9)$& $(263.4,62.7)$ & $(275.9,50.87)$ & $(289.96,33.46)$ & $(302.02,12.53)$ & $(305.60,5.4)$  \\
     
\hline 
  \end{tabular} 
\end{table*}       

\begin{figure*}[htb]
\subfigure[]{
\label{fig9a} 
\begin{minipage}[b]{0.47\textwidth}
\centering \includegraphics[width=\linewidth]{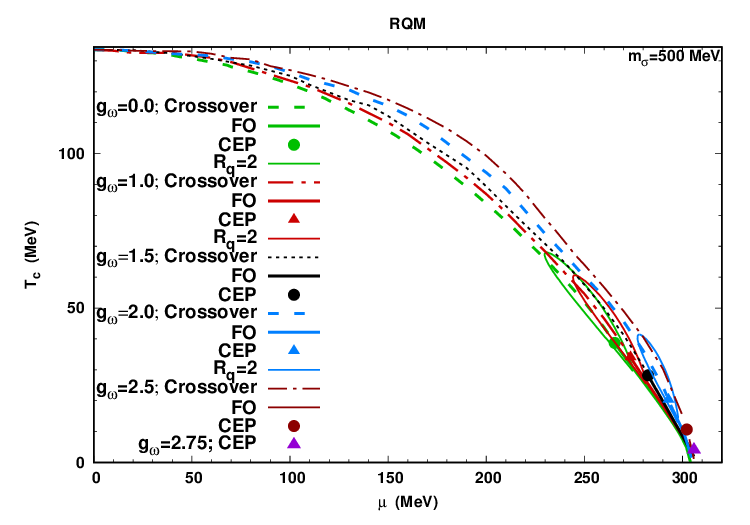}
\end{minipage}}
\hfill
\subfigure[]{
\label{fig9b} 
\begin{minipage}[b]{0.47\textwidth}
\centering \includegraphics[width=\linewidth]{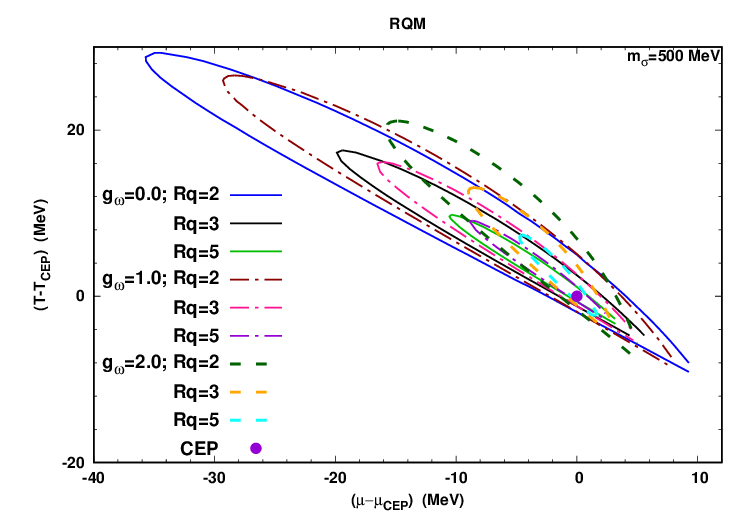}
\end{minipage}}
\caption{The RQM model phase diagrams for different strengths of vector coupling  $g_{\omega}=0.0,1.0,1.5,2.0 \text{ and } 2.5$ are presented in the left panel (a) where the legends and line types are labeled and the position of CEP for $g_{\omega}=2.75$ is also depicted.~$R_{q}=2,3 \text{ and }5$ susceptibility contours  for $g_{\omega}=0.0,1.0 \text{ and } 2.0$ are presented in the right panel (b) for comparison.}
\label{fig:mini:fig9}
\end{figure*}

\subsection{Effects of vector interaction on the phase diagram,~CEP and its critical region}
\label{sec:IIID}

~The repulsive vector type interactions in the NJL model were shown to exert a strong influence on the CEP long before in the Ref.~\cite{Asak}.~The appearance of multiple CEPs, driven by vector interaction during the competition between chiral symmetry restoration and the color superconducting phase transition, has also been addressed in previous studies~\cite{Kitaz, Hastprl}.~Regarding vector interaction strength, even if the value of the vector coupling $G_{V}$ can be fixed in the vacuum~\cite{HAbuki},~it is not clear how much it would be modified in the medium~\cite{fuku08}. Therefore, taking the ratio of the vector to scalar coupling constants, \(r = G_V/G_S\), as a free parameter, several of the NJL and PNJL model studies have varied it between 0 and 1~\cite{fuku08,Kitaz,HAbuki,Kunihiro,Csasaki,KJ4019,Sakai,Bratovic,Blaschke,Hell,HellII} and explored its effect on the position of the CEP in the \(\mu\text{-}T\) plane of the phase diagram. The value \(r = 0.5\) is estimated from perturbative one-gluon exchange~\cite{Kunihiro}, whereas \(r = 0.25\) is based on instanton–anti-instanton molecule model estimates~\cite{Kitaz,Rapp}, and \(r = 1.0\) is obtained by fitting lattice data with the $\text{PNJL}_{8}$ model~\cite{Sakai}.~The CEP and first-order transition disappear when \(r = 0.38\) for the two flavor NJL model~\cite{Kitaz},~and when \(r = 0.206\)  and \(r \geq 0.25\) in Refs.~\cite{fuku08} and~\cite{KJ4019}, respectively,~for the 2+1 flavor PNJL model.~Considering the repulsive vector plus axial-vector interaction terms in the three flavor vector extended NJL model at zero temperature and finite density,~the first-order transition becomes a crossover when \(r = 0.5\) in Ref.~\cite{HAbuki}, whereas  Ref.~\cite{Csasaki} finds that the first-order transition and TCP  disappear when  \(r \geq 0.6\) in the chiral limit of the two flavor NJL model. Considering the curvature of the crossover boundary in the \(\mu\text{--}T\) plane close to \(\mu = 0\) and its dependence on the coupling strength \(G_{V}\), Ref.~\cite{Bratovic} concluded that \(r \ge 0.4\), suggesting that the QCD phase diagram might lack a first-order chiral phase transition. However, by adjusting the strength of the vector coupling to reproduce the lattice-QCD-extracted slope of the pseudocritical temperature for the chiral phase transition at low chemical potential, Ref.~\cite{Blaschke} predicted the existence of a CEP for \(r > 0.555\) in the non-local two-flavor PNJL model, whereas Ref.~\cite{Hell} finds a crossover for \(r \ge 0.5\).~Fixed in different way,~the vector coupling ratio is also taken as  \(0.3 \le G_V/G_S  \le 3.2\) \cite{DutraI,DutraII,Frise}.

The \(\omega \) and \(\phi \) vector mesons are included as explicit degrees of freedom in the Lagrangian of the QM/PQM model, where the vector couplings are defined as \(g_V / 2 = g_\omega = g_\phi / \sqrt{2}\). Given that the scalar Yukawa coupling is \(g\), the coupling ratio for this framework is expressed as \(r = g_V / g = 2(g_\omega / g)\). The first-order transition becomes a crossover when \(r > 0.2\) in Ref. \cite{OhnishiPLB}, where the CEP is probed by its sweep during dynamical black hole formation, and when \(r \ge 0.38\) in Ref. \cite{Ueda} for the two-flavor PQM model. Two- and three-flavor QM models were studied \cite{Beisit} for the properties of quark matter under conditions present in core-collapse supernovae and for the description of compact stars consisting of pure quark matter \cite{zacchi2,Zacchi15}.~Large vector couplings were taken in the range \(g_\omega = 1.0\text{–}9.0\) (which is roughly equivalent to the above-described \(0.3 \le r \le 3.2\)), and their variation showed a significant impact on the equation of state (EoS). Higher coupling values lead to a stiffer EoS, giving rise to maximum mass configurations exceeding two solar masses for quark stars (or neutron/hybrid star matter), which is compatible with recent pulsar mass measurements. In Ref. \cite{zacchi2}, the authors studied how including quark one-loop vacuum fluctuations in the three-flavor quark matter (QM) EoS under the e-MFA influences compact star properties. Unlike the s-MFA, which yields twin stars through a first-order chiral phase transition, the e-MFA produces only crossover transitions in the presence of vector interactions for \(g_{\omega }\) values of 2.5, 3.0, and 3.5.~By applying an (axial) vector meson-extended three-flavor quark-meson model to describe quark matter in the core of neutron stars, the authors of Ref. \cite{KovaNstar} utilized their framework—featuring an inbuilt e-MFA—to demonstrate that the maximum mass of stable hybrid stars depends only weakly on the crossover-type phase transition.~Based on this observation and recent astrophysical constraints, they established a bound of \(2.6 < g_V < 4.3\) (i.e. \(1.31 < g_\omega < 2.15\)) for the quark-vector meson coupling constant. They point out that a nonzero vector coupling is required to fulfill the two-solar-mass criterion for the mass-radius (\(M\text{–}R\)) curves of compact stars; furthermore, within their framework, the existence of a CEP is unlikely if \(g_V \, (g_\omega) \ge 3.1 \, (1.55)\) for \(m_\sigma = 290\) MeV. Recall that the smoothing effect of the quark one-loop vacuum fluctuations on the strength of the chiral transition in the 2+1 flavor QM model is very large in the previously discussed works. This occurs because the effective potential with the vacuum correction term is computed in the $\overline{MS}$ scheme, while the model parameters are fixed using the curvature masses of mesons~\cite{vkkr23,skrvkt24}, as done in the QM model with vacuum term (QMVT) treatment of e-MFA \cite{guptiw, vkkr12, chatmoh1, schafwag12, vkkt13, zacchi2}. In contrast, the renormalized 't Hooft coupling \(c\) becomes significantly stronger, and the explicit symmetry-breaking strengths \(h_{x}\) and \(h_{y}\) weaken due to the consistent renormalization of parameters in the RQM model; this is achieved by matching the counter-terms in the on-shell scheme to those in the MS scheme. Therefore, the softening effect of the quark one-loop vacuum fluctuations is moderate in the e-MFA treatment of the RQM model. Consequently, the RQM model tends to produce a comparatively larger extent of the first-order transition line, shifting the CEP position upward in the \(\mu - T\) plane. In light of this, it will be highly interesting to investigate below how much the CEP shifts toward higher \(\mu \) and lower \(T\) values as the strength of the vector interaction increases, and how it disappears when the transition becomes a smooth crossover across the entire phase diagram at higher values of \(g_{\omega }\).

\begin{table*}[!htbp]
    \caption{The $T$ and $\mu$ axis spread of the $R_{q}=2,\bf {3 \text{ and } 5}$ contours in different model scenarios for different vector coupling strength $g_\omega$.The notation is same as in the table IV.} 
    \label{tab:table6}
      \begin{tabularx}{\textwidth}{l|l|X|X|X|X|X|X} 
      \toprule 
        \hphantom{text} Models &  \multicolumn{7}{c}{Entries below the $\Delta T_{a}$, $\Delta T_{b}$, $\Delta T$, $\Delta\mu_{l}$, $\Delta\mu_{h}$ and $\Delta\mu$ correspond to the susceptibility ratio ${\bf {R_q}}= 2\textbf{(3)(5)}$} \ \\
         &$g_\omega$& $\Delta T_{a}$ & $\Delta T_{b}$ & $ \Delta T=\Delta T_{a}+\Delta T_{b}$ &  $\Delta\mu_{l}$ &  $\Delta \mu_{h}$& $\Delta \mu=\Delta \mu_{l}+\Delta \mu_{h}$ \\      
\hline     
\multirow{2}{*}{RQM}
  & $1$ & $26.6\textbf{(16.1)(9.1)}$ &$8.5\textbf{(5.3)(2.8)}$ &$35.1\textbf{(21.4)(11.9)} $&$29.3\textbf{(16.5)(8.8)}$  &$7.9\textbf{(4.7)(2.4)}$ &$37.2\textbf{(21.2)(11.2)}$ \\[3pt]
  &$2$ & $21.1\textbf{(13.1)(7.4)}$ & $7.3\textbf{(4.3)(2.3)}$ &$28.4\textbf{(17.4)(9.7)}$ &$15.8\textbf{(9.0)(4.7)}$ &$4.9\textbf{(3.0)(1.7)}$ &$20.7\textbf{(12.0)(6.4)}$  \\
\hline

\multirow{2}{*}{Log RPQM}
  & $1$ &$35.0\textbf{(24.0)(14.3)}$ &$15.0\textbf{(10.0)(5.0)}$ &$50.0\textbf{(34.0)(19.3)}$ &$50.6\textbf{(29.1)(15.2)}$ &$11.8\textbf{(7.9)(4.0)}$ &$62.4\textbf{(37.0)(19.2)}$ \\[3pt]
  & $ 2$ &$ 39.3\textbf{(26.1)(15.7)}$ &$ 18.5\textbf{(11.5)(6.5)}$ &$57.8 \textbf{(37.6)(22.2)}$ &$ 25.2\textbf{(13.7)(7.1)}$ & $6.5\textbf{(4.2)(2.6)}$ &$ 31.7\textbf{(17.9)(9.7)}$ \\
\hline

\multirow{2}{*}{\makecell{PolyLog-glue \\	RPQM}}
  & $1$ &$39.3$\textbf{(26.4)(15.6)} &$15.9\textbf{(10.2)(6.7)}$ &$55.2\textbf{(36.6)(22.3)}$ &$45.4\textbf{(27.3)(14.8)}$ &$13.0\textbf{(8.3)(5.3)}$ &$58.4\textbf{(35.6)(20.1)}$ \\[3pt]
  & $2$ & $42.6\textbf{(28.3)(16.7)}$ & $16.0\textbf{(12.0)(7.0)}$ & $ 58.6\textbf{(40.3)(23.7)}$ & $29.2\textbf{(16.9)(9.0)}$ & $8.0\textbf{(5.3)(3.1)}$ & $37.2\textbf{(22.2)(12.1)}$ \\
\hline 
  \end{tabularx} 
\end{table*}

\begin{figure*}[htb]
\subfigure[]{
\label{fig10a} 
\begin{minipage}[b]{0.47\textwidth}
\centering \includegraphics[width=\linewidth]{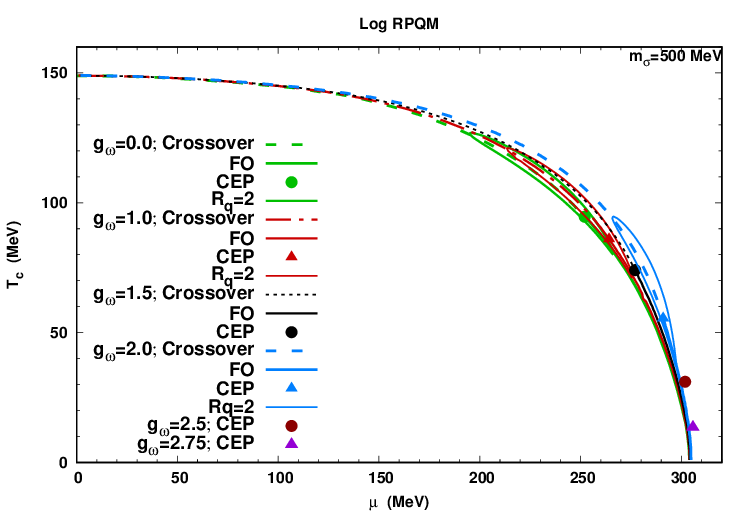}
\end{minipage}}
\hfill
\subfigure[]{
\label{fig10b} 
\begin{minipage}[b]{0.47\textwidth}
\centering \includegraphics[width=\linewidth]{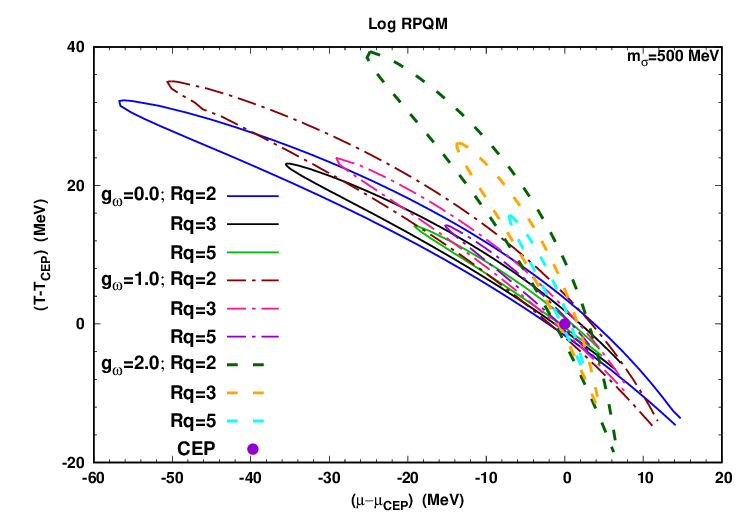}
\end{minipage}}
\caption{The Log RPQM model phase diagrams for different strengths of vector coupling  $g_{\omega}=0.0,1.0,1.5\text{ and }2.0$ are presented in the left panel (a) where the legends and line types are labeled and the position of CEPs for $g_{\omega}=2.5 \text{ an }2.75$ are also depicted.~$R_{q}=2,3 \text{ and }5$ susceptibility contours  for $g_{\omega}=0.0,1.0 \text{ and } 2.0$ are presented in the right panel (b) for comparison.}
\label{fig:mini:fig10}
\end{figure*}

~In this sub-section, we will present the results for the effects of vector interactions on the phase diagrams, positions of critical end point and the extent of critical fluctuations around the CEP in our 2+1 flavor RQM/RPQM  model framework.~We have computed the phase diagrams with the vector interactions taking five different coupling strengths $g_{\omega}=1.0,1.5,2.0, 2.5 \text{ and } 2.75$  when the scalar $\sigma$ mass $m_{\sigma}=500$ MeV.~The phase diagram and the position of the CEP in the absence of vector interaction $g_{\omega}=0$ has been compared with the four different RQM model phase diagrams with the vector interactions in the Fig.~\ref{fig9a}.~The Table~\ref{tab:table5} shows that the $\cep$ position at $(\mu_\cep,T_\cep)=(265.42,38.71)$ with no vector interaction,~shifts respectively to the $(\mu_\cep,\ T_\cep)=(273.5,34.05),\ (282.49,28.23),\ (292.83,20.33)$  $(302.17,10.68) $ and $ (305.84,4.07)$ MeV when the vector interaction strength is  successively increased to the $g_{\omega}=1.0,1.5,2.0, 2.5 \text{ and } 2.75$.~The strength of the first order transition becomes weaker due to the increasing strength of the vector interactions as expected \cite{Asak, Kitaz, Hastprl}.~The first order transition line shrinks at the expanse of the extending crossover transition lines in the phase diagrams of the  Fig.~\ref{fig9a} and the first order transition altogether disappears from the phase diagram when the strength of the vector interaction becomes  $g_{\omega}=2.79$.~Thus the whole phase diagram  turns into a crossover transition line  for the vector to scalar critical coupling ratio $\frac{2g_{\omega}}{g}=0.858$ where the  scalar  coupling $g=6.5$.~The contours for the quark number susceptibility ratio \(R_{q}=2\) around the $\cep$, which serve as a measure of critical fluctuations, are compared in Fig.~\ref{fig9a}, as depicted by solid green, red, and blue lines for the \(g_{\omega}=0, 1, \text{ and } 2\) phase diagrams, respectively.

\begin{figure*}[htb]
\subfigure[]{
\label{fig11a} 
\begin{minipage}[b]{0.47\textwidth}
\centering \includegraphics[width=\linewidth]{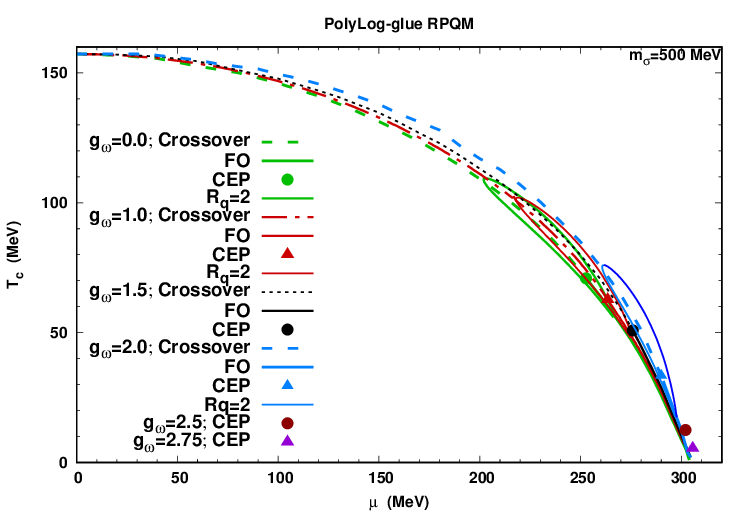}
\end{minipage}}
\hfill
\subfigure[]{
\label{fig11b} 
\begin{minipage}[b]{0.47\textwidth}
\centering \includegraphics[width=\linewidth]{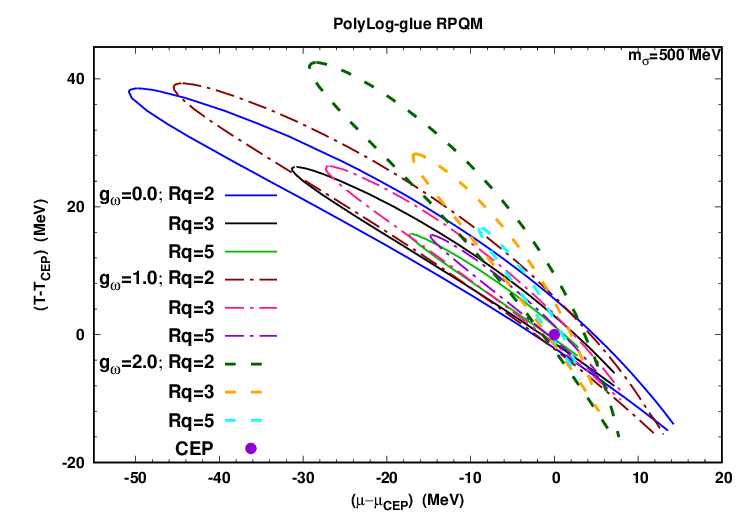}
\end{minipage}}
\caption{The PolyLog-glue RPQM model phase diagrams for different strengths of vector coupling  $g_{\omega}=0.0,1.0,1.5\text{ and }2.0$ are presented in the left panel (a) where the legends and line types are labeled and the position of CEPs for $g_{\omega}=2.5 \text{ an }2.75$ are also depicted.~$R_{q}=2,3 \text{ and }5$ susceptibility contours  for $g_{\omega}=0.0,1.0 \text{ and } 2.0$ are presented in the right panel (b) for comparison}
\label{fig:mini:fig11}
\end{figure*}

The effects of vector interactions on the critical fluctuations around the $\cep$ have been quantified in the Fig.~\ref{fig9b} by drawing the $R_{q}=2,3 \text{ and }5$ susceptibility contours for the $m_{\sigma}=500$ MeV in the RQM model.~The contours drawn for the two different vector coupling strengths $g_{\omega}=1.0 \text{ and } 2.0$,~are presented in the Fig.~\ref{fig9b} for comparison with the earlier obtained contours (in the Fig.~\ref{fig5b}) when the vector interaction is absent.~The line types and legends for different cases are marked and explained in the Figure.~Note that the \(T (\mu)\) axis spreads for the \(R_q = 2, 3,\) and \(5\) contours without the vector interaction are 38.4 (45.0)~MeV, 23.3 (25.5)~MeV, and 13.1 (13.7)~MeV, 
respectively. These values are reduced by the vector interaction to 35.1 (37.2), 
21.4 (21.2), and 11.9 (11.2)~MeV when \(g_\omega = 1\), and to 28.4 (20.7), 
17.4 (12.0), and 9.7 (6.4)~MeV when \(g_\omega = 2\).~Note that as the vector interaction strength increases , the spread of contours decreases by a larger magnitude in the chemical potential direction than in the temperature direction.

Recall that when the physics of statistical color confinement is integrated into the RQM model, the CEP shifts significantly upward along the temperature axis due to the effect of the Polyakov loop potential in the Log RPQM model~\cite{skrvkt24}.~The $\cep$ at $(\mu_\cep,T_\cep)=(252.0,94.6)$ MeV with no vector interaction in the Log RPQM model,~shifts respectively to the $(\mu_\cep,\ T_\cep)=(264.11,86.0),\ (277.02,74.1),\ (290.91,55.5),\ (301.82,31.1) $ and (305.71,13.5) MeV when the vector interaction strength is  successively increased to the $g_{\omega}=1.0,1.5,2.0 ,2.5\text{ and } 2.75$ as shown in Table~\ref{tab:table5}.~Fig.~\ref{fig10a} shows the Log RQPM model phase diagrams for \(g_{\omega}=0, 1.0, 1.5,\) and \(2.0\), along with three \(R_{q}=2.0\) susceptibility contours in green, red, and blue for \(g_{\omega}=0, 1.0,\) and \(2.0\), respectively.~Comparing the RQM model results for the $\cep$ locations with those of the Log RQPM model in Table~\ref{tab:table5}, one finds that the Log form of the Polyakov loop potential causes a significantly large upward shift in the temperature direction.~This shift is reduced successively from 55.89 to 51.95, 45.87, 35.17, and 20.42 MeV as \(g_{\omega }\) increases from 0.0 to 1.0, 1.5, 2.0, and 2.5, respectively.~In contrast,~the downward shift of the $\cep$ along the \(\mu \) axis (toward lower chemical potentials) for the Log RQPM model relative to the RQM model is smaller.~This shift also decreases successively from 13.42 to 9.39, 5.47, 1.92, and 0.35~MeV as \(g_{\omega }\) increases from 0.0 to 1.0, 1.5, 2.0, and 2.5, respectively.


Fig.~\ref{fig10b} depicts the Log RPQM model contours for \(R_{q} = 2, 3,\) and \(5\) with vector coupling strengths \(g_{\omega} = 1.0\) and \(2.0\) when \(m_{\sigma} = 500\) MeV, alongside the corresponding contours without vector interaction from Fig.~\ref{fig6a} for comparison.~Recall that the \(T\) (\(\mu \)) axis spreads for the \(R_q = 2, 3\), and \(5\) contours for the Log RPQM model without the vector interaction are 46.9 (71.4), 29.8 (42.6), and 18.8 (23.8) MeV, respectively. The above values for the \(T\) (\(\mu \)) axis spreads are enhanced (reduced) by the vector interaction to 50.0 (62.4), 34.0 (37.0), and 19.3 (19.2) MeV when \(g_\omega = 1\), and to 57.8 (31.7), 37.6 (17.9), and 22.2 (9.7) MeV, respectively, when \(g_\omega = 2\).~A key result worth emphasizing, in contrast to the RQM model, is the noticeable enhancement of quark number susceptibilities along the \(T\) axis when the vector interaction strength is increased from \(g_{\omega} = 0\) to \(g_{\omega} = 1\) and \(2\) under the influence of the Polyakov loop potential in the Log RPQM model framework.~This increase causes the contour spread to expand along the temperature direction while contracting significantly along the chemical potential direction. Consequently, the \(g_{\omega}=1\) contours in Fig.~\ref{fig10b} are shifted upward in comparison to the \(g_{\omega}=0\) case, while the contours for the \(g_{\omega}=2\) appear much more vertically oriented toward the \(T\) axis.

The PolyLog-glue RPQM model CEP at \((\mu_{\cep}, T_{\cep}) = (252.7, 70.9)\) MeV, found in the absence of vector interactions, shifts to \((\mu_{\cep}, T_{\cep}) = (263.4, 62.7)\), \((275.9, 50.87)\), \((289.96, 33.46)\), \((302.02, 12.53)\) and \((305.60, 5.4)\) MeV respectively, when the vector interaction strength \(g_{\omega }\) is successively increased to \(1.0, 1.5, 2.0, 2.5\), and \(2.75\) as shown in the Table~\ref{tab:table5}.~The PolyLog-glue RPQM model phase diagrams for \(g_{\omega}=0, 1, 1.5,\) and \(2.0\), together with three \(R_{q}=2.0\) susceptibility contours in green, red, and blue for \(g_{\omega}=0, 1,\) and \(2.0\) respectively, are presented in Fig.~\ref{fig11a}.~When the positions of $\cep$ in the PolyLog-glue RQPM model are compared with those in the RQM model in Table~\ref{tab:table5}, one notes that the presence of the quark back-reaction in the PolyLog-glue form of the Polyakov loop potential gives rise to an effect that generates a moderately large upward shift in the temperature direction.~This shift is reduced successively from 32.19 to 28.65, 22.64, 13.13, and 1.85~MeV as \(g_{\omega }\) increases from 0.0 to 1.0, 1.5, 2.0, and 2.5, respectively.~We emphasize that this shift is appreciably smaller than the one discussed above for the Log RPQM model relative to the RQM model.~The downward shifts of the $\cep$ along the \(\mu \) axis (toward lower chemical potentials) relative to the RQM model are nearly the same for both the PolyLog-glue RPQM and Log RPQM models.

The PolyLog-glue RPQM model contours for \(R_{q} = 2\), \(3\), and \(5\) with vector coupling strengths \(g_{\omega} = 1.0\) and \(2.0\) (for \(m_{\sigma} = 500\) MeV), along with the corresponding contours without vector interaction from Fig.~\ref{fig6b}, are presented in Fig.~\ref{fig11b} for comparison. It is worth recalling that for the PolyLog-glue RPQM model without vector interaction, the \(T\) (\(\mu \)) axis spreads for the \(R_q = 2, 3\), and \(5\) contours are 53.5 (65.0), 35.9(40.2), and 21.7 (22.8) MeV, respectively. Turning on the vector interaction enhances the \(T\)-axis spreads but reduces the \(\mu \)-axis spreads. Specifically, at \(g_\omega = 1\), the \(T\) (\(\mu \)) spreads shift to 55.2 (58.4), 36.6 (35.6), and 22.3 (20.1) MeV. At \(g_\omega = 2\), they change further to 58.6 (37.2), 40.3 (22.2), and 23.7 (12.1) MeV, respectively. Notably, the effect of quark back-reaction in the Polyakov loop potential of the PolyLog-glue RPQM model generates a somewhat smaller reduction in the \(\mu \)-direction spread of contours compared to that of the Log RPQM model.~Thus, when the vector interaction strength is increased from \(g_{\omega} = 0\) to \(g_{\omega} = 1\) and \(2\) under the influence of the quark back-reaction, one finds a significant enhancement in the spread of quark number susceptibility contours along the \(T\) axis relative to that of the RQM model; however, the contraction along the chemical potential direction becomes moderate compared to that of the Log RPQM model.~Similar to the behavior observed in the Log RPQM model, the \(g_{\omega}=1\) contours within the PolyLog-glue framework (Fig.~\ref{fig11b}) are shifted upward in comparison to the \(g_{\omega}=0\) case, while the \(g_{\omega}=2\) contours appear  more vertically oriented toward the \(T\) axis.

It is emphasized that the entire transition line becomes a crossover in the RQM model phase diagram for a critical vector-to-scalar coupling ratio of $\frac{2g_{\omega}}{g} = 0.858$ (where the Yukawa coupling is $g = 6.5$ and $g_{\omega} = 2.79$). The Polyakov loop potential has a minuscule effect on this critical ratio; in the Log RPQM model, the first-order phase transition disappears entirely at a ratio of $r = 0.861$ ($g_{\omega} = 2.798$). This result is noticeably different from several NJL/PNJL/EPNJL model studies, where the first-order transition is washed out for $r < 0.6$. We emphasize that the CEP in our RQM/RPQM framework survives a significant vector coupling strength of $g_{\omega}=2.79 \ (g_{V}=5.58)$. This stands in contrast to the three-flavor QM model work of Ref.~\cite{zacchi2} and the (axial) meson-extended three-flavor QM model of Ref.~\cite{KovaNstar}, where the first-order transition disappears for $g_{V} \ge 3.1$. This critical vector coupling strength would have important implications for the compact star equation of state and related astrophysical phenomenology.

\section{Summary}
\label{sec:IV}

Consistent treatment of quark one-loop vacuum fluctuations and parameter renormalization in the renormalized quark-meson (RQM) model--by matching on-shell and $\overline{\text{MS}}$ counterterms--yields a significantly 
stronger 't Hooft coupling $c$ and weaker explicit chiral symmetry breaking ($h_{x}, h_{y}$). Consequently, the critical end point (CEP) shifts upward in the RQM phase diagram~\cite{vkkr23,skrvkt24}. This shift is further 
enhanced in the renormalized Polyakov-quark-meson (RPQM) model upon incorporating the Polyakov loop potential, with or without quark back-reaction. Located higher in the $\mu\text{--}T$ plane than the CEP of the curvature-mass-based PQMVT model~\cite{schafwag12}, the RPQM model CEP~\cite{skrvkt24} aligns closer to recent theoretical estimates of the QCD critical region [$T_{c} \sim 100\text{--}110$~MeV, $\mu_{B} \sim 420\text{--}650$~MeV $\Leftrightarrow \mu_{\text{CEP}} \sim 140\text{--}217$~MeV]~\cite{MishaConf}. Therefore, mapping the critical regions around the CEP in the RQM and RPQM models is essential. It is equally imperative to determine how vector interactions modify the phase diagrams, CEP locations, and surrounding critical fluctuations. Given the RQM model's propensity for a larger first-order transition line, identifying the critical vector coupling strength where the first-order line completely disappears into a smooth crossover is highly interesting. Determining this critical strength has important implications for the compact star equation of state and related astrophysical phenomenology.

To determine the critical temperature $T_{c}$ of the second-order chiral phase transition at $\mu=0$ in the light chiral limit ($m_{\pi}=0$), the temperature variations of the order parameter and its derivative were 
computed for $m_{\sigma}=500$~MeV. In the RPQM model, the Log and PolyLog-glue forms of the Polyakov loop potential were implemented, both with and without quark back-reactions. The modified values of $f_{\pi}$, $f_{K}$, $m_{\eta}$, and $m_{\eta'}$ needed to fix the RQM parameters in this limit, where $(m_{\pi},m_{K})=(0,496)$~MeV, were obtained using large-$N_{c}$ standard chiral perturbation theory inputs from Refs.~\cite{herrPLB,Escribano,vktChpt1}. Addressing our first objective, the light chiral limit phase diagrams for the RQM, Log RPQM, and PolyLog-glue RPQM models (setting $T_{0}=T_{c}^{\text{glue}}=187$~MeV 
for $2+1$ dynamical quark flavors) are computed for $m_{\sigma}=400$ and $500$~MeV, and compared with their physical-point counterparts~\cite{vkkr23,skrvkt24}. For $m_{\sigma}=500\,(400)$~MeV, the RQM trictirical point (TCP) at $(\mu_{\tcp},T_{\tcp})=(220.15,71.52)\,[(194.6,65.9)]$~MeV and critical end point (CEP) at $(\mu_{\cep},T_{\cep})=(265.42,38.71)\,[(243.12,37.03)]$~MeV shift to $(\mu_{\tcp},T_{\tcp})=(194.3,126.53)\,[(164.21,126.41)]$~MeV and $(\mu_{\cep},T_{\cep})=(252.0,94.6)\,[(230.5,93.8)]$~MeV in the Log RPQM model. In the PolyLog-glue RPQM model, the shifted TCP and CEP are located at $(\mu_{\tcp},T_{\tcp})=(202.2,108.14)\,[(172.4,107.28)]$~MeV and $(\mu_{\cep},T_{\cep})=(252.7,70.9)\,[(227.6,73.6)]$~MeV, respectively.

Critical regions around the CEP are mapped by drawing contours of constant ratios ($R_{q}=2, 3$, and $5$) of quark number susceptibility, normalized by the free quark gas susceptibility. These contours are evaluated for the QM, PolyLog-glue PQM, RQM, Log RPQM, and PolyLog-glue RPQM models (with $T_{0}=T_{c}^{\text{glue}}=187$~MeV). In the QM model under the no-sea mean-field approximation, the critical regions for $R_{q}=2$ and $3$ are highly compressed along the temperature axis, while the Polyakov loop potential in the PolyLog-glue PQM model yields extremely pinched and minuscule regions. Conversely, the RQM model produces large, smooth, and symmetric critical regions for $m_{\sigma}=400$ and $500$~MeV; its consistent on-shell renormalization 
of vacuum corrections significantly broadens the regions perpendicular to the crossover line. Without quark back-reaction, the Log RPQM model shows extensively large critical regions with a wider spread along the $\mu$ axis, albeit with somewhat compressed contour widths. Incorporating the quark back-reaction in the PolyLog-glue RPQM model generates broader, rounder, and smoother critical regions characterized by larger spreads along the $T$ axis and less width compression. Across the RQM and RPQM (Log and PolyLog-glue) models, the $R_{q}=2,3,5$ 
critical regions are noticeably larger for $m_{\sigma}=500$~MeV than for $m_{\sigma}=400$~MeV. This occurs because the $m_{\sigma}=500$~MeV CEPs lie at higher chemical potentials $\mu_{\cep}$, where the phase boundary 
curvature is more pronounced. Furthermore, for $m_{\sigma}=500$~MeV in the RQM model, the TCP lies closer to the boundary of the $R_{q}=2$ critical region, whereas for $m_{\sigma}=400$~MeV, it lies noticeably away. This 
indicates that critical fluctuations around the CEP are likely to be influenced by the proximity of the TCP when $m_{\sigma}=500$~MeV. Similarly, in the Log and PolyLog-glue RPQM models with $m_{\sigma}=500$~MeV, the TCP is just inside the boundary of the $R_{q}=2$ critical region, while remaining far 
outside for $m_{\sigma}=400$~MeV. Consequently, critical fluctuations around the RPQM CEPs are distinctly affected by the nearby TCP at higher sigma mass. This comprehensive analysis fully addresses the first objective of the present work.

To analyze how different treatments of quark one-loop vacuum fluctuations affect the phase diagram, CEP location, and critical fluctuations, the phase diagrams of the on-shell renormalized RQM, RPQM-I, and RPQM-II models for $m_{\sigma}=400$~MeV are compared with those of the curvature-mass-based parameterized QMVT, PQMVT-I, and PQMVT-II models from Schaefer et~al. in Ref.~\cite{schafwag12}. In Ref.~\cite{schafwag12}, the PQMVT-I model denotes the baseline case with $T_{0}=270$~MeV for the pure gauge theory in the Log Polyakov loop potential, while PQMVT-II incorporates quark back-reaction via a chemical-potential-dependent $T_{0}(\mu)$ [with $T_{0}(\mu=0)=270$~MeV]. Correspondingly, for our $m_{\sigma}=400$~MeV study, RPQM-I signifies the Log 
RPQM model setup, and RPQM-II represents the PolyLog-glue RPQM scenario, both evaluated with $T_{0}=270$~MeV.
 The resulting CEP locations reveal distinct shifts between the two approaches. The RQM model CEP is located at $(\mu_{\cep},T_{\cep})=(243.12,37.03)$~MeV, whereas the QMVT CEP lies at $(\mu_{\cep},T_{\cep})=(286.0,32.0)$~MeV.~For the Polyakov-loop-extended frameworks without back-reaction, the RPQM-I model CEP is 
at $(\mu_{\cep},T_{\cep})=(237.8,104.23)$~MeV compared to the PQMVT-I model CEP at $(\mu_{\cep},T_{\cep})=(283.0,90.0)$~MeV. Finally, including quark back-reaction shifts the RPQM-II model CEP to $(\mu_{\cep},T_{\cep})=(228.54,86.47)$~MeV, while the PQMVT-II model CEP is situated at 
$(\mu_{\cep},T_{\cep})=(280.0,83.0)$~MeV.

The RQM model critical regions are smooth, noticeably broad perpendicular to the crossover line, and share a similar total spread with the QMVT model contours in Ref.~\cite{schafwag12}. However, while the QMVT model contours have a large spread above the CEP, they compress into a narrow, neck-like structure below $T_{\cep}$ because its CEP lies at a higher chemical potential ($\mu_{\cep}=286$~MeV) where the first-order line curvature is large. This narrowing is practically absent in the RQM model contours due to its significantly lower  CEP position in $\mu$ direction ($\mu_{\cep}=243.12$~MeV) where the boundary curvature is much smaller. For the Polyakov-loop-extended frameworks without quark back-reaction, the total $T$-axis spread of the PQMVT-I critical region (signified by the $R_{q}=2$ contour) is more than double that of the RPQM-I model, whereas its $\mu$-axis spread is only about 17~MeV larger. Like the QMVT case, the large PQMVT-I model critical region narrows drastically below $T_{\cep}$ into a compressed neck because its CEP is situated at a high chemical potential ($\mu_{\cep}=283$~MeV). Conversely, the RPQM-I model CEP is located noticeably higher and further left at a smaller chemical potential ($\mu_{\cep}=237.8$~MeV), yielding smooth, somewhat symmetrical critical regions with negligible contour compression. Incorporating quark back-reaction, the $R_{q}=2$ critical region for the RPQM-II model exhibits nearly symmetric spreads along the $T$ axis (48.3~MeV) and $\mu$ axis (49.6~MeV). The RPQM-II model contours are larger, smoother, and broader than their RPQM-I model counterparts, lacking any narrow neck below $T_{\cep}$. Similarly, the PQMVT-II model critical regions show a reduced narrowing below $T_{\cep}$ compared to PQMVT-I model due to the back-reaction effect. This comparative analysis completely addresses our second objective.

We have investigated how vector interactions modify the phase diagrams, shift the position of the CEP, and alter the corresponding critical regions by increasing the vector coupling strength in steps from $g_{\omega} =0.0, 1.0$ to $2.75$. The first-order transition weakens as the vector coupling strength increases. In the RQM model, the entire transition line becomes a crossover at a critical vector-to-scalar coupling ratio of $\frac{2g_{\omega}}{g} = 0.858$ (where the Yukawa coupling is $g = 6.5$ and $g_{\omega} = 2.79$). The Polyakov loop potential has a minuscule effect on this critical ratio; in the Log RPQM model, the first-order phase transition disappears entirely at a ratio of $r = 0.861$ ($g_{\omega} = 2.798$). This result is noticeably different from several NJL/PNJL/EPNJL model studies, where the first-order transition is washed out for $r < 0.6$. We emphasize that the CEP in our RQM/RPQM framework survives a significant vector coupling strength of $g_{\omega}=2.79 \ (g_{V}=5.58)$. This stands in contrast to the three-flavor QM model work of Ref.~\cite{zacchi2} and the (axial) meson-extended three-flavor QM model of Ref.~\cite{KovaNstar}, where the first-order transition disappears for $g_{V} \ge 3.1$. This critical vector coupling strength could have important implications for the compact star equation of state and related astrophysical phenomenology.

The critical regions in the RQM model shrink along both the temperature ($T$) and chemical potential ($\mu$) axes with increasing vector coupling, whereas this shrinking is slightly more pronounced in the $\mu$ direction. As $g_{\omega}$ increases from $1.0$ to $2.0$ in the Log RPQM model, the contour spread reduces much more significantly along the $\mu$-axis than the $T$-axis. This trend is moderated in the PolyLog-glue RPQM model due to the effects of quark back-reaction. As the vector interaction strengthens in the Log and PolyLog-glue RPQM models, the critical regions are drastically suppressed in the chemical potential direction while elongating along the temperature direction. This simultaneous reduction along $\mu$ and elongation along $T$ causes the contours to develop a vertical tilt in the $T-\mu$ plane.~This analysis fulfills our third objective.

\section*{Acknowledgments} 
Akanksha Tripathi acknowledges the support of CSIR Junior Research Fellowship.~We are very much thankful to Pooja Kumari for valuable discussions and reading of the manuscript.    \\ \\ 
{\bf{Data Availability}}
The data supporting the results of this article are openly
available in the ancillary files of \cite{Akakn}



\begin{thebibliography}{189}

\bibitem{Cabibbo75}
N.~Cabibbo and G.~Parisi,
\href{https://doi.org/10.1016/0370-2693(75)90158-6}{Phys.Lett.\textbf{B 59},67-69(1975)}.

\bibitem{SveLer}
L. D. McLerran and B. Svetitsky, \href{https://doi.org/10.1103/PhysRevD.24.450}{Phys. Rev. D 24, 450
(1981)}; B. Svetitsky, \href{https://doi.org/10.1016/0370-1573(86)90014-1}{Phys. Rep. 132, 1 (1986)}.
\bibitem{Mull}
B. Muller, \href{https://doi.org/10.1088/0034-4885/58/6/002}{Rep. Prog. Phys. 58, 611 (1995)}.
\bibitem{Ortms}
H. Meyer-Ortmanns, \href{https://doi.org/10.1103/RevModPhys.68.473}{Rev. Mod. Phys. 68, 473 (1996)}.
\bibitem{Riske}
D. H. Rischke, \href{https://doi.org/10.1016/j.ppnp.2003.09.002}{Prog. Part. Nucl. Phys. 52, 197 (2004)}.
\bibitem{AliKhan:2001ek}
A.Ali~Khan et~al.
\href{https://doi.org/10.1103/PhysRevD.64.074510}{Phys. Rev. {\bf D 64}, 074510 (2001)}.
\bibitem{Digal:01}
S. Digal, E. Laermann and H. Satz,
\href{https://doi.org/10.1007/s100520000538}{Eur. Phys. J. {\bf C 18}, 583 (2001)}.

\bibitem{Fodor:03}
Z. Fodor, S. D. Katz, and K. K. Szabo,
\href{https://doi.org/10.1016/j.physletb.2003.06.011}{Phys. Lett. {\bf B 568}, 73 (2003)}.
\bibitem{Allton:05}
C. R. Allton, M. Doring, S. Ejiri, S. J. Hands, O. Kaczmarek, F. Karsch, 
E Laermann and K. Redlich,
\href{https://doi.org/10.1103/PhysRevD.71.054508}{Phys. Rev. {\bf D 71}, 054508 (2005)}.
\bibitem{Karsch:05}
F. Karsch,
\href{https://doi.org/10.1088/0954-3899/31/6/002}{J. Phys. {\bf G 31}, S633 (2005)}.  
\bibitem{Aoki:06}
Y. Aoki, Z. Fodor, S. D. Katz and K. K. Szabo,
\href{https://doi.org/10.1016/j.physletb.2006.10.021}{Phys. Lett. {\bf B 643}, 46 (2006)}.
\bibitem{Cheng:06}
M. Cheng et al., 
\href{https://doi.org/10.1103/PhysRevD.74.054507}{Phys. Rev. {\bf D 74}, 054507 (2006)}.
\bibitem{Cheng:08}
M. Cheng et al., 
\href{https://doi.org/10.1103/PhysRevD.77.014511}{Phys. Rev. {\bf D 77}, 014511 (2008)}.
\bibitem{JLange}
J. Langelage, S. Lottini, and O. Philipsen, \href{https://doi.org/10.1007/JHEP02(2011)057}{J. High Energy Phys. {\bf 02} (2011) 057}.
\bibitem{ABAZ}
A. Bazavov et al., 
\href{https://doi.org/10.1103/PhysRevD.80.014504}{Phys. Rev. D 80, 014504 (2009)}; 
\bibitem{Ejiri.Lat}
S. Ejiri et al.,
\href{ https://doi.org/10.1103/PhysRevD.80.094505}{ Phys. Rev. D 80, 094505 (2009)}
\bibitem{Karsch:02}
F. Karsch,\href{https://doi.org/10.1007/3-540-45792-5_6}{
Lect. Notes Phys. {\bf 583}, 209 (2002)}
\bibitem{Alf}
M. G. Alford, A. Schmitt and K .Rajagopal, 
\href{https://doi.org/10.1103/RevModPhys.80.1455}{Rev. Mod. Phys. {\bf 80}, 1455 (2008)}.
\bibitem{Fukhat}
K. Fukushima, and T. Hatsuda,
\href{https://doi.org/10.1088/0034-4885/74/1/014001}{Rep. Prog. Phys. {\bf 74}, 014001 (2011)}. 
\bibitem{FejosI}
G. Fejos,\href{https://doi.org/10.1103/PhysRevD.92.036011} {Phys. Rev. D {\bf 92}, 036011 (2015)}.
\bibitem{FejosII}
G. Fejos, A. Hosaka,\href{https://doi.org/10.1103/PhysRevD.94.036005}{Phys. Rev. D {\bf 94}, 036005 (2016)}.
\bibitem{FbRenk}
Fabian Rennecke, Bernd-Jochen Schaefer,\href{https://doi.org/10.1103/PhysRevD.96.016009} {Phys, Rev. D {\bf 96}, 016009 (2017)}.
\bibitem{Fejos3}
G. Fejos, A. Hosaka,\href{https://doi.org/10.1103/PhysRevD.98.036009}{Phys. Rev. D {\bf 98}, 036009 (2018)}.
\bibitem{Fejos4}
G. Fejos, A. Patkos,\href{https://doi.org/10.1103/PhysRevD.105.096007}{Phys. Rev. D {\bf 105}, 096007 (2022)}.


\bibitem{tHooft:76prl}
G. 't Hooft,\href{https://link.aps.org/doi/10.1103/PhysRevLett.37.8} 
{Phys. Rev. Lett. {\bf 37}, 8 (1976)}

\bibitem{Schaefer:09}
B. J. Schaefer and M. Wagner,
\href{https://doi.org/10.1103/PhysRevD.79.014018}{Phys. Rev. {\bf D 79}, 014018 (2009)}.

\bibitem{Ortman}
H. Meyer-Ortmanns and B. J. Schaefer,
\href{https://doi.org/10.1103/PhysRevD.53.6586}{Phys. Rev. {\bf D 53}, 6586 (1996)}

\bibitem{Rischke:00}
J. T. Lenaghan, D. H. Rischke and J. Schaffner-Bielich,
\href{https://doi.org/10.1103/PhysRevD.62.085008}{Phys. Rev {\bf D 62}, 085008 (2000)}.\\
J. T. Lenaghan, D. H. Rischke,
\href{https://doi.org/10.1088/0954-3899/26/4/309}{J. Phys. {\bf G 26}, 431 (2000)}.
\bibitem{Lenagh}
J. T. Lenaghan, 
\href{https://doi.org/10.1103/PhysRevD.63.037901}{Phys. Rev.{\bf D 63}, 037901 (2001)}.

\bibitem{costaA}
Pedro Costa, M. C. Ruivo, C. A. de Sousa and Yu. L. Kalinovsky,
\href{https://doi.org/10.1063/1.1961054}{AIP Conf. Proc. 775, 173–181 (2005)}.

\bibitem{costaB}
Pedro Costa, M. C. Ruivo, C. A. de Sousa and Yu. L. Kalinovsky,
\href{https://doi.org/10.1103/PhysRevD.71.116002}{Phys. Rev. {\bf D 71}, 116002 (2005)}.

\bibitem{fuku08}
K. Fukushima,
\href{https://doi.org/10.1103/PhysRevD.77.114028}{Phys. Rev. {\bf D 77}, 114028 (2008); {\bf  78}, 039902(E) (2008) }.

\bibitem{Polyakov:78plb}
A. M. Polyakov,
\href{https://doi.org/10.1016/0370-2693(78)90737-2}{Phys.\ Lett. {\bf B 72}, 477 (1978)}.

\bibitem{benji}
B. Svetitsky and L. G. Yaffe,
\href{https://doi.org/10.1016/0550-3213(82)90172-9}{Nucl. Phys. B {\bf 210}, 423 (1982)}.

\bibitem{BankUka}
T. Banks and A. Ukawa,  
\href{https://doi.org/10.1016/0550-3213(83)90016-0}{Nucl. Phys. B {\bf 225}, 145 (1983)}.

\bibitem{Pisarski:00prd}
R. D. Pisarski,
\href{https://doi.org/10.1103/PhysRevD.62.111501}{Phys. Rev. {\bf D 62}, 111501(R) (2000)}.

\bibitem{fuku}
K. Fukushima, 
\href{https://doi.org/10.1016/j.physletb.2004.04.027}{Phys. Lett. B {\bf 591}, 277 (2004)}.

\bibitem{Vkt:06}
B. Layek, A. P. Mishra, A. M. Srivastava and V. K. Tiwari, 
\href{https://doi.org/10.1103/PhysRevD.73.103514}{Phys. Rev. {\bf D 73}, 103514 (2006)}.

\bibitem{ratti}
C. Ratti, M. A. Thaler, and W. Weise, 
\href{https://doi.org/10.1103/PhysRevD.73.014019}{Phys. Rev. D {\bf 73},
014019 (2006)}.

\bibitem{Roesnr}
S. Roessner, C. Ratti and W. Weise, 
\href{https://doi.org/10.1103/PhysRevD.75.034007}{Phys. Rev. D {\bf 75},
034007 (2007)}.


\bibitem{SchaPQM2F}
B. J. Schaefer, J. M. Pawlowski, and J. Wambach, \href{https://doi.org/10.1103/PhysRevD.76.074023}{Phys.
Rev. {\bf D 76}, 074023 (2007)}.

\bibitem{SchaPQM3F}
B. J. Schaefer, M. Wagner, and J. Wambach, \href{https://doi.org/10.1103/PhysRevD.81.074013}{Phys. Rev. {\bf D 81},
074013 (2010)}.


\bibitem{Schaefer:09wspax}
B. J. Schaefer, M. Wagner and J. Wambach, CPOD(2009)017, arXiv:0909.0289


\bibitem{Mao}
H. Mao, J. Jin, and M. Huang, \href{https://doi.org/10.1088/0954-3899/37/3/035001}{J. Phys. {\bf G 37}, 035001}.

\bibitem{TiPQM3F}
U. S. Gupta and V. K. Tiwari, \href{https://doi.org/10.1103/PhysRevD.81.054019}{Phys. Rev. {\bf D 81}, 054019
(2010)}.


\bibitem{Haas}
L. M. Haas, R. Stiele, J. Braun, J. M. Pawlowski and J. Schaffner-Bielich, \href{https://doi.org/10.1103/PhysRevD.87.076004}{Phys. Rev. {\bf D 87}, 076004 (2013)}.
\bibitem{Redlo}
P. M. Lo, B. Friman, O. Kaczmarek, K. Redlich, and C. Sasaki, \href{https://doi.org/10.1103/PhysRevD.88.074502}{Phys. Rev. {\bf D 88}, 074502 (2013)}.
\bibitem{BielichP}
R. Stiele and J. Schaffner-Bielich, \href{https://doi.org/10.1103/PhysRevD.93.094014}{Phys. Rev. {\bf D 93}, 094014 (2016)}.
\bibitem{THerbst2}
T. K. Herbst, J. M. Pawlowski, and B.-J. Schaefer, \href{https://doi.org/10.1016/j.physletb.2010.12.003}{Phys. Lett. B {\bf 696}, 58 (2011)}.

\bibitem{Herbst}
T. K. Herbst, J. M. Pawlowski, and B.-J. Schaefer, \href{https://doi.org/10.1103/PhysRevD.88.014007}{Phys.
Rev. D {\bf 88}, 014007 (2013)}.


\bibitem{Wupertal2010}
S. Borsányi, Z. Fodor, C. Hoelbling, S. D. Katz, S. Krieg, C. Ratti, and K. K. Szabó, \href{https://doi.org/10.1007/JHEP09(2010)073}{J. High Energy Phys. 09 (2010) 73}.

\bibitem{WB2014}
S.~Borsanyi, G.~Endrodi, Z.~Fodor, A.~Jakovac, S.~D.~Katz, S.~Krieg, C.~Ratti and K.~K.~Szabo,
\href{https://doi.org/10.1007/JHEP11(2010)077}{JHEP \textbf{11},  (2010) 077}. ; \  \ \  
S. Borsanyi, Z. Fodor, C. Hoelbling, S. D. Katz, S. Krieg and K. K. Szabo,\href{https://doi.org/10.1016/j.physletb.2014.01.007}{Phys. Lett. B 730 (2014) 99-104}.

\bibitem{HotQCD2014}
A. Bazavov et al., \href{https://doi.org/10.1103/PhysRevD.85.054503}{Phys. Rev. {\bf D 85}, 054503 (2012)}; \ 
\href{https://doi.org/10.1103/PhysRevD.90.094503}{Phys. Rev. D {\bf 90}, 094503 (2014)}

\bibitem{rob} 
R. D. Pisarski and F. Wilczek, \href{https://doi.org/10.1103/PhysRevD.29.338}{Phys. Rev. D {\bf 29}, 338 (1984)}.

\bibitem{Asak}
M. Asakawa and K. Yazaki, \href{https://doi.org/10.1016/0375-9474(89)90002-X} {Nucl. Phys. {\bf A 504}, 668, (1989)}.
\bibitem{Bard}
A. Barducci, R. Casalbuoni, S. De Curtis, R. Gatto, and G.
Pettini, \href{https://doi.org/10.1016/0370-2693(89)90695-3}{Phys. Lett. {\bf B 231}, 463 (1989)}; 
\href{https://doi.org/10.1103/PhysRevD.41.1610}{Phys. Rev. {\bf D 41}, 1610 (1990)}.
\bibitem{Berg}
J. Berges and K. Rajagopal, \href{https://doi.org/10.1016/S0550-3213(98)00620-8}{Nucl. Phys. {\bf B 538}, 215 (1999)}.
\bibitem{Hatta}
Y. Hatta and T. Ikeda, \href{https://doi.org/10.1103/PhysRevD.67.014028}{Phys. Rev. {\bf D 67}, 014028 (2003)}.
\bibitem{Fuji}
H. Fujii, \href{https://doi.org/10.1103/PhysRevD.67.094018}{Phys. Rev. {\bf D 67}, 094018 (2003)}.
\bibitem{KFuku}
Fukushima, Kenji
\href{https://doi.org/10.1134/S1547477111080097}{Physics of Particles and Nuclei Letters. {\bf 8} 838-844, (2011)}

\bibitem{columb} 
F.~R.~Brown, F.~P.~Butler, H.~Chen, N.~H.~Christ, Z.-h.~Dong, W.~Schaffer, L.~I.~Unger and A.~Vaccarino, \href{https://doi.org/10.1103/PhysRevLett.65.2491}{\emph{Phys. Rev. Lett.} \textbf{65} (1990) 2491}.

\bibitem{vktChpt1}
V. K. Tiwari, \href{https://doi.org/10.1103/7yqv-v754}{Phys. Rev. D {\bf 111}, 114014 (2025)}
\bibitem{vktChpt2}
V. K. Tiwari, \href{https://doi.org/10.1103/9zzh-d65x}{Phys. Rev. D {\bf 112}, 054046 (2025)}




\bibitem{Cute}
F.~Cuteri, O.~Philipsen and A.~Sciarra, \href{https://doi.org/10.1007/JHEP11(2021)141}{\emph{JHEP} \textbf{11} (2021) 141}.

\bibitem{bernhardt23} 
J.~Bernhardt and C.-S.~Fischer, \href{https://doi.org/10.1103/PhysRevD.108.114018}{\emph{Phys. Rev. D} \textbf{108} (2023) 114018}.

\bibitem{kousvos22} 
S.~R.~Kousvos and A.~Stergiou, \href{https://doi.org/10.21468/SciPostPhys.15.2.075}{\emph{SciPost Phys.} \textbf{15} (2023) 075}.


\bibitem{Son}
D. T. Son and M. A. Stephanov,\href{https://doi.org/10.1103/PhysRevD.70.056001}{Phys. Rev. {\bf D 70}, 056001 (2004)}; 
Y. Aoki, G. Androdi, Z. Fodor, S. D. Katz, and
K. K. Szabo, \href{https://doi.org/10.1038/nature05120}{Nature (London) {\bf 443}, 675 (2006)}; 
S. Gupta,X. Luo, B. Mohanty, H. G. Ritter, and N. Xu, \href{DOI: 10.1126/science.1204621}{Science {\bf 332},1525 (2011)}.
\bibitem{Misha3}
M. A. Stephanov, K. Rajagopal, and E. V. Shuryak, \href{https://doi.org/10.1103/PhysRevD.60.114028}{Phys.Rev. D 60, 114028 (1999)}; \href{https://doi.org/10.1103/PhysRevLett.81.4816}{Phys. Rev. Lett. {\bf 81}, 4816 (1998)}.
\bibitem{Kraja2}
B. Berdnikov and K. Rajagopal, 
\href{https://doi.org/10.1103/PhysRevD.61.105017}{Phys. Rev. {\bf D 61}, 105017 (2000)}.
\bibitem{Jeon}
S. Jeon and V. Koch,\href{https://doi.org/10.1103/PhysRevLett.85.2076}{Phys. Rev. Lett. {\bf 85}, 2076 (2000)}.
\bibitem{Ejik}
S. Ejiri, F. Karsch, and K. Redlich,
\href{https://doi.org/10.1016/j.physletb.2005.11.083}{Phys. Lett. {\bf B 633}, 275 (2006)}.
\bibitem{AdamAg}
J. Adams et al. (STAR Collaboration), \href{https://doi.org/10.1016/j.nuclphysa.2005.03.085} {Nucl. Phys. \bf{A 757}, 102 (2005)}; M. Aggarwal et al. (STAR Collaboration),
\href{https://link.aps.org/doi/10.1103/Phys.Rev.Lett.105.022302}{Phys. Rev. Lett. \bf{105}, 022302 (2010)}.
\bibitem{Odyniec:2019XZ}
G.Odyniec; RHIC Beam Energy Scan Program: Phase I and II, \href{10.22323/1.185.0043(2013).}{PoS CPOD 2013, 043(2013)} and 
\href{10.22323/1.347.0151}{PoS {\bf CORFU2018} 151 (2019)}.
\bibitem{Cliu}
C. Liu et al.
\href{doi:10.18429/JACoW-IPAC2022-WEPOPT032}{13th Int. Particle Acc. Conf. WEPOPT032}
\bibitem{APandav}
Pandav, Ashish; \href{https://doi.org/10.1051/epjconf/202429601016}{EPJ Web Conf.{ \bf 296}, 01016, (2024)} ; A. Pandav (STAR collaboration), plenary talk at CPOD 2024, \href{https://conferences.lbl.gov/event/1376/contributions/8772/}.
\bibitem{MishaConf}
M. Stephanov, \href{https://doi.org/10.1051/epjconf/202431400042}{EPJ Web of Conferences 314, 00042 (2024)}
\bibitem{Bzda}
A. Bzdak, S. Esumi, V. Koch, J. Liao, M. Stephanov, N. Xu, 
\href{ https://doi.org/10.1016/j.physrep.2020.01.005}{Phys. Rept. 853, 1 (2020)}.
\bibitem{Ldu}
L. Du, A. Sorensen, M. Stephanov
\href{https://doi.org/10.1142/S021830132430008X}{Int. J. Mod. Phys. \bf{E 33} (2024) 07, 2430008}.


\bibitem{MishaPRL}
M.A. Stephanov, 
\href{https://doi.org/10.1103/PhysRevLett.102.032301}{Phys. Rev. Lett. {\bf 102}, 032301 (2009)}.
\bibitem{Krajamisha}
C. Athanasiou, K. Rajagopal, and M. Stephanov,
\href{https://doi.org/10.1103/PhysRevD.82.074008}{Phys. Rev. {\bf D 82}, 074008 (2010)}.
\bibitem{bedang}
B. Mohanty, \href{https://doi.org/10.1016/j.nuclphysa.2009.10.132}{Nucl. Phys. \bf{ A 830}, 899c (2009)}.; X. Luo, B. Mohanty, H. G. Ritter, and N. Xu,\href{https://doi.org/10.1134/S1063778812060348}{Phys. At. Nucl. \bf{75}, 676 (2012)}.
\bibitem{Misha11}
M. A. Stephanov,
\href{https://link.aps.org/doi/10.1103/Phys.Rev.Lett.107.052301}{Phys. Rev. Lett.\bf{ 107}, 052301 (2011)}.


\bibitem{xin}
Xin An, Marcus Bluhm, Lipei Du and Gerald V. Dunne et. All
 \href{https://doi.org/10.1016/j.nuclphysa.2021.122343}{Nuclear Physics A,1017,(2022)}.
\bibitem{Akam}
Akamatsu Yukinao and Teaney Derek and Yan Fanglida and Yin Yi
\href{https://link.aps.org/doi/10.1103/PhysRevC.100.044901}{Phys. Rev. \bf {C 100}, 044901 (2019)}.
\bibitem{vovchen}
V. Vovchenko, V. Koch and  C. Shen, 
\href{https://doi.org/10.1103/PhysRevc.105.014904}{Phys. Rev. C 105, 014904
(2022)}.
\bibitem{xan}
X. An, G. Basar, M. Stephanov, H.U. Yee, 
\href{https://doi.org/10.1103/PhysRevLett.127.072301}{Phys. Rev. Lett. {\bf 127}, 072301 (2021)}.
\href{https://doi.org/10.1103/PhysRevC.108.034910}{Phys. Rev. {\bf C 108}, 034910 (2023)}.

\bibitem{Pradeep1}
M. Pradeep, K. Rajagopal, M. Stephanov, Y. Yin, Freezing out fluctuations in Hy-
dro+ near the QCD critical point, \href{https://doi.org/10.1103/PhysRevD.106.036017}{Phys. Rev. {\bf D 106}, 036017 (2022)}; \ M.S. Pradeep, M. Stephanov, 
\href{https://doi.org/10.1103/PhysRevLett.130.162301}{Phys. Rev. Lett. {\bf 130}, 162301 (2023)}. 

\bibitem{Parot}
P. Parotto, M. Bluhm, D. Mroczek, M. Nahrgang, J. Noronha-Hostler, K. Rajagopal,
C. Ratti, T. Schäfer, M. Stephanov, 
\href{//doi.org/10.1103/PhysRevC.101.034901}{Phys. Rev. {\bf C 101}, 034901 (2020)}.
\bibitem{Pradeep2}
J.M. Karthein, M.S. Pradeep, K. Rajagopal, M. Stephanov, Y. Yin, 
Equilibrium expectations for non-Gaussian fluctuations near a QCD critical point, in 21st International
Conference on Strangeness in Quark Matter 2024 \href{ 	
https://doi.org/10.48550/arXiv.2409.16249}{arXiv:2409.16249 (nucl-th)}


\bibitem{Roder}
D. Roder,J. Ruppert and D. H. Rischke,
\href{https://doi.org/10.1103/PhysRevD.68.016003}{Phys. Rev. {\bf D 68}, 016003 (2003)}.
\bibitem{fuku11}
K. Fukushima,K. Kamikado and B. Klein,
\href{https://doi.org/10.1103/PhysRevD.83.116005}{Phys. Rev. {\bf D 83}, 116005 (2011)}.
\bibitem{grahl}
M. Grahl  and D. H. Rischke,
\href{https://doi.org/10.1103/PhysRevD.88.056014}{Phys. Rev. {\bf D 88}, 056014 (2013)}.

\bibitem{jakobi}A. Jakovac, A. Patkos, Z. Szep, and P. Szepfalusy, 
\href{https://doi.org/10.1016/j.physletb.2004.01.008}{Phys. Lett. {\bf B 582}, 179 (2004)}.

\bibitem{Herpay:05}
T. Herpay, A. Patk\'{o}s, Zs. Sz\'{e}p and P. Sz\'{e}pfalusy,
\href{https://doi.org/10.1103/PhysRevD.71.125017}{Phys. Rev. {\bf D 71}, 125017 (2005)}.


\bibitem{Herpay:06}
T. Herpay and Zs. Sz\'{e}p, 
\href{https://doi.org/10.1103/PhysRevD.74.025008}{Phys. Rev. {\bf D 74}, 025008 (2006)}.


\bibitem{Herpay:07}
P. Kov\'acs and Zs. Sz\'{e}p,
\href{https://doi.org/10.1103/PhysRevD.75.025015}{Phys. Rev. {\bf D 75}, 025015 (2007)}.


\bibitem{Kovacs:2006ym}
P.~Kovacs and Zs. Szep,
Phys. Rev. {\bf D 75}, 025015 (2007).

\bibitem{kahara}
T. Kahara and K. Tuominen, \href{https://doi.org/10.1103/PhysRevD.78.034015}{Phys. Rev. {\bf D 78}, 034015
(2008)}; \href{https://doi.org/10.1103/PhysRevD.80.114022}{80, 114022 (2009)}; \href{https://doi.org/10.1103/PhysRevD.82.114026}{82, 114026 (2010)}.

\bibitem{Bowman:2008kc}
E.~S. Bowman and J.~I. Kapusta,
\href{https://doi.org/10.1103/PhysRevC.79.015202}{Phys. Rev. {\bf C 79}, 015202 (2009)};
J.~I. Kapusta, and E.~S. Bowman,
\href{https://doi.org/10.1016/j.nuclphysa.2009.10.118}{Nucl.\ Phys.\  {\bf A 830}, 721C (2009)}.

\bibitem{Fejos}
G. Fejos, A. Patkos,
\href{https://doi.org/10.1103/PhysRevD.82.045011}{Phys. Rev. {\bf D 82}, 045011 (2010)}.


\bibitem{Jakovac:2010uy}
A.~Jakovac and Zs. Szep, 
\href{https://doi.org/10.1103/PhysRevD.82.125038}{Phys. Rev. {\bf D 82}, 125038, (2010)}. 

\bibitem{koch}
L. Ferroni, V. Koch, and M. B. Pinto, 
\href{https://doi.org/10.1103/PhysRevC.82.055205}{Phys. Rev. {\bf C 82}, 055205 (2010)}.

\bibitem{marko}
G. Marko and Zs. Szep, \href{https://doi.org/10.1103/PhysRevD.82.065021}{Phys. Rev. {\bf D 82}, 065021 (2010)}.

\bibitem{GFejo}
Gergely Fejős and András Patkós, \href{https://doi.org/10.1103/t723-m9y3}{Phys. Rev. {\bf D 112}; 7, 076013 (2025)}.
        
\bibitem{scav}	
O. Scavenuius, A. Mocsy, I. N. Mishustin, and D. H. Rischke,
\href{https://doi.org/10.1103/PhysRevC.64.045202}{Phys. Rev. C {\bf 64}, 045202 (2001)}.


\bibitem{mocsy} A. Mocsy, I. N. Mishustin, and P. J. Ellis, 
\href{https://doi.org/10.1103/PhysRevC.70.015204}{Phys. Rev. {\bf C 70}, 015204 (2004)}.

\bibitem{bj}
B.-J. Schaefer and J. Wambach, 
\href{https://doi.org/10.1016/j.nuclphysa.2005.04.012}{Nucl. Phys. {\bf A 757}, 479 (2005)}. 

\bibitem{Schaefer:2006ds}
B.-J. Schaefer and J. Wambach,
\href{https://doi.org/10.1103/PhysRevD.75.085015}{Phys. Rev. {\bf D 75}, 085015 (2007)}.



\bibitem{hjss}
A. Halasz, A. D. Jackson, R. E. Shrock, M. A. Stephanov,
and J. J. M. Verbaarschot, \href{https://doi.org/10.1103/PhysRevD.58.096007}{Phys. Rev. D {\bf 58}, 096007 (1998)}.


\bibitem{vac} 	
V. Skokov, B. Friman, E. Nakano, K. Redlich, and 
B.-J. Schaefer, 
\href{https://doi.org/10.1103/PhysRevD.82.034029}{Phys. Rev. D {\bf 82}, 034029 (2010)}.

\bibitem{Fraga1}
L. F. Palhares and E. S. Fraga, \href{https://doi.org/10.1103/PhysRevD.78.025013}{Phys. Rev. D {\bf 78}, 025013 (2008)}.
\bibitem{Fraga2}
E. S. Fraga, L. F. Palhares, and M. B. Pinto, \href{https://doi.org/10.1103/PhysRevD.79.065026}{Phys. Rev. D {\bf 79}, 065026 (2009)}.
\bibitem{Fraga3}
L. F. Palhares and E. S. Fraga, \href{https://doi.org/10.1103/PhysRevD.82.125018} {Phys. Rev. D {\bf 82}, 125018
(2010)}.




\bibitem{Gatto}
R. Gatto, M. Ruggieri,
\href{https://doi.org/10.1103/PhysRevD.82.054027}{Phys. Rev. D {\bf 82}, 054027 (2010)}.



\bibitem{Anna} 
A. J. Mizher, M. N. Chernodub, and E. S. Fraga, 
\href{https://doi.org/10.1103/PhysRevD.82.105016}{Phys. Rev. D {\bf 82}, 105016 (2010)}.

\bibitem{lars}
R. Khan and L. T. Kyllingstad, 
\href{https://doi.org/10.1063/1.3575076}{AIP  Conf. Proc. {\bf 1343}, 504 (2011)}.

\bibitem{guptiw}
U. S. Gupta, V. K. Tiwari,
\href{https://doi.org/10.1103/PhysRevD.85.014010}{Phys. Rev. D {\bf 85},  014010 (2012)}.

\bibitem{chatmoh1}
S. Chatterjee and K. A. Mohan, \href{https://doi.org/10.1103/PhysRevD.85.074018}{Phys. Rev. D {\bf 85}, 074018 (2012)}.

\bibitem{vkkr12}
V. K. Tiwari, \href{https://doi.org/10.1103/PhysRevD.86.094032}{Phys. Rev. D {\bf 86}, 094032 (2012)}.

\bibitem{schafwag12}
B.-J. Schaefer and M. Wagner, \href{https://doi.org/10.1103/PhysRevD.85.034027}{Phys. Rev. D {\bf 85}, 034027 (2012)}.

\bibitem{TranAnd}
J. O. Andersen and A. Tranberg, \href{https://doi.org/10.1007/JHEP08(2016)045}{J. High Energy Phys. {\bf 08}
(2012) 002}.

\bibitem{Dima}
D. Kharzeev, K. Landsteiner, A. Schmitt, H.-U. Yee (Eds.),Strongly Interacting Matter in Magnetic Fields,
Lecture Notes in Phys. 871,Spriner (2013)

\bibitem{vkkt13}
V. K. Tiwari, \href{https://doi.org/10.1103/PhysRevD.88.074017}{Phys. Rev. D {\bf 88}, 074017 (2013)}.


\bibitem{Weyrich}
J. Weyrich, N. Strodthoff, and L. von Smekal, \href{https://doi.org/10.1103/PhysRevC.92.015214}{Phys. Rev. C
{\bf 92}, 015214 (2015)}.


\bibitem{Weise1}
M. Drew, W. Weise, \href{https://doi.org/10.1103/PhysRevC.91.035802}{Phys. Rev. C {\bf 91}, 035802 (2015)}

\bibitem{kovacs}
P. Kovács, Zs Szép, Gy Wolf,\href{https://doi.org/10.1103/PhysRevD.93.114014}{Phys. Rev. D {\bf 93}, 114014 (2016)}.


\bibitem{zacchi1}
Andreas Zacchi and Jürgen Schaffner-Bielich, \href{https://doi.org/10.1103/PhysRevD.97.074011}{Phys. Rev. D {\bf 97}, 074011 (2018)}. 
\bibitem{zacchi2}

Andreas Zacchi and Jürgen Schaffner-Bielich, \href{https://doi.org/10.1103/PhysRevD.100.123024}{Phys. Rev. D {\bf 100}, 0123024 (2019)}.

\bibitem{Rai}
S. K. Rai and V. K. Tiwari, \href{https://doi.org/10.1140/epjp/s13360-020-00851-5}{Eur. Phys. J. Plus {\bf 135:844}, (2020)}.

\bibitem{Weise3}
L. Brandes, N. Kaiser, W. Weise, \href{https://doi.org/10.1140/epja/s10050-021-00528-2} {Eur. Phys. J. A {\bf 57:243}, (2021)}.



\bibitem{Kobes}
R. Kobes, G. Kunstatter, and A. Rebhan, \href{https://doi.org/10.1103/PhysRevLett.64.2992}{Phys. Rev. Lett.
{\bf 64}}, 2992 (1990); \href{https://doi.org/10.1016/0550-3213(91)90300-M}{Nucl. Phys. B {\bf 355}, 1 (1991)}.

\bibitem{Rebhan}
A. K. Rebhan, \href{https://doi.org/10.1103/PhysRevD.48.R3967}{Phys. Rev.  D {\bf 48}, R3967 (1993)}.
\bibitem{laine}
K. Kajantie, M. Laine, K. Rummukainen, and M. E. Shaposhnikov
\href{https://doi.org/10.1016/0550-3213(95)00549-8}{Nucl. Phys. B {\bf 458}, 90 (1996)}.

\bibitem{BubaCar}
S. Carignano, M. Buballa and B-J Schaefer 
\href{https://doi.org/10.1103/PhysRevD.90.014033}{Phys. Rev. D {\bf 90}, 014033 (2014)}.

\bibitem{fix1}
S. Carignano, M. Buballa, and W. Elkamhawy,
\href{https://doi.org/10.1103/PhysRevD.94.034023}{Phys. Rev. D {\bf 94}, 034023 (2016)}.

\bibitem{Naylor}
J. O. Andersen, W. R. Naylor, and A. Tranberg, \href{https://doi.org/10.1103/RevModPhys.88.025001}{Rev. Mod.
Phys. {\bf 88}, 025001 (2016)}.


\bibitem{Adhiand1}
P. Adhikari, J. O. Andersen and P. Kneschke, \href{https://doi.org/10.1103/PhysRevD.95.036017}{Phys. Rev. D {\bf 95}, 036017 (2017)}.
\bibitem{Adhiand2}
P. Adhikari, J. O. Andersen and P. Kneschke, \href{https://doi.org/10.1103/PhysRevD.96.016013}{Phys.Rev.D {\bf 96}, 016013 (2017)}.
\bibitem{Adhiand3}
P. Adhikari, J. O. Andersen and P. Kneschke, \href{https://doi.org/10.1103/PhysRevD.98.074016}{Phys.Rev.D {\bf 98}, 074016 (2018)}.
\bibitem{asmuAnd}
A.Folkestad, J. O. Andersen, \href{https://doi.org/10.1103/PhysRevD.99.054006}{Phys.Rev.D {\bf 99}, 054006 (2019)}. 
\bibitem{RaiTiw22}
S. K. Rai and V. K. Tiwari, \href{https://doi.org/10.1103/PhysRevD.105.094010}{Phys.Rev.D {\bf 105}, 094010 (2022)}.
\bibitem{raiti23}
S. K. Rai and V. K. Tiwari, \href{https://doi.org/10.1103/PhysRevD.108.074014}{Phys.Rev.D {\bf 108}, 074014 (2023)}.
\bibitem{vkkr23}
V. K. Tiwari, \href{https://doi.org/10.1103/PhysRevD.108.074002}{Phys. Rev. D {\bf 108}, 074002 (2023)}
\bibitem{skrvkt24}
S. K. Rai and V. K. Tiwari, 
\href{https://doi.org/10.1103/PhysRevD.109.034025}{Phys. Rev. D {\bf 109}, 034025 (2024)}



\bibitem{Joand}
Jens O. Andersen and Mathias P. Nødtvedt, arXiv:2506.02941.
\bibitem{Gholami}
Hosein Gholami, Kurth Lennart, Ugo Mire, Michael
Buballa, and Bernd-Jochen Schaefer, arXiv:2505.22542.

\bibitem{Clark}
D. A. Clarke, P. Dimopoulos, F. Di Renzo, J. Goswami, C. Schmidt, S. Singh, K. Zambello
\href{https://doi.org/10.48550/arXiv.2405.10196}{arXiv:2405.10196 [hep-lat]}
\bibitem{Basar}
G. Basar,
\href{https://doi.org/10.1103/PhysRevC.110.015203}{Phys Rev {\bf C 110}, 015203, (2024)}
\bibitem{Hipp}
M. Hippert, J. Grefa, T. A. Manning, J. Noronha, J. Noronha-Hostler, I. P. Vazquez, C. Ratti, R. Rougemont, M. Trujillo.
\href{https://doi.org/10.1103/PhysRevD.110.094006}{Phys. Rev. {\bf D 110}, 094006 (2024)}.
\bibitem{Lugao}
Y. Lu, F. Gao, Y.X. Liu, J.M. Pawlowski. 
\href{https://doi.org/10.1103/PhysRevD.110.014036}{Phys. Rev. \bf {D 110}, 014036 (2024)}
\bibitem{herrPLB}
P. Herrera-Sikl\'{o}dy, J. I. Latorre, P. pascual and J. Taron,
\href{https://doi.org/10.1016/S0370-2693(97)01507-4}{Phys. Lett. B {\bf 419}, 326 (1998)}.

\bibitem{Escribano}
R. Escribano, F. S. Ling, M. H. G. Tytgat,
\href{https://doi.org/10.1103/PhysRevD.62.056004}{Phys. Rev. D {\bf 62}, 056004 (2000)}


\bibitem{Hatsvect}
N. Ishii, S. Aoki, and T. Hatsuda
\href{https://doi.org/10.1103/PhysRevLett.99.022001}{Phys. Rev. Lett. {\bf 99} 022001 (2007)}.

\bibitem{Kitaz}
M. Kitazawa, T. Koide, T. Kunihiro, and Y. Nemoto.
\href{https://doi.org/10.1143/PTP.108.929}{Prog. Theor. Phys. {\bf 108}, 929 (2002)}.

\bibitem{Hastprl}
T. Hatsuda, Motoi Tachibana, Naoki Yamamoto, and Gordon Baym.
\href{https://doi.org/10.1103/PhysRevLett.97.122001}{Phys. Rev. Lett. {\bf 97} 122001 (2006)}.


\bibitem{HAbuki}
H. Abuki, R. Gatto and M. Ruggieri
\href{https://doi.org/10.1103/PhysRevD.80.074019}{Phy. Rev. \bf { D 80}, 074019 (2009)}


\bibitem{Kunihiro}
T. Hatsuda, T. Kunihiro 
\href{https://doi.org/10.1143/PTP.74.765}{Prog. Theor. Phys. {\bf 74}, 765 (1985)}


\bibitem{Csasaki}
C. Sasaki,B. Friman,and K. Redlich
\href{https://doi.org/10.1103/PhysRevD.75.054026}{Phys. Rev. \bf{D 75}, 054026 (2007)}




\bibitem{KJ4019}
Kenji Fukushima
\href{https://doi.org/10.1103/PhysRevD.78.114019}{Phy. Rev. \bf { D 78}, 114019 (2008)}

\bibitem{Sakai}
Yuji Sakai, Kouji Kashiwa, Hiroaki Kouno, Masayuki Matsuzaki, and Masanobu Yahiro
\href{https://doi.org/10.1103/PhysRevD.78.076007}{Phys. Rev. \bf{D 78}, 076007 (2008)}




\bibitem{Bratovic}
N.M. Bratovic, T. Hatsuda, W. Weise
\href{https://doi.org/10.1016/j.physletb.2013.01.003.}{Phys. Lett. \bf{B 719}, 131 (2013)}

\bibitem{Blaschke}
G. A. Contrera, A.G. Grunfeld, D. B. Blaschke.
\href{https://doi.org/10.1134/S1547477114040128}{Physics of Particles and Nuclei Letters \bf{ \text{Vol. }11 ,\text{No.\ 4}} 342-351, (2014)}


\bibitem{Hell}
Thomas Hell, Kouji Kashiwa and Wolfram Weise
\href{http://dx.doi.org/10.4236/jmp.2013.45093}{Journal of Modern Physics,\bf{4}, 644-650 (2013)}

\bibitem{HellII}
Kouji Kashiwa, Thomas Hell and  Wolfram Weise
\href{https://doi.org/10.1103/PhysRevD.84.056010}{Phys. Rev. \bf{D 84}, 056010 (2011)}

\bibitem{Feroni}
L. Ferroni, V. Koch, \href{https://doi.org/10.1103/PhysRevC.83.045205}{Phys. Rev. {\bf C 83}, 045205 (2011)}

\bibitem{Beisit}
Thomas Beisitzer, R. Stiele and J. Schaffner-Bielich, \href{https://doi.org/10.1103/PhysRevD.90.085001}{Phys. Rev. {\bf D 90}, 085001 (2014)}

\bibitem{RosenNPA}
S. Rößner, T. Hell, C. Ratti, W. Weise.
\href{https://doi.org/10.1016/j.nuclphysa.2008.10.006}{Nucl. Phys. A {\bf 814}, 118 (2008)}.

\bibitem{Mintz}
B. W. Mintz, R. Stiele, Rudnei O. Ramos, and J. Schaffner-Bielich, \href{https://doi.org/10.1103/PhysRevD.87.036004}{Phys. Rev. {\bf D 87}, 036004 (2012)}.

\bibitem{Hansen:2019lnf}
 Hubert Hansen, Rainer Stiele and Pedro Costa, 
\href{https://doi.org/10.1103/PhysRevD.101.094001}{Phys. Rev. \bf{D 101}, 094001 (2020)}



\bibitem{Hidaka}
Kenji Fukushima and Yoshimasa Hidaka
\href{https://doi.org/10.1103/PhysRevD.75.036002}{Phys. Rev. \bf{D 75}, 036002 (2007)}

\bibitem{KarshPRL}
F. Karsch and H. W. Wyld
\href{https://doi.org/10.1103/PhysRevLett.55.2242}{Phys. Rev. Lett. \bf{55}, 2242(1985)}

\bibitem{Dumitru}
A. Dumitru, R. D. Pisarski, and D. Zschiesche, Phys. Rev.
D 72, 065008 (2005).
\href{https://doi.org/10.1103/PhysRevD.72.065008}{Phys. Rev. \bf{D 72}, 065008 (2005)}

\bibitem{NishimuraI}
Hiromichi Nishimura,Michael C. Ogilvie and Kamal Pangeni
\href{https://doi.org/10.1103/PhysRevD.90.045039}{Phys. Rev. \bf{D 90}, 045039 (2014)}

\bibitem{NishimuraII}
Hiromichi Nishimura,Michael C. Ogilvie and Kamal Pangeni
\href{https://doi.org/10.1103/PhysRevD.91.054004}{Phys. Rev. \bf{D 91}, 054004 (2015)}

\bibitem{Tanizaki}
Yuya Tanizaki, Hiromichi Nishimura, and Kouji Kashiwa
\href{https://doi.org/10.1103/PhysRevD.91.101701}{Phys. Rev. \bf{D 91}, 0101701(R) (2015)}

\bibitem{Mori}
Kouji Kashiwa,Yuto Mori, and Akira Ohnishi
\href{https://doi.org/10.1103/PhysRevD.99.114005}{Phys. Rev. \bf{D 99}, 114005 2019}


\bibitem{Pelaez}
J.~R.~Pelaez, \href{https://doi.org/10.1016/j.physrep.2016.09.001}{\emph{Phys. Rep.} \textbf{658} (2016) 1}.

\bibitem{Ghosh:2014}S.~K. Ghosh, A.~Lahiri, S.~Majumder, M.~G. Mustafa, S.~Raha, and R.~Ray
\href{https://doi.org/10.1103/PhysRevD.90.054030}{Phys. Rev. \bf{D 90}, 054030 (2014)}

\bibitem{Gholami:2025}
Hosein Gholami \href{https://doi.org/10.48550/arXiv.2501.05192}{arXiv:2501.05192 [hep-ph]}.




\bibitem{xiong}
Juan Xiong, Meng Jin, and Jiarong Li
\href{https://doi.org/10.1103/PhysRevC.83.025204}{Phys. Rev. \bf {C 83}, 025204 (2011)}


\bibitem{Rapp}
R. Rapp T. Schäfer E. Shuryak and M. Velkovsky.
\href{https://doi.org/10.1103/PhysRevLett.81.53}{Phys. Rev. Lett. {\bf 81} 53 (1998)}.

\bibitem{DutraI}
O. Lourenço, M. Dutra,T. Frederico, A. Delfino, and M. Malheiro
\href{10.1103/PhysRevD.85.097504}{Phys. Rev. \bf{D 85}, 097504 (2012)}

\bibitem{DutraII}
M. Dutra,O. Lourenço, A. Delfino,T. Frederico and M. Malheiro
\href{10.1103/PhysRevD.88.114013 }{Phys. Rev. \bf{D 88}, 0114013 (2013)}

\bibitem{Frise}
A. V. Friesen, Yu. L. Kalinovsky and V. D. Toneev
\href{ https://doi.org/10.1142/S0217751X1550089X}{International Journal of Modern Physics \bf{A 30}, No. 16 1550089 (2015)}



\bibitem{OhnishiPLB}
A. Ohnishi, H. Ueda, T.Z. Nakano, M. Ruggieri, K. Sumiyoshi
\href{https://doi:10.1016/j.physletb.2011.09.018}{Phys. Lett. \bf{B 709}, 284 (2011)}

\bibitem{Ueda}
H. Ueda,T. Z. Nakano, A. Ohnishi, M. Ruggieri, and K. Sumiyoshi
\href{https://doi.org/10.1103/PhysRevD.88.074006}{Phys. Rev. \bf{D 88}, 074006 (2013)}


\bibitem{Zacchi15}
A. Zacchi, R. Stiele, J. Schaffner-Bielich, 
\href{https://doi.org/10.1103/PhysRevD.92.045022 }{Phys. Rev.\bf{ D 92}, 045022 (2015)}






\bibitem{KovaNstar}
Péter Kovács, János Takátsy, Jürgen Schaffner-Bielich and György Wolf
\href{https://doi.org/10.1103/PhysRevD.105.103014}{Phys.Rev. \bf{D 105}, 103014 (2022)}



\bibitem{Akakn}
Akanksha Tripathi, Suraj Kumar Rai, and Vivek Kumar Tiwari
https://doi.org/10.48550/arXiv.2512.13398


















\end{thebibliography}


\end{document}